\documentclass[%
notitlepage,
twocolumn,
superscriptaddress,
 amsmath,amssymb,
longbibliography
]{revtex4-2}

\usepackage{graphicx}% Include figure files
\usepackage{dcolumn}% Align table columns on decimal point
\usepackage{bm}% bold math
\usepackage{units}

\usepackage{cancel}

\usepackage{mleftright}

\usepackage{subfigure}
\usepackage{xcolor}
\usepackage{mathtools}
\usepackage{bbold}
\usepackage{rotating}
\usepackage{framed}
\colorlet{shadecolor}{gray!20}

\usepackage{empheq}

\usepackage[most]{tcolorbox}

\tcbset{colback=yellow!10!white, colframe=black!50!black, 
        highlight math style= {enhanced, %<-- needed for the ’remember’ options
            colframe=black,colback=red!10!white,boxsep=0pt}
        }

\newcommand*{\Pol}[1]{P_{#1}^{}}
\newcommand*{\Poltwo}[2]{P{\vphantom{P}}_{#1}^{#2}}
\newcommand*{\Polstar}[1]{P_{#1}^{*}}
\newcommand*{\Polstartwo}[2]{{P^*_{}}{\vphantom{P}}_{#1}^{#2}}

\newcommand*{\Nex}[1]{N_{#1}^{}}
\newcommand*{\Bex}[1]{B_{#1}^{}}
\newcommand*{\Bextwo}[2]{B\vphantom{B}_{#1}^{#2}}
\newcommand*{\Nextwo}[2]{N\vphantom{N}_{#1}^{#2}}

\newcommand*{\Rex}[1]{Z_{#1}^{}}
\newcommand*{\Rextwo}[2]{Z\vphantom{Z}_{#1}^{#2}}

\newcommand*{\vdag}[1]{\hat v^{\dagger}_{#1}}
\newcommand*{\vndag}[1]{\hat v_{#1}^{\vphantom{\dagger}}}
\newcommand*{\cdag}[1]{\hat c^{\dagger}_{#1}}
\newcommand*{\cndag}[1]{\hat c_{#1}^{\vphantom{\dagger}}}

\newcommand*{\vdagtwo}[2]{{\hat v^{\dagger}_{}}{\vphantom{v}}_{#1}^{#2}}
\newcommand*{\vndagtwo}[2]{{\hat v_{}^{\vphantom{\dagger}}}{\vphantom{v}}_{#1}^{#2}}
\newcommand*{\cdagtwo}[2]{{\hat c_{}^{\dagger}}{\vphantom{c}}_{#1}^{#2}}
\newcommand*{\cndagtwo}[2]{{\hat c_{}^{\vphantom{\dagger}}}{\vphantom{c}}_{#1}^{#2}}

\newcommand{\ExWFtwo}[2]{\varphi_{#1}^{#2}}
\newcommand{\ExWFstartwo}[2]{\varphi^*{\vphantom{\varphi}}_{#1}^{#2}}

\newcommand{\ExWF}[1]{\varphi_{#1}^{}}

\newcommand{\BiexWFtwo}[2]{\Phi_{#1}^{#2}}
\newcommand{\BiexWFstartwo}[2]{\Phi^*{\vphantom{\varphi}}_{#1}^{#2}}

\usepackage{placeins}
\usepackage{appendix}
\setcitestyle{numbers,square}

\definecolor{custom}{RGB}{15,45,150}
\usepackage[urlcolor=blue]{hyperref}
\hypersetup{colorlinks=true,linktoc=page,citecolor = blue,linkcolor=blue}

\begin{document}
%\newgeometry{textwidth=10cm, left = 1.5cm,top=2cm,bottom=2cm}

\preprint{APS/123-QED}
\title{Excitonic Theory of the Ultrafast Optical Response of 2D-Quantum-Confined Semiconductors at Elevated Densities}
\author{Henry Mittenzwey}
\email{henry.mittenzwey@uni-giessen.de}
\affiliation{Institut für Theoretische Physik und Zentrum für Materialforschung, Justus-Liebig-Universität Gie{\ss}en, 
%Heinrich-Buff-Ring 16, 
35392 Gie{\ss}en, Germany}
\affiliation{Nichtlineare Optik und Quantenelektronik, Institut f\"ur Physik und Astronomie (IFPA), Technische Universit\"at Berlin, 10623 Berlin, Germany}
\author{Oliver Voigt}
\affiliation{Nichtlineare Optik und Quantenelektronik, Institut f\"ur Physik und Astronomie (IFPA), Technische Universit\"at Berlin, 10623 Berlin, Germany}
\author{Andreas Knorr}
\affiliation{Nichtlineare Optik und Quantenelektronik, Institut f\"ur Physik und Astronomie (IFPA), Technische Universit\"at Berlin, 10623 Berlin, Germany}

\begin{abstract}
    %\centering\begin{minipage}{\dimexpr\paperwidth-13cm}
    An excitonic approach to the ultrafast optical response of confined semiconductors at elevated densities below the Mott transition is presented. The theory is valid from the coherent regime, where coherent excitonic transitions and biexcitons dominate, to the incoherent regime, where excitonic occupations dominate.
    Numerical simulations of the $1s$ exciton dynamics during intense circularly polarized pump pulses in two different Coulomb-interaction regimes are performed for two-dimensional semiconductors: Moderate Coulomb interaction is compared with dominating Coulomb interaction with respect to the light-matter interaction strength. 
    The different many-body contributions are disentangled and it is found, that excitonic Rabi oscillations in the Coulomb-dominated regime are considerably less strong. By also comparing circular and linear excitation in a MoSe$_2$ monolayer, it is found, that linear excitation creates a regime, where excitonic Rabi oscillations are almost completely suppressed.
    %\end{minipage}
\end{abstract}
\date{\today}
\maketitle

%\tableofcontents
\section{Introduction}

Rabi oscillations \cite{rabi1937space}, while first having been observed in atomic two-level systems \cite{knight1980rabi}, later in quantum dots \cite{danckwerts2006theory,forstner2003phonon} or assisted by optical cavities \cite{dominici2014ultrafast}, also occur in confined semiconductors such as quantum wells without an optical cavity \cite{schulzgen1999direct,giessen1998self,cundiff1994rabi,kuhn1992analysis,binder1990ultrafast}. 
In these structures, however, Rabi oscillations are not only determined by the usual Pauli blocking but also by Coulomb interactions between optically excited electrons and holes.
Rabi oscillations in quantum wells have been successfully predicted by the semiconductor Bloch equations in Hartree-Fock limit (SBE) and observed in experiments \cite{lindberg1988effective,schulzgen1999direct,binder1999many}. However, if -- in materials exhibiting strong electron-hole interactions -- higher correlations beyond the Hartree-Fock limit and incoherent interaction with the phonon bath need to be taken into account, the numerical effort of the SBEs quickly increases \cite{steinhoff2025wannier}. 
Also, the Hartree-Fock-limit lacks any possibility to resolve correlated exciton-exciton effects \cite{takayama2002t,buck2004light,steinhoff2018biexciton,katsch2019theory,katsch2020exciton,tang2025theoretical} 
or an accurate description of intraband processes \cite{dignam2014excitonic,parks2013excitonic,axt1996intraband}. 
Therefore, for material parameters, where Coulomb interaction dominates over the light-matter interaction, such as atomically thin transition metal dichalcogenide monolayers, an excitonic theory, i.e., a formulation in electron-hole pair operators \cite{katsch2018theory,axt1994dynamics,victor1995hierarchy}, is better suited.

In the past, several excitonic theories valid in the regime of elevated densities have been proposed: By using a momentum-independent formulation using the Usui transformation \cite{usui1960excitations,rochat2000excitonic,wang2009excitonic,wang2007excitonic}, by using a localized Wannier basis \cite{axt2001evidence}, by employing a strict coherent limit \cite{ostreich1994nonperturbative,axt2001influence,katsch2020exciton,katsch2019theory} or via a bosonic non-equilibrium Green's function method focusing on the dynamics of optically inactive excitons via nonlinear light-matter interaction \cite{yang2004ultrafast}. Other approaches regarding exciton dynamics in optical cavities introduce additional Hamiltonians to account for fermionic Pauli-blocking in an otherwise bosonic excitonic description \cite{takemura2015dephasing,takemura2016coherent,combescot2007polariton} or stay in a strict coherent limit \cite{rose2025microscopic,denning2022efficient,denning2022bichromatic}. 

Hence, a comprehensive and %rigorous theory
microscopic momentum-resolved theory for Wannier-Mott excitons \cite{wannier1937structure} developing a consistent description of simultaneous Coulomb, light-matter and exciton-phonon interaction in an undoped quantum-confined semiconductor, which is able to bridge the gap between the coherent \cite{takayama2002t,buck2004light,katsch2019theory,katsch2020exciton,deckert2025coherent} and incoherent regime \cite{thranhardt2000quantum,selig2018dark,brem2020phonon,katzer2023exciton,policht2023time} in the low-density limit as well as in the elevated-density regime beyond the third-order susceptibility \cite{katzer2024fermionic,katzer2023exciton,erkensten2021dark,steinhoff2021microscopic}, is still missing.

In this manuscript, we develop an excitonic approach for Wannier excitons in a few-band effective-mass model at the symmetry points of the Brillouin zone up to fourth-order dynamics-controlled truncation (DCT) \cite{axt1994dynamics,victor1995hierarchy} for up to three-exciton correlations needed to describe the optical response from the coherent to the incoherent regime at elevated exciton densities.

The manuscript is organized as follows: In Sec.~\ref{sec:Hamiltonians}, we lay out the Hamiltonians of the considered interaction mechanisms. 
In Sec.~\ref{sec:excitonic_equations_of_motion}, we develop the exciton Bloch equations and discuss all contributions up to fourth order DCT. 
In Sec.~\ref{sec:rabi_oscillations}, we study the exciton Rabi-flopping dynamics within intense pump pulses of extended duration for two different Coulomb regimes in circular excitation: A GaAs quantum well (QW) and a h-BN encapsulated MoSe$_2$ monolayer (ML) in circular excitation and additionally compare circular and linear excitation in a MoSe$_2$ ML.

\section{Hamiltonians}
\label{sec:Hamiltonians}
The total Hamiltonian $\hat H$ reads \cite{lindberg1988effective,haug2009quantum,jahnke1997linear}:
\begin{multline}
    \hat H = \hat H_0 + \hat H_{\text{Coul-eh}} + \hat H_{\text{Coul-ee}} + \hat H_{\text{Coul-hh}} + \hat H_{\text{l-m}}.%\\
    %+ \hat H_{\text{e-phon}}.
    \label{eq:hamiltonian_total}
\end{multline}
$\hat H_0$ is the free Hamiltonian:
\begin{align}
    \hat H_0 = \sum_{\mathbf k,\xi}\left(E_{v,\mathbf k}^{\xi}\vdagtwo{\mathbf k}{\xi}\vndagtwo{\mathbf k}{\xi} + E_{c,\mathbf k}^{\xi}\cdagtwo{\mathbf k}{\xi}\cndagtwo{\mathbf k}{\xi}\right),
\end{align}
where $E_{v/c,\mathbf k}^{\xi}$
$E_{v/c,\mathbf k}$ is the valence/conduction band dispersion in effective-mass approximation $E_{v/c,\mathbf k}^{\xi} = E_{v/c}^{\xi} \mp \frac{\hbar^2\mathbf k^2}{2m_{h/e}^{\xi}}$ with valence/conduction band edge $E_{v/c}^{\xi}$ and hole/electron masses $m_{h/e}^{\xi}$ at momentum $\mathbf k$ and spin-valley index $\xi$. 
$v^{(\dagger)}$/$c^{(\dagger)}$ are the valence/conduction band annihilation (creation) operators. The Coulomb Hamiltonian \cite{jahnke1997linear} consists of an electron-hole contribution $\hat H_{\text{Coul-eh}}$:
\begin{align}
     \hat H_{\text{Coul-eh}} = \sum_{\mathbf k,\mathbf k^{\prime},\mathbf q,\xi,\xi^{\prime}} V_{\mathbf q}\vdagtwo{\mathbf k+\mathbf q}{\xi}\cdagtwo{\mathbf k^{\prime}-\mathbf q}{\xi^{\prime}}\cndagtwo{\mathbf k^{\prime}}{\xi^{\prime}}\vndagtwo{\mathbf k}{\xi},
\end{align}
an electron-electron contribution $\hat H_{\text{Coul-ee}}$:
\begin{align}
     \hat H_{\text{Coul-ee}} = \frac{1}{2} \sum_{\mathbf k,\mathbf k^{\prime},\mathbf q,\xi,\xi^{\prime}} V_{\mathbf q}\cdagtwo{\mathbf k+\mathbf q}{\xi}\cdagtwo{\mathbf k^{\prime}-\mathbf q}{\xi^{\prime}}\cndagtwo{\mathbf k^{\prime}}{\xi^{\prime}}\cndagtwo{\mathbf k}{\xi},
\end{align}
and a hole-hole contribution $\hat H_{\text{Coul-hh}}$:
\begin{align}
     \hat H_{\text{Coul-hh}} = \frac{1}{2}\sum_{\mathbf k,\mathbf k^{\prime},\mathbf q,\xi,\xi^{\prime}} V_{\mathbf q}\vdagtwo{\mathbf k+\mathbf q}{\xi}\vdagtwo{\mathbf k^{\prime}-\mathbf q}{\xi^{\prime}}\vndagtwo{\mathbf k^{\prime}}{\xi^{\prime}}\vndagtwo{\mathbf k}{\xi}.
\end{align}
Here, $V_{\mathbf q} = \frac{e^2}{\mathcal A}\int\mathrm dz\,\mathrm dz^{\prime}\,\left|\zeta(z)\right|^2G_{\mathbf q}\mleft(z,z^{\prime}\mright)\left|\zeta(z^{\prime})\right|^2$ is the screened quantum-confined Coulomb potential, where $e$ is the elementary charge, $\mathcal A$ is the sample area, $\zeta(z)$ is the confinement wave function and $G_{\mathbf q}(z,z^{\prime})$ is the Green's function solving the generalized Poisson equation of the corresponding sample geometry including the dielectric environment, cf.\ Eq.~(S68) in the Supplementary Material (SM). Coulomb electron-hole exchange interaction -- not to be confused with the usual contributions due to fermionic exchange -- \cite{qiu2015nonanalyticity,yu2014valley,combescot2023ab}, which causes corrections to the exciton \cite{qiu2015nonanalyticity} or biexciton dispersion \cite{kwong2021effect}, intervalley scattering \cite{selig2019ultrafast,selig2020suppression} and the biexciton fine structure \cite{steinhoff2018biexciton,torche2021biexcitons}, as well as Auger scattering \cite{steinhoff2021microscopic,erkensten2021dark} and Dexter interaction \cite{dogadov2026diss,berghauser2018inverted,bernal2018exciton}, are neglected, since we focus on the most dominant (direct electron-hole, electron-electron and hole-hole) Coulomb interaction processes needed to explain recent experiments \cite{schafer2025distinct}. However, these interaction processes can be straightforwardly included by adding the corresponding Hamiltonians.

The light-matter interaction Hamiltonian in dipole approximation reads:
\begin{align}
     \hat H_{\text{l-m}} = - \sum_{\mathbf k,\mathbf Q,\xi} \left(\mathbf E_{\mathbf Q}\cdot\mathbf d^{cv}\cdagtwo{\mathbf k+\mathbf Q}{\xi}\vndagtwo{\mathbf k}{\xi}+\text{h.a.}\right).
     \label{eq:light_matter_hamiltonian}
\end{align}
Here, $\mathbf E_{\mathbf Q} = \frac{1}{\mathcal A}\int\mathrm dz\,|\zeta(z)|^2\mathbf E_{\mathbf Q}(z)$ is the quantum-confined optical field, where $\mathbf E_{\mathbf Q}(z)$ solves the wave equation in momentum space of the corresponding sample geometry \cite{knorr1996theory} 
%, cf.\ Eq.~(S94) in the SM, 
and $\mathbf d^{cv}$ is the transition dipole moment in low-wavenumber approximation \cite{aversa1995nonlinear,gu2013relation}. As we excite excitonic states just below the free-particle band gap, we neglect any intraband contributions \cite{huttner2017ultrahigh}. Note, that intraband light-matter interaction can induce light-driven transport of optically excited electron-hole pairs \cite{oliaei2021transition}, intraexcitonic transitions \cite{rice2013observation,stroucken2021magnetic} or dynamical localization \cite{meier2000coherent,meier1995dynamic}, even throughout the total Brillouin zone if the exciting pulse is strong enough \cite{langer2018lightwave}, constituting the key ingredient for lightwave electronics \cite{borsch2023lightwave}. However, in such a regime, the effective-mass approximation and, hence, a projection on excitonic states solving the Wannier equation, cf.~Eq.~\eqref{eq:P}, is not possible anymore, since we would have to take into account the full connected band structure throughout the total Brillouin zone. 

We note, that we disregard any exciton-phonon scattering causing incoherent scattering and thermalization \cite{thranhardt2000quantum,selig2018dark,brem2020phonon,katzer2023exciton,katzer2024fermionic}, since the goal of this work is to study the impact of exciton-exciton interactions via Coulomb interaction in its pure form.

\section{Excitonic Equations of Motion}
\label{sec:excitonic_equations_of_motion}

The excitonic equations of motion are derived as follows. 
We start by deriving Heisenberg's equations of motion $\mathrm i\hbar\partial_t \hat A = [\hat A,\hat H]$ 
for the electron-hole pair operators $\hat A$ of interest, which can contain an arbitrary but equal number of conduction/valence band annihilation (creation) operators $c^{(\dagger)}$/$v^{(\dagger)}$. However, after evaluating the commutator $[\hat A,\hat H]$, the appearing operator products on the right-hand side of the corresponding equation of motion do not necessarily contain an equal number of electron and hole operators. Therefore, we project all operator products, which contain an unequal number of electron and hole operators onto electron-hole pairs via the unit-operator method \cite{katsch2018theory,ivanov1993self}. E.g., electron and hole occupations are expanded according to (using a compound index $i = \{\mathbf k_i,\xi_i\}$):
\begin{align}
\begin{split}
    \cdag{1}\cndag{2} = \sum_3\cdag{1}\vndag{3}\vdag{3}\cndag{2} \,- &\,\frac{1}{2}\sum_{3,4,5} \cdag{1}\vndag{3}\cdag{4}\vndag{5}\vdag{5}\cndag{4}\vdag{3}\cndag{2}\\
    &+ \mathcal O(Na_{\text{B}}^2)^3,\\
    \vndag{1}\vdag{2} = \sum_3\cdag{3}\vndag{1}\vdag{2}\cndag{3} \,- &\,\frac{1}{2}\sum_{3,4,5}\cdag{3}\vndag{1}\cdag{4}\vndag{5}\vdag{5}\cndag{4}\vdag{2}\cndag{3}\\
    &+ \mathcal O(Na_{\text{B}}^2)^3.
    \end{split}
    \label{eq:unit_operator_method_example}
\end{align}
Note, that the unit-operator method corresponds exactly to the contraction theorem in Ref.~\cite{victor1995hierarchy}. 
Then, we evaluate the expectation values via the correlation expansion technique \cite{fricke1996transport,kira2006many}. 
Within this framework, the expectation values of the operator products are expanded as follows ($ \hat A_i$ is an arbitrary fermionic creation or annihilation operator):
\begin{align}
    \begin{split}
    \langle \hat A_1\hat A_2\rangle = &\, \langle \hat A_1\hat A_2\rangle_{\text{c}}^{}\quad\text{(singlet)},\\
    \langle  \hat A_1 \hat A_2  \hat A_3 \hat A_4\rangle = &\, \langle  \hat A_1 \hat A_2  \hat A_3 \hat A_4\rangle_{\text{c}}^{} \quad \text{(doublet)}\\
    &+ \langle  \hat A_1 \hat A_2\rangle_{\text{c}}^{}\langle \hat A_3  \hat A_4\rangle_{\text{c}}^{} \\
    &- \langle  \hat A_1 \hat A_3\rangle_{\text{c}}^{}\langle \hat A_2  \hat A_4 \rangle_{\text{c}}^{}\\
    &+ \langle  \hat A_1 \hat A_4\rangle_{\text{c}}^{}\langle \hat A_2  \hat A_3\rangle_{\text{c}}^{}\\
    \langle  \hat A_1 \hat A_2  \hat A_3 \hat A_4 \hat A_5 \hat A_6\rangle = &\, \langle  \hat A_1 \hat A_2  \hat A_3 \hat A_4 \hat A_5 \hat A_6\rangle_{\text{c}} \quad \text{(triplet)}\\
    &+ \dots
    \end{split}
    \label{eq:correlation_expansion}
\end{align}
Here, the subscript ``$\text{c}$'' denotes the correlated part of the expectation value on the corresponding operator level, which cannot be factorized in lower contributions anymore. The sign of each term is determined by the amount of commutations necessary to disentangle the operator products. The correlated expectation values behave antisymmetric with respect to pairwise exchange:
\begin{align}
    \langle \dots  \hat A_i \hat  A_j \dots\rangle_{\text{c}}^{} = -\langle \dots  \hat A_j  \hat A_i \dots\rangle_{\text{c}}^{}.
\end{align}
We formulate our theory always in the truly correlated quantities $\langle\hat A\rangle_{\text{c}}^{}$, which are obtained by subtracting all lower contributions from the total expectation value \cite{fricke1996transport,hoyer2002pair}:
\begin{align}
    \langle\hat A\rangle_{\text{c}}^{} = \langle\hat A\rangle - \text{(all lower correlations)}.
\end{align}
This definition provides the clearest physical picture, since it ensures, that the corresponding quantum kinetics only include fully correlated contributions on the corresponding level (singlet, doublet, triplet etc.). In particular, it ensures a clear separation of coherent and incoherent excitonic effects. 
To determine the relevant microscopic quantities, 
we perform a dynamics-controlled truncation (DCT) \cite{axt1994dynamics,victor1995hierarchy}. The DCT is a method to systematically truncate the quantum-mechanical hierarchy problem, where a specific correlation order is linked to a certain order of the optical field $\mathbf E$. Notably, the Heisenberg-equations-of-motion method, Eq.~\eqref{eq:unit_operator_method_example}, provides the possibility to classify the dynamics of an occurring expectation value of an arbitrary operator $\hat{A}^{(n)}$ containing a total number $n$ of valence or conduction band creation and annihilation operators (whichever number is higher) in powers of the optical field $m$:
\begin{align}
    \mathrm i\hbar\partial_t\langle\hat{A}^{(n)}\rangle = \mathcal O(\mathbf E^m),
    \label{eq:DCT_general_theorem}
\end{align}
where $m\geq n$. More specifically, $m$ is the \textit{dynamical} field order of a given equation of motion of $\hat A^{(n)}$ and $n$ is a measure of the \textit{electron-hole pair} order of $\hat A^{(n)}$. $m$ and $n$ are not necessarily equal. Therefore, if the overall optical response of a semiconductor up to a certain order $m$ in a perturbation approach of nonlinear optics within the optical field is required, one first has to fix the desired dynamical order $m$ and subsequently collects all occurring (correlated) expectation values of order $n\leq m$. To determine the order $m$ of the dynamical equation of a specific expectation value, one counts all occurring electron-hole pairs and optical fields on the right-hand side of the corresponding Heisenberg equations of motion.

For example, if we are interested in the second-order optical response, we fix $m=2$ and collect all correlated expectation values of order $n\leq 2$, which is the excitonic transition $P = \langle \vdag{}\cndag{}\rangle_{\text{c}}^{}$ with $n=1$, the excitonic occupation $N=\langle \cdag{}\vndag{} \vdag{}\cndag{}\rangle_{\text{c}}^{}$ with $n=2$ and the two-exciton transition $B=\langle \vdag{}\cndag{}\vdag{}\cndag{}\rangle_{\text{c}}^{}$ with $n=2$, where we use a short-hand notation without indices: The full expressions can be found in Eq.~\eqref{eq:P}, Eq.~\eqref{eq:N} and Eq.~\eqref{eq:B}, respectively. Then, we calculate the corresponding equations of motion of $P$, $B$ and $N$, which we truncate by using the condition $m=2$ on the right-hand side. Schematically, the equations of motion read:
\begin{align}
\begin{split}
    \mathrm i\hbar \partial_t P = &\, EP - \hbar\Omega,\\
    \mathrm i\hbar\partial_t N = &\, 0,\\
    \mathrm i\hbar\partial_t B = &\, E_{B}B + VP^2,
    \end{split}
    \label{eq:dct_second_order_example}
\end{align}
where $E$ is the excitonic energy, $\Omega$ is the Rabi frequency induced by the exciting optical field $\mathbf E$, $E_B$ is the two-exciton energy and $V$ is the Coulomb-induced exciton-exciton interaction. Again, the corresponding full versions of Eq.~\eqref{eq:dct_second_order_example} can be found in Eq.~\eqref{eq:P_EOM_first_order}, Eq.~\eqref{eq:N_EOM_second_order} and Eq.~\eqref{eq:B_EOM_second_order}. Since the optical response is fully characterized by the dynamics of the excitonic transitions $P$ and higher-order correlations $N$ and $B$ do not couple back to the excitonic transitions $P$ in the first line in Eq.~\eqref{eq:dct_second_order_example}, the optical response in second order is identical to the first order.

After we set up all equations of motion in the electron-hole picture, we perform an expansion in excitonic wave functions afterwards. This is crucial, since approaches, which apply an excitonic picture already on the Hamiltonian level are prone to miss important 
corrections due to the fermionic substructure of the excitons \cite{combescot2007exciton} within higher orders of the optical field.

\begin{figure}
    \centering
    \includegraphics[width=0.49\linewidth]{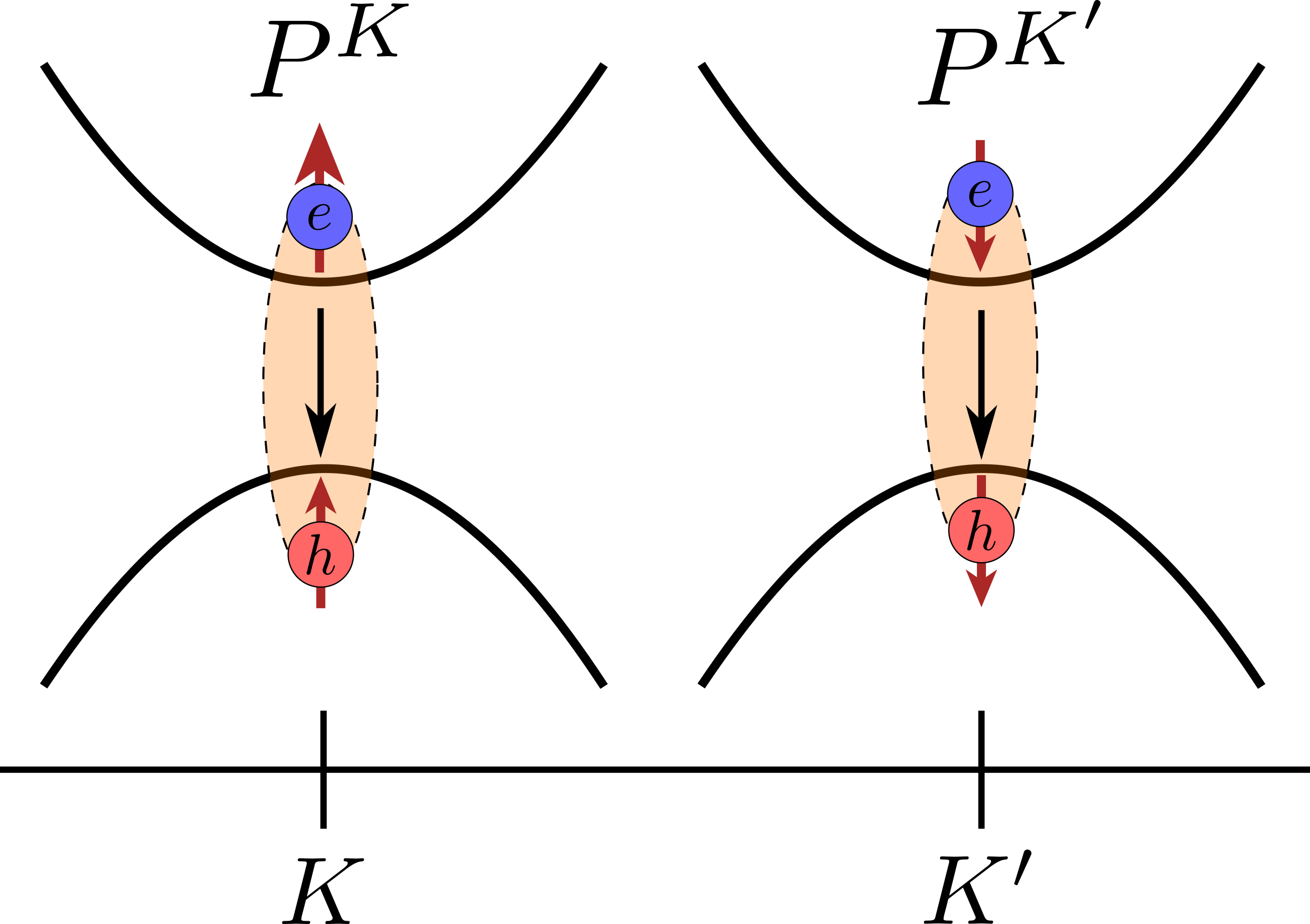}
    \includegraphics[width=0.49\linewidth]{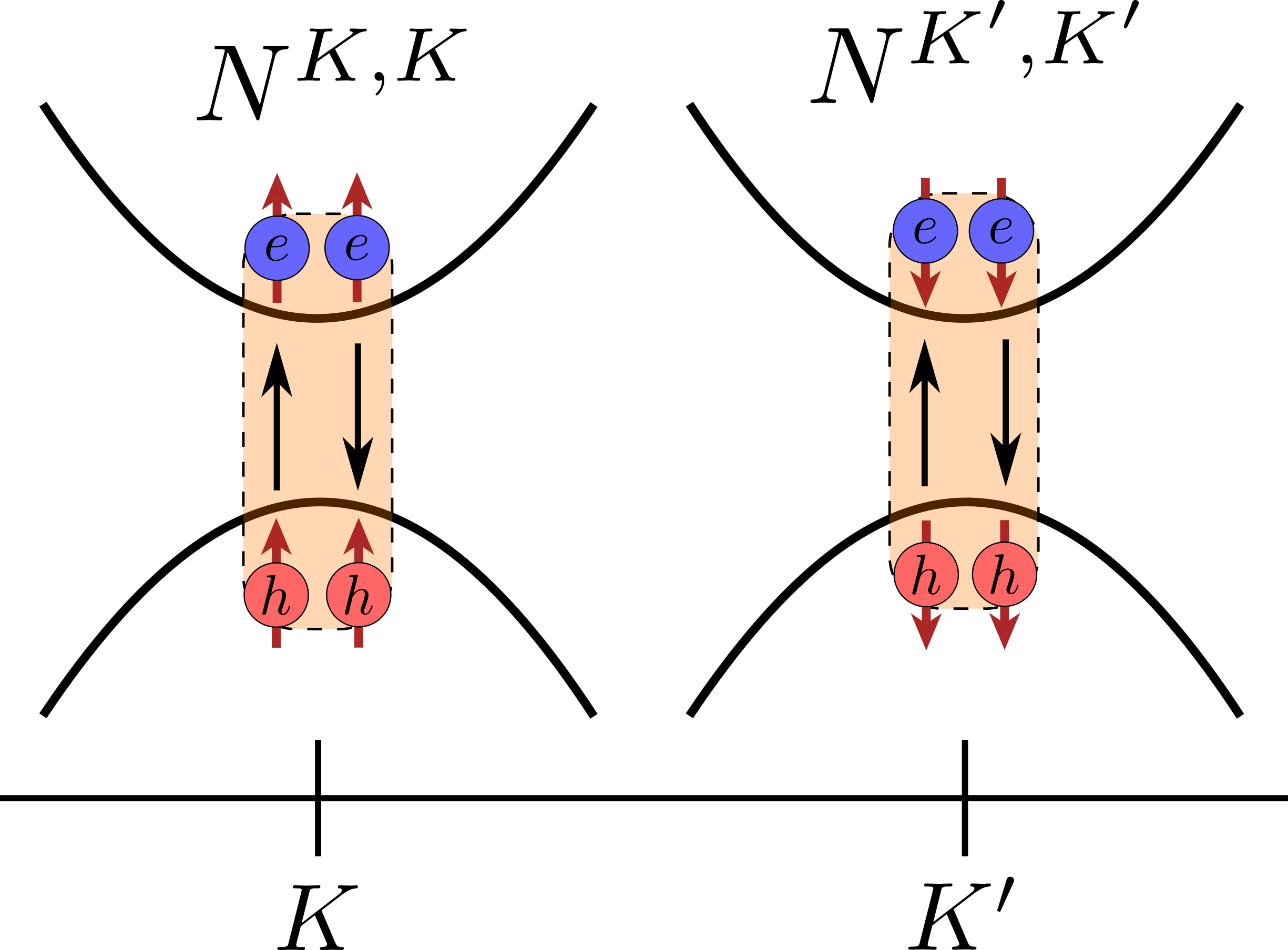}
    \caption{Scheme of excitonic transitions in Eq.~\eqref{eq:P} at the $K$ and $K^{\prime}$ valley (left) and excitonic occupations from Eq.~\eqref{eq:N} in intravalley configuration at the $K$ and $K^{\prime}$ valleys (right) in a TMDC ML.}
    \label{fig:P_N_scheme}
\end{figure}

In this work, we develop the excitonic equations of motion up to fourth order, i.e., $m\leq4$, in the optical field up to triplets or six-particle correlations and neglect quadruplets, i.e., eight-particle correlations. In this limit, the relevant quantities are the excitonic transition $P$, cf.\ Fig.~\ref{fig:P_N_scheme}(left) of order $n=1$:
\begin{align}
    P \sim \langle \hat A^{(1)} \rangle_{\text{c}}^{} = \langle \vdag{1}\cndag{2} \rangle_{\text{c}}^{}  = \langle \vdag{1}\cndag{2} \rangle,
\end{align}
which in the excitonic picture explicitly reads:
\begin{align}
    \Poltwo{\mu}{\xi} = \sum_{\mathbf q}\ExWFstartwo{\mu,\mathbf q}{\xi,\xi}\langle \vdagtwo{\mathbf q}{\xi}\cndagtwo{\mathbf q}{\xi} \rangle_{\text{c}}^{},
    \label{eq:P}
\end{align}
where $\ExWFtwo{\mu,\mathbf q}{\xi,\xi^{\prime}}$ is the excitonic wave function solving the Wannier equation with excitonic quantum number $\mu$, relative momentum $\mathbf q$, hole configuration $\xi$ and electron configuration $\xi^{\prime}$ denoting different spins and symmetry points in the Brillouin zone, such as $K$, $K^{\prime}$ in atomically thin semiconductors, cf.\ Eq.~(S1). Note, that we only consider optically excited excitonic transitions with a center-of-mass momentum of zero.

The excitonic occupations/intraexcitonic coherences ($\mu=\nu$/$\mu\neq\nu$) of order $n=2$, cf.\ Fig.~\ref{fig:P_N_scheme} (right), are defined as:
\begin{align}
\begin{split}
    N \sim \langle \hat A^{(2)} \rangle_{\text{c}}^{} &= \langle \cdag{1}\vndag{2}\vdag{3}\cndag{4} \rangle_{\text{c}}^{}  \\
    &= \langle \cdag{1}\vndag{2}\vdag{3}\cndag{4} \rangle - \langle \cdag{1}\vndag{2}\rangle \langle\vdag{3}\cndag{4} \rangle,
    \end{split}
\end{align}
which in the excitonic picture explicitly read:
\begin{multline}
    \Nextwo{\mu,\nu,\mathbf Q}{\xi,\xi^{\prime}} = \sum_{\mathbf q,\mathbf q^{\prime}}\ExWFtwo{\mu,\mathbf q-\alpha_{\xi,\xi^{\prime}}\mathbf Q}{\xi,\xi^{\prime}}\ExWFstartwo{\nu,\mathbf q^{\prime}+\beta_{\xi,\xi^{\prime}}\mathbf Q}{\xi,\xi^{\prime}}\\
    \times\langle\cdagtwo{\mathbf q}{\xi^{\prime}}\vndagtwo{\mathbf q-\mathbf Q}{\xi}\vdagtwo{\mathbf q^{\prime}}{\xi}\cndagtwo{\mathbf q^{\prime}+\mathbf Q}{\xi^{\prime}}\rangle_{\text{c}}^{},
    \label{eq:N}
\end{multline}
where $\mathbf Q$ is the center-of-mass momentum and $\alpha_{\xi,\xi^{\prime}}$ and $\beta_{\xi,\xi^{\prime}}$ are the effective-mass ratios of the corresponding excitonic configuration $\xi,\xi^{\prime}$:
\begin{align}
    \alpha_{\xi,\xi^{\prime}} = \frac{m_e^{\xi^{\prime}}}{m_h^{\xi}+m_e^{\xi^{\prime}}},\quad \beta_{\xi,\xi^{\prime}} = \frac{m_h^{\xi}}{m_h^{\xi}+m_e^{\xi^{\prime}}}.
\end{align}
Throughout this work, we restrict the excitonic occupations/intraexcitonic coherences to carry total momenta of zero, since we assume spatially homogeneous optical excitation, where no net momentum is injected into the excitonic system. Note, that the total momentum of $\Nextwo{\mu,\nu,\mathbf Q}{\xi,\xi^{\prime}}$ is not equal to the center-of-mass momentum of an electron-hole pair $\mathbf Q$ but rather corresponds to the center-of-mass momentum of a pair of two excitons, which is zero.

\begin{figure}
    \centering
    \includegraphics[width=0.49\linewidth]{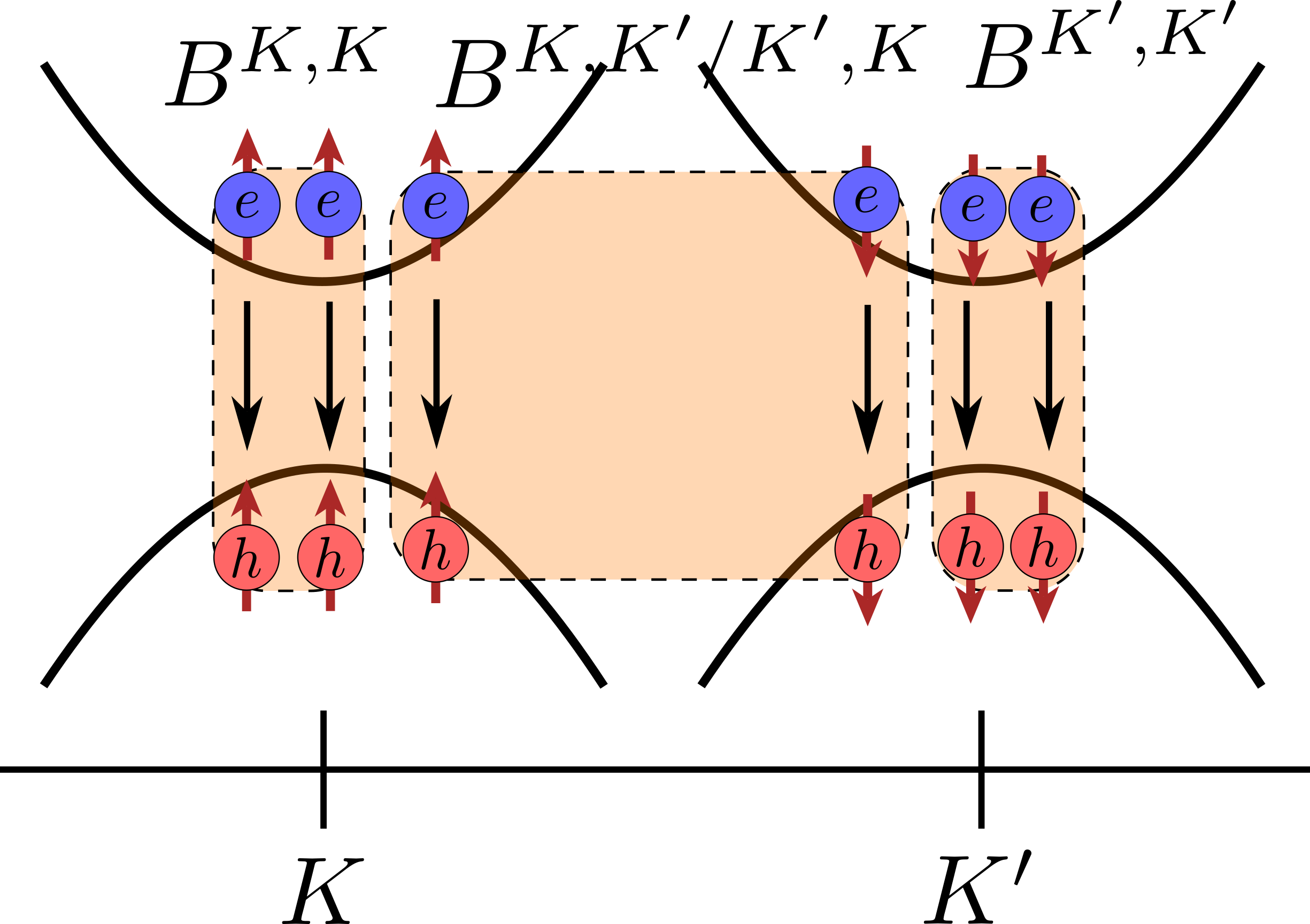}
    \includegraphics[width=0.49\linewidth]{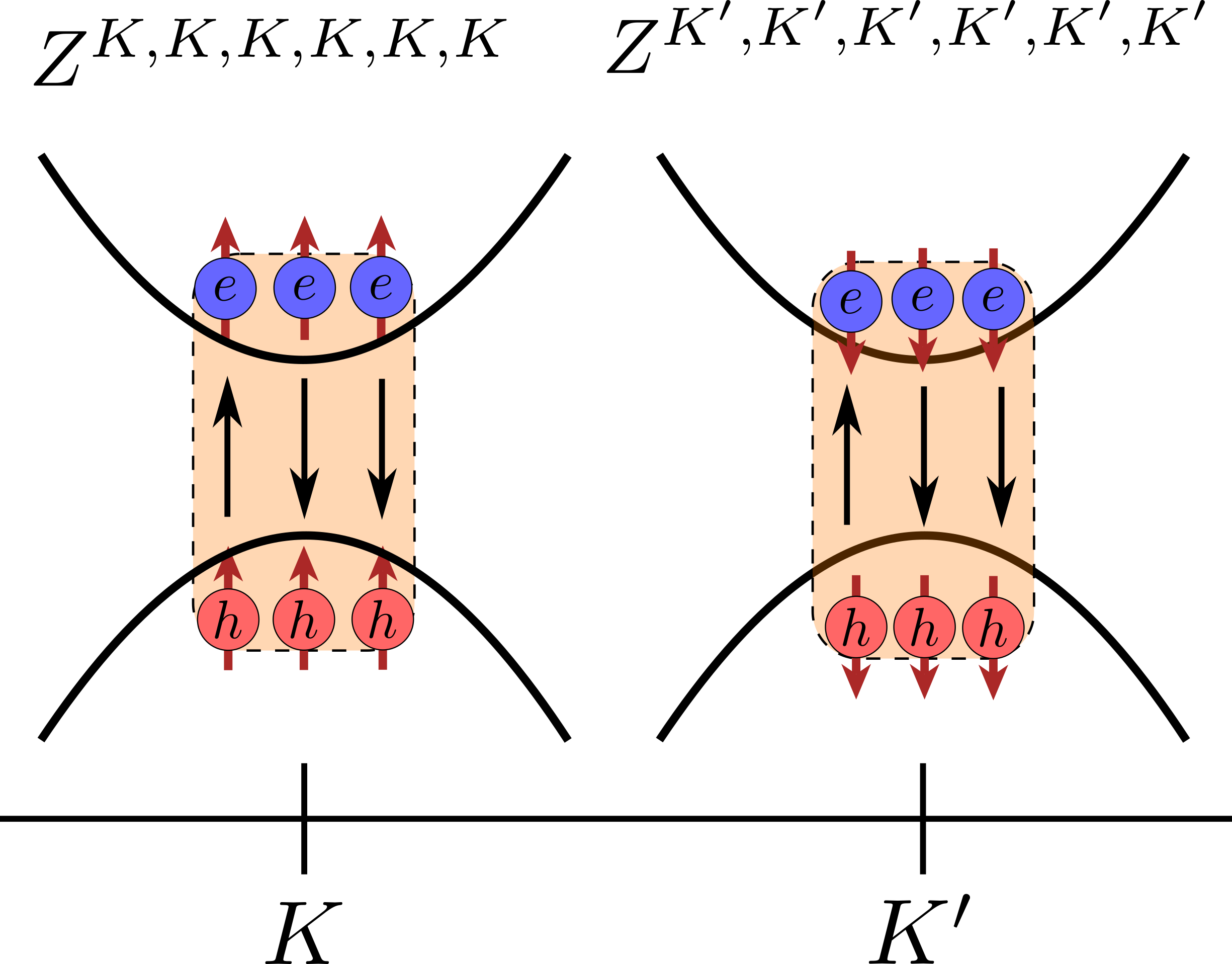}
    \caption{Scheme of two-exciton transitions from Eq.~\eqref{eq:B} in the intravalley configuration ($K,K$/$K^{\prime},K^{\prime}$) and in the intervalley configuration $K,K^{\prime}/K^{\prime},K$ (left) and exciton-two-exciton transitions from Eq.~\eqref{eq:Z} in the intravalley configuration (right).}
    \label{fig:B_scheme}
\end{figure}

As introduced by \textit{Moskalenko} \cite{moskalenko1958theory} and \textit{Lampert} \cite{lampert1958mobile} and later explicitly calculated by \textit{Ivanov} using a variational approach \cite{ivanov1978ground}, the two-exciton transition of order $n=2$ are defined as \cite{katsch2019theory,schumacher2006coherent,takayama2002t,schafer1996femtosecond}, cf.\ Fig.~\ref{fig:B_scheme}(left):
\begin{align}
\begin{split}
    B \sim  &\, \langle \hat A^{(2)}\rangle_{\text{c}}^{} =\langle \vdag{1}\cndag{2}\vdag{3}\cndag{4}\rangle_{\text{c}}^{}\\
    =&\, \langle \vdag{1}\cndag{2}\vdag{3}\cndag{4}\rangle - \langle \vdag{1}\cndag{2}\rangle \langle\vdag{3}\cndag{4}\rangle + \langle \vdag{1}\cndag{4}\rangle \langle\vdag{3}\cndag{2}\rangle,
    \end{split}
\end{align}
which, after an exciton expansion and a consecutive biexciton expansion, explicitly read:
\begin{multline}
    \Bextwo{\pm,\zeta}{\xi_1,\xi_2,\xi_3,\xi_4} = \sum_{\mu,\nu,\mathbf Q}\BiexWFstartwo{B,\pm,\zeta,\mu,\nu,\mathbf Q}{\text{L},\xi_1,\xi_2,\xi_3,\xi_4}\\
    \times\sum_{\rho,\eta,\mathbf Q^{\prime}}\left(S^{-1}_{\pm}\right)_{\mu,\nu,\mathbf Q,\rho,\eta,\mathbf Q^{\prime}}^{\xi_1,\xi_2,\xi_3,\xi_4}(\mathbf 0)\\
    \times\sum_{\mathbf q,\mathbf q^{\prime}}\ExWFstartwo{\rho,\mathbf q+\beta_{\xi_1,\xi_2/\xi_4}\mathbf Q^{\prime}}{\xi_1,\xi_2/\xi_4}\ExWFstartwo{\eta,\mathbf q^{\prime}+\alpha_{\xi_3,\xi_2/\xi_4}\mathbf Q^{\prime}}{\xi_3,\xi_2/\xi_4}\\
    \times\frac{1}{2}\left(\langle\vdagtwo{\mathbf q}{\xi_1}\cndagtwo{\mathbf q^{\prime}}{\xi_2}\vdagtwo{\mathbf q^{\prime}+\mathbf Q}{\xi_3}\cndagtwo{\mathbf q+\mathbf Q}{\xi_4} \rangle_{\text{c}}^{}\pm \langle\vdagtwo{\mathbf q}{\xi_1}\cndagtwo{\mathbf q^{\prime}}{\xi_4}\vdagtwo{\mathbf q^{\prime}+\mathbf Q}{\xi_3}\cndagtwo{\mathbf q+\mathbf Q}{\xi_2} \rangle_{\text{c}}^{}\right),
    \label{eq:B}
\end{multline}
where $\BiexWFstartwo{B,\pm,\zeta,\mu,\nu,\mathbf Q}{\text{L},\xi_1,\xi_2,\xi_3,\xi_4}$ is the two-exciton wave function solving the two-exciton Schr\"odinger equation with quantum number $\zeta$ and configuration $\pm$, cf.~Eq.~(S18). $(S_{\pm}^{})_{\mu,\nu,\mathbf Q,\rho,\eta,\mathbf Q^{\prime}}^{\xi_1,\xi_2,\xi_3,\xi_4}(\mathbf 0)$ is the projection matrix element arising due to the expansion in symmetric ($+$) and antisymmetric ($-$) configurations with respect to electron exchange, cf.~Eq.~(S23). The subscript ``$\xi/\xi^{\prime}$'' with respect to the electron configuration in the excitonic wave functions denotes an averaging over the electron configurations $\xi$ and $\xi^{\prime}$, if they are different, i.e., the corresponding excitonic wave functions obey the averaged Wannier equation, cf.~Eq.~(S1). 
This averaging procedure is necessary to expand the two-exciton correlations in excitonic wave functions. Throughout this work, we use the term ``two-exciton transition'' to generally denote all bound and unbound two-exciton states with $\zeta=b$ or $\zeta\neq b$, while we use the term ``two-exciton continuum'', if we specifically designate the continuum states with $\zeta\neq b$, and we use the term ``bound biexciton'', if we specifically designate the bound two-exciton state $\zeta=b$. Note, since we only consider optically excited excitonic transitions at zero center-of-mass momentum $\mathbf Q=\mathbf 0$, the two-exciton states are restricted to a total momentum of zero, which greatly reduces the relevant two-exciton phase space.

\begin{figure}
    \centering
    \includegraphics[width=0.49\linewidth]{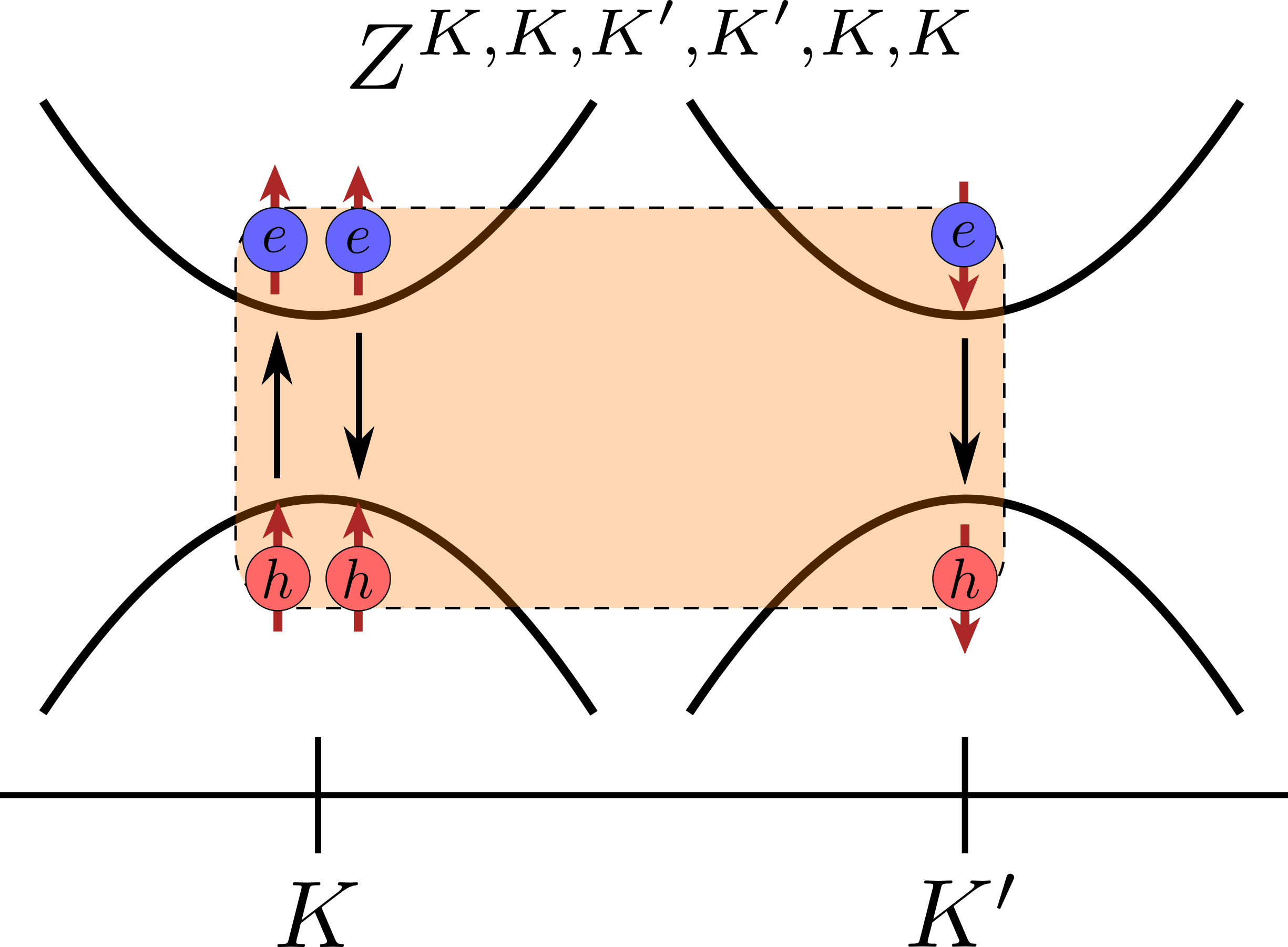}
    \includegraphics[width=0.49\linewidth]{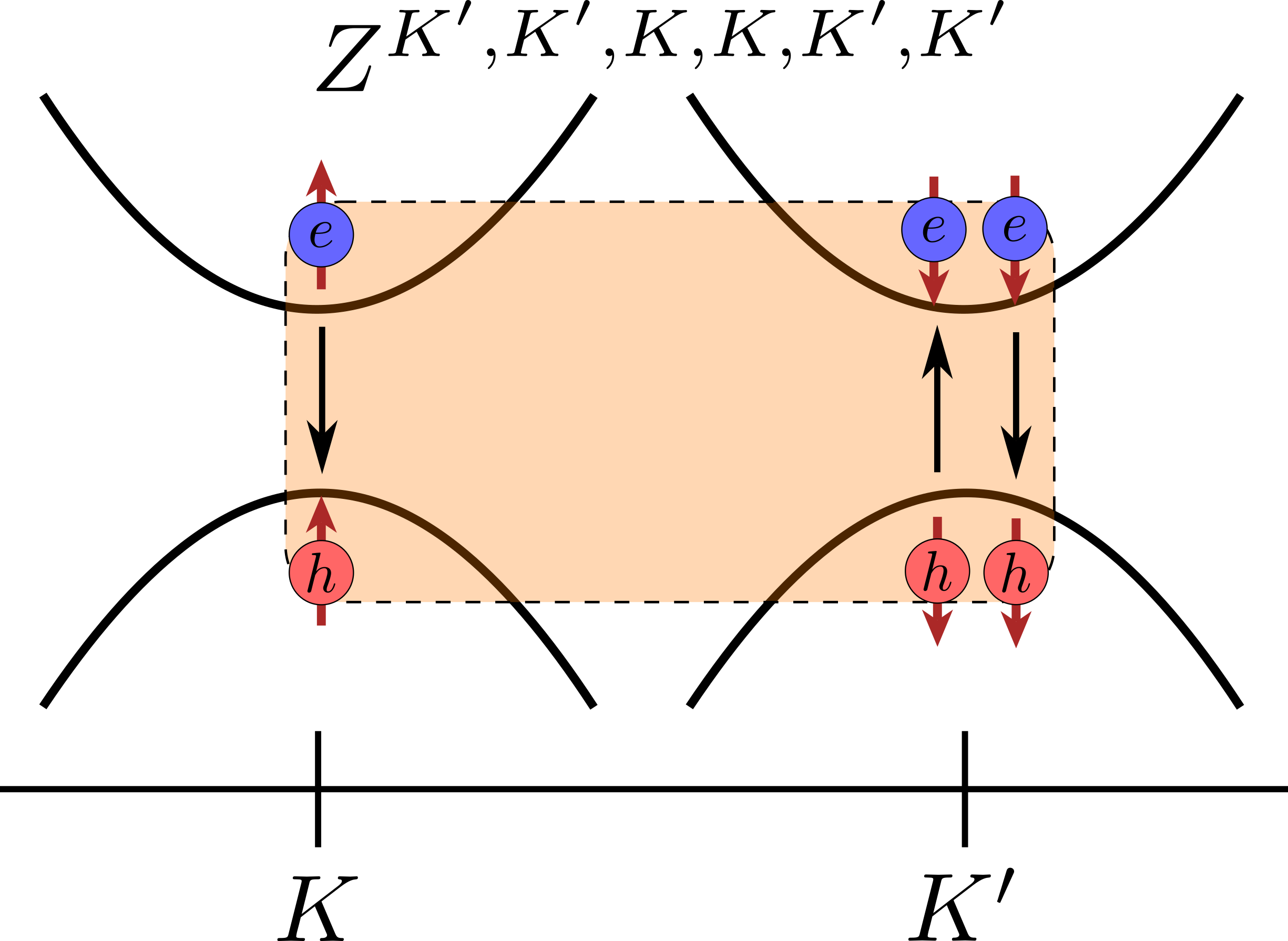}
    \caption{Scheme of exciton-two-exciton transitions from Eq.~\eqref{eq:Z} in the two possible intervalley configurations.}
    \label{fig:Z_scheme}
\end{figure}

First rigorously calculated by \textit{Axt} \cite{axt1994dynamics}, the exciton-two-exciton transitions of order $n=3$ are defined as \cite{axt2001evidence,bolton2000demonstration,meier1999excitons,bartels1997coherent}, cf.\ Fig.~\ref{fig:B_scheme}(right) and Fig.~\ref{fig:Z_scheme}:
\begin{align}
\begin{split}
    \Rextwo{}{} \sim &\, \langle \hat A^{(3)}_{}\rangle_{\text{c}}^{} =  \langle\cdag{1}\vndag{2}\vdag{3}\cndag{4}\vdag{5}\cndag{6}\rangle_{\text{c}}^{}\\
    =&\, \langle\cdag{1}\vndag{2}\vdag{3}\cndag{4}\vdag{5}\cndag{6}\rangle - \langle\cdag{1}\vndag{2}\rangle\langle\vdag{3}\cndag{4}\vdag{5}\cndag{6}\rangle_{\text{c}}^{}\\
    &\,- \langle \vdag{3}\cndag{4}\rangle \langle \cdag{1}\vndag{2}\vdag{5}\cndag{6}\rangle_{\text{c}}^{} + \langle\vdag{3}\cndag{6}\rangle\langle\cdag{1}\vndag{2}\vdag{5}\cndag{4}\rangle_{\text{c}}^{}\\
    &\, + \langle\vdag{5}\cndag{4}\rangle \langle\cdag{1}\vndag{2}\vdag{3}\cndag{6}\rangle_{\text{c}}^{} + \langle\vdag{5}\cndag{6}\rangle\langle \cdag{1}\vndag{2}\vdag{3}\cndag{4}\rangle_{\text{c}}^{}\\
    &\, - \langle \cdag{1}\vndag{2}\rangle \langle \vdag{3}\cndag{4}\rangle\langle\vdag{5}\cndag{6}\rangle + \langle \cdag{1}\vndag{2}\rangle \langle \vdag{3}\cndag{6}\rangle \langle \vdag{5}\cndag{4}\rangle,
    \end{split}
\end{align}
which, after an exciton and exciton-two-exciton expansion, explicitly read:
\begin{multline}
    \Rextwo{\pm,\zeta,\mu,\mathbf Q}{\xi_1,\xi_2,\xi_3,\xi_4,\xi_5,\xi_6} = \sum_{\nu,\rho,\mathbf Q^{\prime}}\BiexWFstartwo{Z,\pm,\zeta,\mu,\nu,\rho,\mathbf Q^{\prime}}{\text{L},\xi_1,\xi_2,\xi_3,\xi_4,\xi_5,\xi_6}(\mathbf Q) \\
    \times\sum_{\eta,\lambda,\mathbf Q^{\prime\prime}}\left(S^{-1}_{\pm}\right)_{\nu,\rho,\mathbf Q^{\prime},\eta,\lambda,\mathbf Q^{\prime\prime}}^{\xi_3,\xi_4,\xi_5,\xi_6}(\mathbf Q)\sum_{\mathbf q}\ExWFtwo{\mu,\mathbf q-\alpha_{\xi_2,\xi_1}\mathbf Q}{\xi_2,\xi_1}\\
    \times\sum_{\mathbf q^{\prime},\mathbf q^{\prime\prime}}
    \ExWFstartwo{\eta,\mathbf q^{\prime}+\beta_{\xi_3,\xi_4/\xi_6}\mathbf Q^{\prime\prime}}{\xi_3,\xi_4/\xi_6}\ExWFstartwo{\lambda,\mathbf q^{\prime\prime}-\alpha_{\xi_3,\xi_2/\xi_4}(\mathbf Q-\mathbf Q^{\prime\prime})}{\xi_3,\xi_2/\xi_4}\\
    \times\frac{1}{2}\left(\langle\cdagtwo{\mathbf q}{\xi_1}\vndagtwo{\mathbf q-\mathbf Q}{\xi_2}\vdagtwo{\mathbf q^{\prime}}{\xi_3}\cndagtwo{\mathbf q^{\prime\prime}}{\xi_4}\vdagtwo{\mathbf q^{\prime\prime}+\mathbf Q^{\prime\prime}-\mathbf Q}{\xi_5}\cndagtwo{\mathbf q^{\prime}+\mathbf Q^{\prime\prime}}{\xi_6} \rangle_{\text{c}}^{}\right.\\
    \left. \pm \langle\cdagtwo{\mathbf q}{\xi_1}\vndagtwo{\mathbf q-\mathbf Q}{\xi_2}\vdagtwo{\mathbf q^{\prime}}{\xi_3}\cndagtwo{\mathbf q^{\prime\prime}}{\xi_6}\vdagtwo{\mathbf q^{\prime\prime}+\mathbf Q^{\prime\prime}-\mathbf Q}{\xi_5}\cndagtwo{\mathbf q^{\prime}+\mathbf Q^{\prime\prime}}{\xi_4} \rangle_{\text{c}}^{}\right),
    \label{eq:Z}
\end{multline}
where $\BiexWFstartwo{Z,\pm,\zeta,\mu,\nu,\rho,\mathbf Q^{\prime}}{\text{L},\xi_1,\xi_2,\xi_3,\xi_4,\xi_5,\xi_6}(\mathbf Q)$ is the exciton-two-exciton wave function solving the exciton-two-exciton Schr\"odinger equation with quantum number $\zeta$ and configuration $\pm$, cf.\ Eq.~(S19). We use the term ``exciton-two-exciton transition'' to generally denote all bound and unbound exciton-two-exciton states, while we use the term ``exciton-two-exciton continuum'', if we specifically designate the continuum states, and we use the term ``bound exciton-biexciton'', if we specifically designate the bound exciton-two-exciton state. Note, that the exciton-two-exciton transition is a different many-body correlation than a triexciton, as described in Refs.~\cite{meier2003signatures,bartels1995chi}, which is at least relevant in a fifth-order response. 
Also note, that the main difference to the two-exciton transitions $\Bex{}$ in Eq.~\eqref{eq:B} is the additional center-of-mass-$\mathbf Q$-dependence, as their source is of partly incoherent nature. 
Also note, that there is some ambiguity in the interpretation of Eq.~\eqref{eq:Z}, since $\Rex{}\sim \langle \cdag{}\vndag{}\vdag{}\cndag{}\vdag{}\cndag{}\rangle_{\text{c}}^{}$ can be equally interpreted as a correlated coherence between an excitonic transition and a two-exciton transition $\langle (\cdag{}\vndag{})(\vdag{}\cndag{}\vdag{}\cndag{})\rangle_{\text{c}}^{}$ or as an occupation-assisted excitonic transition $\langle (\cdag{}\vndag{}\vdag{}\cndag{})(\vdag{}\cndag{})\rangle_{\text{c}}^{}$. However, both of these contributions are taken into account.

Throughout this work, we disregard correlated eight-operator expectation values of the form $\langle \cdag{1}\vndag{2}\cdag{3}\vndag{4}\vdag{5}\cndag{6}\vdag{7}\cndag{8}\rangle_{\text{c}}^{}$ of order $n=4$ (quadruplets), which describe two-exciton occupations or density-density correlations giving rise to excitonic incoherent Coulomb scattering in the sense of thermalization-causing exciton-exciton collisions \cite{schmitt2001bose,schmitt1999exciton}. 
While it has been shown, that Coulomb scattering in the electron-hole picture beyond the Hartree-Fock limit \cite{banyai1998ultrafast} significantly suppresses Rabi oscillations in GaAs QWs, they can be recovered if the electron and hole occupation dynamics are formulated in a dressed-state basis \cite{ciuti2000strongly}. This is the reason, why the semiconductor Bloch equations applied to material systems with weak Coulomb interaction already in Hartree-Fock limit successfully reproduced actual measured Rabi oscillations \cite{schulzgen1999direct,cundiff1994rabi}. Therefore, the neglection of incoherent Coulomb scattering in the study of Rabi oscillations in confined semiconductors such as conventional GaAs-like QWs, where light-matter interaction is of similar strength compared to Coulomb interaction, is justified as a first approximation. Future studies might shed more light on this issue. 

We note, that Figs.~\ref{fig:P_N_scheme}--\ref{fig:Z_scheme} only display a selection of possible configurations, as we restrict the visualization to spin-bright A excitons, which are the primary focus of this manuscript. However, the excitonic theory developed here is equally able to resolve spin-dark A and spin-bright/spin-dark B excitons. 

In Tab.~\ref{tab:excitonic_quantities}, we summarize all excitonic quantities considered in this work.

\begin{table}[]
    \centering
    \begin{tabular}{l}
        $\Poltwo{}{}$ 
        -- Excitonic transition \\
        $\Nextwo{}{}$ 
        -- Excitonic occupation/intraexcitonic coherence\\
        $\Bextwo{}{}$ 
        -- Two-exciton transition\\
        $\Rextwo{}{}$ 
        -- Exciton-two-exciton transition
    \end{tabular}
    \caption{Excitonic quantities considered in this work.}
    \label{tab:excitonic_quantities}
\end{table}

\subsection*{First-Order Contributions}
\begin{figure}
    \centering
    \includegraphics[width=1.0\linewidth]{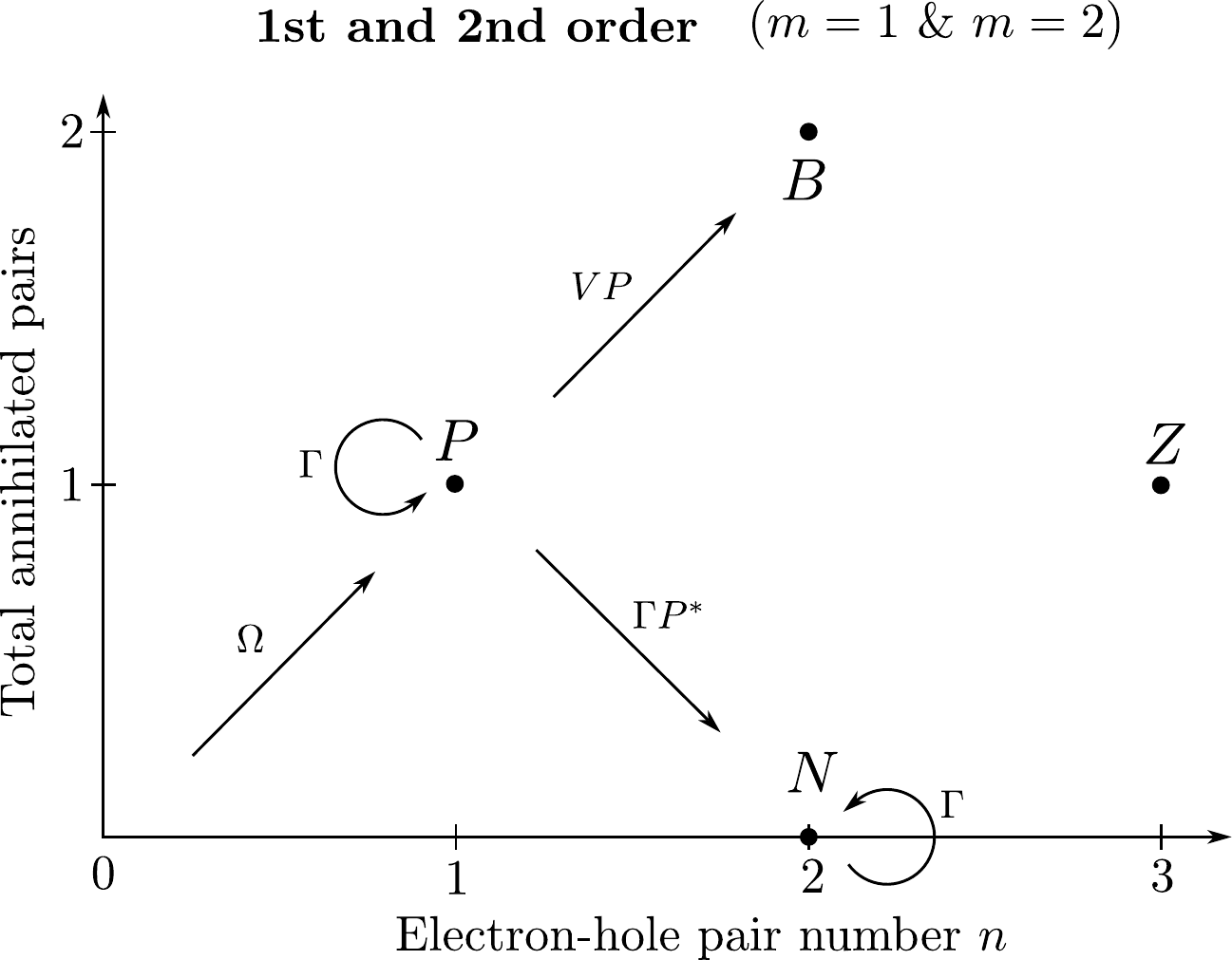}
    \caption{Scheme of the interaction processes via optical ($\Omega\overset{\wedge}{=}\Omega^{cv}$) and Coulomb ($V$) interaction up to second order DCT for the correlated expectation values in our theory. Additionally, we depict the exciton-phonon interaction in second-order Born-Markov approximation ($\Gamma$)}
    \label{fig:1st2nd_order_scheme}
\end{figure}
In first order $m=1$, which is the regime of linear optics, only excitonic transitions $\Poltwo{\mu}{\xi}$ occur:
\begin{multline}
\mathrm i\hbar\partial_t\Poltwo{\mu}{\xi}\Big|_{\mathbf E^1} =\\
\left(E_{\mu}^{\xi}-\hbar\omega_{\text{P}}-\mathrm i\hbar\gamma_{\text{nrad}}\right)\Poltwo{\mu}{\xi}
- \hbar\Omega_{}^{cv,\xi}\sum_{\mathbf q}\ExWFstartwo{\mu,\mathbf q}{\xi,\xi}.
\label{eq:P_EOM_first_order}
\end{multline}
Here, $E_{\mu}^{\xi} = E_{\mu,\mathbf Q=\mathbf 0}^{\xi}$ is the excitonic energy with excitonic quantum number $\mu$ at zero center-of-mass momentum $\mathbf Q$ and spin-valley index $\xi$, $\hbar\omega_{\text{P}}$ is the center frequency of the exciting optical pulse in a rotating frame, $\gamma_{\text{nrad}}$ is the total non-radiative homogeneous dephasing $\gamma_{\text{nrad}} = \gamma_{\text{X-phon}} + \gamma_{\text{res}}$ by exciton-phonon interaction $\gamma_{\text{X-phon}}$ \cite{selig2016excitonic,lengers2020theory}, cf.\ circled arrow denoted by ``$\Gamma$'' in Fig.~\ref{fig:1st2nd_order_scheme}, and by other residual contributions $\gamma_{\text{res}}$ via, e.g., scattering with uniformly distributed impurities \cite{ma2014charge,jena2007enhancement}. Inhomogeneous contributions are not considered, as their evaluation entails an ensemble average over many spatially dependent and randomly fluctuating slab- \cite{glutsch1994theory,thranhardt2003interplay} or substrate thicknesses \cite{raja2019dielectric} or non-uniformly distributed impurities, which greatly increases the numerical complexity. $\hbar\Omega^{cv} = \left(\hbar\Omega^{vc}\right)^*$ is the Rabi energy:
\begin{align}
    \hbar\Omega^{cv,\xi}(t) = \frac{\mathbf d^{cv,\xi}\cdot \mathbf E^{\sigma_{\pm}}}{2},
\end{align}
with transition dipole moment $\mathbf d^{cv,\xi}$ \cite{aversa1995nonlinear,pfalz2005optical,mkrtchian2019theory} and optical field in circular basis in rotating wave approximation:
\begin{align}
    \mathbf E^{\sigma_{\pm}}(t) = \mathbf e_{\pm} \tilde{E}(t),
\end{align}
with Jones vectors $\mathbf e_{\pm}$ and Gaussian envelope:
\begin{align}
    \tilde{E}(t) = \frac{E_0}{\sqrt{2\pi}\sigma_{\text{P}}}\mathrm e^{-\frac{t^2}{2\sigma_{\text{P}}^2}},
\end{align}
where $\sigma_{\text{P}}$ is the pulse duration determined by the intensity FWHM:
\begin{align}
    \sigma_{\text{P}} = \frac{\text{FWHM}}{2\sqrt{\ln(2)}}.
\end{align}
To keep the parameter set as small as possible, we introduce the pulse area $\Theta$ as a tuning knob of the optical excitation strength:
\begin{align}
    \Theta = \int_{-\infty}^{\infty}\mathrm dt\,\Omega^{cv}(t),
    \label{eq:PulseArea}
\end{align}
and neglect any contributions due to reradiation \cite{knorr1996theory}.

Throughout the manuscript, we assume normal incidence with $\mathbf E_{\mathbf Q} = \mathbf E\delta_{\mathbf Q,\mathbf 0}$, i.e., we always assume spatial in-plane homogeneity and neglect any diffusion \cite{hess1996maxwell_1,hess1996maxwell_2}, as such effect is negligible during the ultrafast timescales considered in this manuscript \cite{wagner2021nonclassical,rosati2021non,zipfel2020exciton,zhao2003spatiotemporal}.

In Fig.~\ref{fig:1st2nd_order_scheme}, we depict the action of the optical pulse ($\Omega$) annihilating an excitonic transition $P$ in first order and the coupling to $B$ and $N$ in second order: Here, the $y$-axis denotes the total annihilated electron-hole pairs described by the corresponding correlation $P$, $B$, $N$ etc. and the $x$-axis denotes the total electron-hole pair number $n$, cf. also Eq.~\eqref{eq:DCT_general_theorem}. Note again, that -- in contrast to, e.g., Refs.~\cite{victor1995hierarchy,axt1994dynamics,axt1996intraband} -- we always use the correlated excitonic operator expectation values throughout this work.

\subsection*{Second-Order Contributions}
In second order $m=2$, the equations of motion for the excitonic occupations/intraexcitonic coherences read:
\begin{multline}
    \mathrm i\hbar\partial_t\Nextwo{\sigma,\lambda,\mathbf Q}{\xi,\xi^{\prime}}\Big|_{\mathbf E^2} = \left(E_{\lambda,\mathbf Q}^{\xi,\xi^{\prime}} - E_{\sigma,\mathbf Q}^{\xi,\xi^{\prime}}\right)\Nextwo{\sigma,\lambda,\mathbf Q}{\xi,\xi^{\prime}} \\
    + \mathrm i\hbar\partial_t\Nextwo{\sigma,\lambda,\mathbf Q}{\xi,\xi^{\prime}}\Big|_{\mathbf E^2,\text{X-phon}},
    \label{eq:N_EOM_second_order}
\end{multline}
where the first line denotes the free excitonic contribution. We note, that, on this level, exciton-phonon interaction (second line in Eq.~\eqref{eq:N_EOM_second_order}) provides an additional important contribution, which, in second-order Born-Markov approximation, can be written as \cite{axt1996influence,thranhardt2000quantum,selig2018dark}:
\begin{multline}
    \mathrm i\hbar\partial_t\Nextwo{\sigma,\lambda,\mathbf Q}{\xi,\xi^{\prime}}\Big|_{\mathbf E^2,\text{X-phon}} = \sum_{\mu,\nu,\mathbf K}\Gamma_{\sigma,\lambda,\mathbf Q,\mu,\nu,\mathbf 0,-\mathbf Q}^{\text{in},\xi,\xi^{\prime}}\Polstartwo{\mu}{\xi}\Poltwo{\nu}{\xi}\\
    + \sum_{\mu,\nu,\mathbf K}\left(\Gamma_{\sigma,\lambda,\mathbf Q,\mu,\nu,\mathbf Q+\mathbf K,\mathbf K}^{\text{in},\xi,\xi^{\prime}} \Nextwo{\mu,\nu,\mathbf Q+\mathbf K}{\xi,\xi^{\prime}}\right.\\
    \left.- \Gamma_{\sigma,\lambda,\mathbf Q,\mu,\nu,\mathbf Q+\mathbf K,\mathbf K}^{\text{out},\xi,\xi^{\prime}} \Nextwo{\sigma,\lambda,\mathbf Q}{\xi,\xi^{\prime}}\right).
    \label{eq:N_EOM_xphon}
\end{multline}
Here, the first line denotes the phonon-assisted formation of excitonic occupations/intraexcitonic coherences, process ``$\Gamma P^*$'' in Fig.~\ref{fig:1st2nd_order_scheme}, while the remaining terms denote phonon-assisted scattering and thermalization, process ``$\Gamma$'' denoted by a circled arrow. The phonon-assisted scattering rates $\Gamma^{\text{in/out}}_{}$ can be deduced from, e.g., Refs.~\cite{selig2018dark,thranhardt2000quantum}. Since we focus on light-matter- and Coulomb interaction at elevated densities during intense optical fields beyond the $\chi^{(3)}$-limit, we neglect incoherent exciton-phonon scattering in the following. Nevertheless, we want to emphasize, that exciton-phonon scattering can introduce important corrections in the incoherent limit at nonlinear exciton densities. For a detailed analysis of the occurring fermionic, bosonic and exchange contributions we refer to Refs.~\cite{katzer2023exciton,katzer2024fermionic}. 
Direct optical-field pump-contributions to the excitonic occupations/intraexcitonic coherences, denoted by ``$\Omega$'', vanish: $\mathrm i\hbar\partial_t\Nextwo{\sigma,\lambda,\mathbf Q}{\xi,\xi^{\prime}}\Big|_{\mathbf E^2,\Omega} = 0$. Note, that if we additionally considered intraband light-matter interaction \cite{huttner2017ultrahigh}, intraexcitonic coherences and excitonic drift can occur already in second-order via, e.g., terahertz fields \cite{kira2001exciton,stroucken2021magnetic,kira2003microscopic}. 

The equations of motion for the two-exciton transitions read \cite{katsch2019theory,katsch2020exciton,schumacher2006coherent}:
\begin{multline}
\mathrm i\hbar\partial_t\Bextwo{\pm,\zeta}{\xi,\xi^{\prime}}\Big|_{\mathbf E^2} = \left(E_{B,\pm,\zeta}^{\xi,\xi^{\prime}}-\mathrm i\hbar\gamma_{B}\right)\Bextwo{\pm,\zeta}{\xi,\xi^{\prime}}\\
+\frac{1}{2}\sum_{\sigma,\eta,\mathbf K}\BiexWFstartwo{B,\pm,\zeta,\sigma,\eta,\mathbf K}{\text{L},\xi,\xi,\xi^{\prime},\xi^{\prime}}\sum_{\rho,\lambda,\mathbf Q}\left(S^{-1}\right)_{\sigma,\eta,\mathbf K,\rho,\lambda,\mathbf Q}^{\xi,\xi,\xi^{\prime},\xi^{\prime}}\mleft(\mathbf 0\mright)\\
\times\sum_{\mu,\nu}
\left(V_{1,\rho,\lambda,\mathbf Q,\mu,\nu,\mathbf 0}^{\xi,\xi,\xi^{\prime},\xi^{\prime}}\mp V_{2,\rho,\lambda,\mathbf Q,\mu,\nu,\mathbf 0}^{\xi,\xi,\xi^{\prime},\xi^{\prime}}\mleft(\mathbf 0\mright)
\right)\\
\times\Poltwo{\mu}{\xi}\Poltwo{\nu}{\xi^{\prime}}\left(\delta_{\xi,\xi^{\prime}}\pm 1\right).
\label{eq:B_EOM_second_order}
\end{multline}
Here, $E_{B,\pm,\zeta}^{\xi,\xi^{\prime}}$ is the two-exciton dispersion with two-exciton quantum number $\zeta$, configuration $\pm$ and two-exciton dephasing $\gamma_B$, which we set as $\gamma_B = 2\gamma_{\text{nrad}}$. 
Note, that the dephasing of bound biexcitons can differ from this assumption \cite{langbein2000dephasing}. 
$\BiexWFstartwo{B,\pm,\zeta,\sigma,\eta,\mathbf K}{\text{L},\xi,\xi,\xi^{\prime},\xi^{\prime}}$ is the two-exciton wave function as a solution of the two-exciton Schr\"odinger equation in Eq.~(S18), $S$ is the projection matrix in Eq.~(S23) and $V_1$ and $V_2$ are exciton-exciton matrix elements, cf.~Eq.~(S24) and Eq.~(S25). The ``$+$'' configuration, which can occur within equal valleys $\xi=\xi^{\prime}$ or distinct valleys $\xi\neq\xi^{\prime}$, hosts only continuum states. In contrast, the ``$-$'' configuration, where only two excitonic transitions from different spin-valleys $\xi\neq\xi^{\prime}$ are involved, as the contribution within equal valleys $\xi=\xi^{\prime}$ vanishes, hosts one bound state, the bound biexciton, and a continuum. In general, two-exciton transitions are induced via a product of two excitonic transitions $P$ via the exciton-exciton scattering encoded in the matrix elements $V_{1/2}$. 
The corresponding process is depicted in Fig.~\ref{fig:1st2nd_order_scheme} via ``$VP$''.

We note, that at this level, no impact of two-exciton transitions or excitonic occupations/intraexcitonic coherences on the optical response occurs, as the two-exciton transitions $\Bex{}$ and the excitonic occupations/intraexcitonic coherences $\Nex{}$ do not couple back to the excitonic transitions $\Pol{}$, cf.~Fig.~\ref{fig:1st2nd_order_scheme}. However, at this level, exciton dynamics can still be calculated, if exciton-phonon interaction is considered. Since there is no interaction between excitonic occupations/intraexcitonic coherences and two-exciton transitions, the exciton dynamics are purely bosonic up to second order.

\subsection*{Third-Order Contributions}
\begin{figure}
    \centering
    \includegraphics[width=1.0\linewidth]{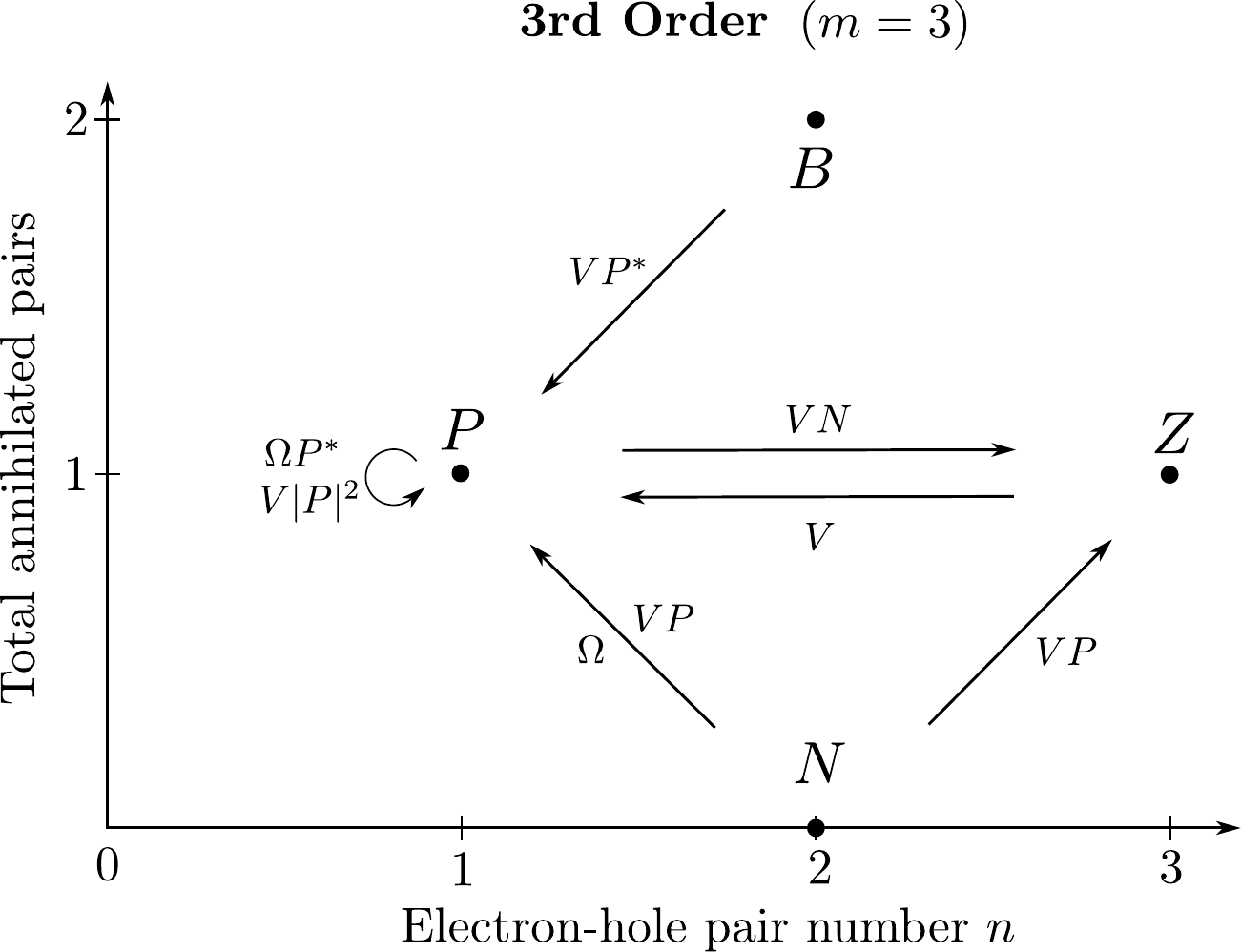}
    \caption{Scheme of the interaction processes via optical ($\Omega\overset{\wedge}{=}\Omega^{cv}$) and Coulomb ($V$) interaction on third order DCT for the correlated expectation values considered in our theory.}
    \label{fig:3rd_order_scheme}
\end{figure}
In third order in the optical field $m=3$, the following contributions to the excitonic transition need to be distinguished: Fully coherent contributions (COH), correlated contributions (CORR) and exciton-exciton (XX) contributions. 
Fully coherent (COH) contributions only involve products of the excitonic transitions $\Poltwo{}{}$, correlated contributions (CORR) involve products of excitonic occupations/intraexcitonic coherences $\Nextwo{}{}$, i.e., they constitute correlated coherent (intraexcitonic coherences) or fully incoherent (excitonic occupations) contributions, and exciton-exciton (XX) contributions involve coherent two-exciton transitions $\Bextwo{}{}$ and coherent-incoherent exciton-two-exciton transitions $\Rextwo{}{}$. In Fig.~\ref{fig:3rd_order_scheme}, we depict all interaction processes in third order schematically.

The fully coherent contributions to the excitonic transition $\Pol{}$ read:
\begin{multline}
\mathrm i\hbar\partial_t\Poltwo{\mu}{\xi}\Big|_{\mathbf E^3,\text{COH}}\\
=
\sum_{\nu,\rho,\eta}
\frac{1}{2}\left(U_{3,\mu,\nu,\rho,\eta,\mathbf 0}^{\xi,\xi} + U_{4,\mu,\nu,\rho,\eta,\mathbf 0}^{\xi,\xi}\right)
\Polstartwo{\rho}{\xi}\Poltwo{\eta}{\xi}\Poltwo{\nu}{\xi}\\
 + \hbar\Omega_{}^{cv,\xi}\sum_{\substack{\nu,\rho}}\left(D_{1,\mu,\nu,\rho,\mathbf 0}^{\xi,\xi}+D_{2,\mu,\nu,\rho,\mathbf 0}^{\xi,\xi}\right)\Polstartwo{\nu}{\xi}\Poltwo{\eta}{\xi},
\label{eq:P_EOM_COH_third_order}
\end{multline}
where the first line encodes coherent band-gap reduction and excitonic phase-space filling, cf.~process ``$V|P|^2$'' in Fig.~\ref{fig:3rd_order_scheme}, mediated via the Coulomb matrix element $U_{3/4}$, cf.~Eq.~(S40) and Eq.~(S41), and the second line describes coherent Pauli-blocking, cf.~process ``$\Omega P^*$'' in Fig.~\ref{fig:3rd_order_scheme}, via the matrix elements $D_{1/2}$, cf.~Eq.~(S28) and Eq.~(S29). 
Note, that the coherent Pauli-blocking term can also be viewed as a light-induced dressing contribution, since we can always factor out one excitonic transition $\Pol{}$.

The contributions to the excitonic transition $\Pol{}$ due to excitonic occupations and intraexcitonic coherences read:
\begin{multline}
\mathrm i\hbar\partial_t\Poltwo{\mu}{\xi}\Big|_{\mathbf E^3,\text{CORR}} = \sum_{\nu,\rho,\eta,\mathbf Q}U_{2,\mu,\nu,\rho,\eta,\mathbf Q}^{\xi,\xi}\Nextwo{\rho,\eta,\mathbf Q}{\xi,\xi}\Poltwo{\nu}{\xi}\\
 + \sum_{\nu,\rho,\eta,\mathbf Q,\xi^{\prime}}\left(U_{3,\mu,\nu,\rho,\eta,\mathbf Q}^{\xi,\xi^{\prime}}\Nextwo{\rho,\eta,\mathbf Q}{\xi,\xi^{\prime}} + U_{4,\mu,\nu,\rho,\eta,\mathbf Q}^{\xi^{\prime},\xi}\Nextwo{\rho,\eta,\mathbf Q}{\xi^{\prime},\xi}
\right)\Poltwo{\nu}{\xi}\\
+\hbar\Omega_{}^{cv}\sum_{\substack{\mathbf Q,\nu,\rho,\xi^{\prime}}}\left(D_{1,\mu,\nu,\rho,\mathbf Q}^{\xi,\xi^{\prime}}\Nextwo{\nu,\rho,\mathbf Q}{\xi,\xi^{\prime}} + D_{2,\mu,\nu,\rho,\mathbf Q}^{\xi^{\prime},\xi}\Nextwo{\nu,\rho,\mathbf Q}{\xi^{\prime},\xi}\right),%\\
\label{eq:P_EOM_CORR_third_order}
\end{multline}
where the first two lines encode incoherent band-gap reduction and excitonic phase-space filling, cf.~process ``$VP$'' denoted by an arrow from $N$ to $P$ in Fig.~\ref{fig:3rd_order_scheme}, mediated via the Coulomb matrix element $U_{2/3/4}$, cf.~Eqs.~(S39)--(S41), the third line describes incoherent Pauli-blocking, cf.~process ``$\Omega$'' in Fig.~\ref{fig:3rd_order_scheme}, via the matrix elements $D_{1/2}$, cf.~Eq.~(S28) and Eq.~(S29).

The exciton-exciton contributions read:
\begin{multline}
\mathrm i\hbar\partial_t\Poltwo{\mu}{\xi}\Big|_{\mathbf E^3,\text{XX}} = \\
\sum_{\substack{\nu,\rho,\lambda,\mathbf Q,\\\xi^{\prime},\zeta,\pm}}\left(\pm \mleft(V_{1,\rho,\lambda,\mathbf Q,\mu,\nu,\mathbf 0}^{\xi,\xi,\xi^{\prime},\xi^{\prime}}\mright)^* - \left(V_{2,\rho,\lambda,\mathbf Q,\mu,\nu,\mathbf 0}^{\xi,\xi,\xi^{\prime},\xi^{\prime}}\mleft(\mathbf 0\mright)\right)^*%\right.\\
\right)\\
\times
\BiexWFtwo{\pm,\zeta,\rho,\lambda,\mathbf Q}{\text{R},\xi,\xi,\xi^{\prime},\xi^{\prime}}\mleft(\mathbf 0\mright)\Bextwo{\pm,\zeta}{\xi,\xi^{\prime}}\Polstartwo{\nu}{\xi^{\prime}}\\
+ \sum_{\substack{\nu,\rho,\lambda,\mathbf Q,\mathbf Q^{\prime},\\\xi^{\prime},\xi^{\prime\prime},\zeta,\pm}}\left(\pm \mleft(V_{1,\rho,\lambda,\mathbf Q^{\prime};\mu,\nu,\mathbf 0}^{\xi,\xi,\xi^{\prime},\xi^{\prime\prime}}\mright)^* - \left(V_{2,\rho,\lambda,\mathbf Q^{\prime};\mu,\nu,\mathbf 0}^{\xi,\xi,\xi^{\prime},\xi^{\prime\prime}}\mleft(\mathbf Q\mright)\right)^*
\right)\\
\times
\BiexWFtwo{\pm,\zeta,\rho,\lambda,\mathbf Q^{\prime}}{\text{R},\xi,\xi,\xi^{\prime},\xi^{\prime\prime}}\mleft(\mathbf Q\mright)\Rextwo{\pm,\zeta,\nu,\mathbf Q}{\xi^{\prime\prime},\xi^{\prime},\xi,\xi,\xi^{\prime},\xi^{\prime\prime}}
\label{eq:P_EOM_XX_third_order}
\end{multline}
where the first two lines describe the interaction with two-exciton transitions $\Bextwo{}{}$, cf.~processes ``$VP^*$'' in Fig.~\ref{fig:3rd_order_scheme}, and the last two lines describe the interaction with exciton-two-exciton transitions $\Rextwo{}{}$, cf.~processes ``$V$'' with the exciton-exciton matrix elements $V_{1/2}$, cf.~Eq.~(S24) and Eq.~(S25).

Also, in third order, the equations of motion for the exciton-two-exciton transitions $\Rex{}$ occur:
\begin{multline}
    \mathrm i\hbar\partial_t\Rextwo{\pm,\zeta,\rho,\mathbf Q}{\xi_1,\xi_2,\xi_3,\xi_4,\xi_5,\xi_6}\Big|_{\mathbf E^3} = \left(E_{R,\pm,\zeta,\rho,\mathbf Q}^{\xi_1,\xi_2,\xi_3,\xi_4,\xi_5,\xi_6} - \mathrm i\hbar\gamma_{Z}\right)\\
    \times\Rextwo{\pm,\zeta,\rho,\mathbf Q}{\xi_1,\xi_2,\xi_3,\xi_4,\xi_5,\xi_6} + \frac{1}{2}\sum_{\mu,\delta,\mathbf K}\BiexWFstartwo{R,\pm,\zeta;\mu,\delta,\mathbf K}{\text{L},\xi_3,\xi_4,\xi_5,\xi_6}\mleft(\mathbf Q\mright)\\
    \times\sum_{\lambda,\sigma,\mathbf Q^{\prime}}\left(S_{\pm}^{-1}\right)_{\mu,\delta,\mathbf K,\lambda,\sigma,\mathbf Q^{\prime}}^{\xi_3,\xi_4,\xi_5,\xi_6}\mleft(\mathbf Q\mright)\\
    \times\sum_{\nu}\left(\left(V_{1,\lambda,\sigma,\mathbf Q^{\prime},\nu,\rho,\mathbf 0}^{\xi_3,\xi_4,\xi_5,\xi_6} \mp V_{4,\lambda,\sigma,\mathbf Q^{\prime},\nu,\rho,\mathbf Q}^{\xi_3,\xi_4,\xi_5,\xi_6}\right)\right.\\
    \times\Poltwo{\nu}{\xi_3}\left(\delta_{\xi_3,\xi_6}\delta_{\xi_1,\xi_4} \pm \delta_{\xi_3,\xi_4}\delta_{\xi_1,\xi_6}\right)\delta_{\xi_2,\xi_5}\\
     + \left.\left(V_{1,\lambda,\sigma,\mathbf Q^{\prime},\nu,\rho,\mathbf Q}^{\xi_3,\xi_4,\xi_5,\xi_6} \mp V_{3,\lambda,\sigma,\mathbf Q^{\prime},\nu,\rho,\mathbf Q}^{\xi_3,\xi_4,\xi_5,\xi_6}\right)\right.\\
     \left.\times \Poltwo{\nu}{\xi_5}\left(\delta_{\xi_5,\xi_4}\delta_{\xi_1,\xi_6} \pm \delta_{\xi_5,\xi_6}\delta_{\xi_1,\xi_4}\right)\delta_{\xi_2,\xi_3}\right)\Nextwo{\rho,\mathbf Q}{\xi_2,\xi_1}.
     \label{eq:Z_EOM_third_order}
\end{multline}
Here, $E_{R,\pm,\zeta,\rho,\mathbf Q}^{\xi_1,\xi_2,\xi_3,\xi_4,\xi_5,\xi_6}$ is the exciton-two-exciton dispersion for configuration $\pm$, exciton-two-exciton quantum number $\zeta$, momentum $\mathbf Q$ and $\gamma_Z$ is the exciton-biexciton dephasing, which we set as $\gamma_Z = 3\gamma_{\text{nrad}}$. 
Note, that the dephasing of the bound exciton-biexciton may differ from this assumption \cite{langbein2000dephasing}. 
Similar to Eq.~\eqref{eq:B}, the ``$+$'' configuration, which exists for intra- and intervalley configurations, hosts exciton-two-exciton continuum states, while the ``$-$'' configuration, which vanishes for intravalley interactions, hosts one bound state and a continuum. The main difference to the two-exciton transitions in Eq.~\eqref{eq:B} is, that the exciton-two-exciton transitions exhibit a partly incoherent nature, as they are induced by a product of intraexcitonic coherences/excitonic occupations $\Nex{\mathbf Q}$ and excitonic transitions $P$ mediated via the exciton-exciton matrix elements $V_{1/3/4}$. 
The corresponding process ``$VN$'' is denoted by an arrow from $P$ to $Z$. We note, that ``$VP$'', denoted by an arrow from $N$ to $Z$ in Fig.~\ref{fig:3rd_order_scheme} denotes the same process, but from a different viewpoint. For that matter, the exciton-two-exciton wave functions $\BiexWFstartwo{R,\pm,\zeta;\mu,\delta,\mathbf K}{\text{L},\xi_3,\xi_4,\xi_5,\xi_6}\mleft(\mathbf Q\mright)$ solving the exciton-two-exciton Schr\"odinger equation in Eq.~(S19) acquire an additional dependence on the center-of-mass momentum $\mathbf Q$ corresponding to the exciton-two-exciton-inducing intraexcitonic coherence/incoherent excitonic occupation $\Nex{\mathbf Q}{}$. The exciton-exciton matrix elements $V_{1/3/4}$ entering Eq.~\eqref{eq:Z_EOM_third_order} are given in Eq.~(S24), Eq.~(S26) and Eq.~(S27), respectively. 

Compared to second order, incoherent occupations/intraexcitonic coherences and two-exciton transitions couple back to the excitonic transitions via many-body optical and/or Coulomb interaction processes, cf.~Fig.~\ref{fig:3rd_order_scheme}. Additionally, the partially incoherent exciton-two-exciton transitions, cf.~Eq.~\eqref{eq:Z}, which emerge in third order, cf.~Eq.~\eqref{eq:Z_EOM_third_order}, and, at the same time, couple back to the excitonic transitions, cf.~Eq.~\eqref{eq:P_EOM_XX_third_order}, serve as an important bridge between the coherent and incoherent dynamics. In particular, in two-pulse pump-probe experiments, they adopt the role of the fully coherent two-exciton transitions $\Bex{}$ in Eq.~\eqref{eq:B}, when all pump-induced coherent excitonic transitions $P$ have been decayed and the system is in a fully incoherent state $\Nex{\mathbf Q}{}$. 
Hence, if we combine the equations of motion up to third order with exciton-phonon scattering (and density-independent Coulomb scattering such as intervalley exchange \cite{selig2019ultrafast,selig2020suppression,yu2014valley,qiu2015nonanalyticity} or Dexter interaction \cite{dogadov2026diss,berghauser2018inverted}), a comprehensive description of pump-probe experiments from the coherent to the incoherent regime valid for small to moderate pump fluences can be obtained.

\subsection*{Fourth-Order Contributions}
\begin{figure}
    \centering
    \includegraphics[width=1.0\linewidth]{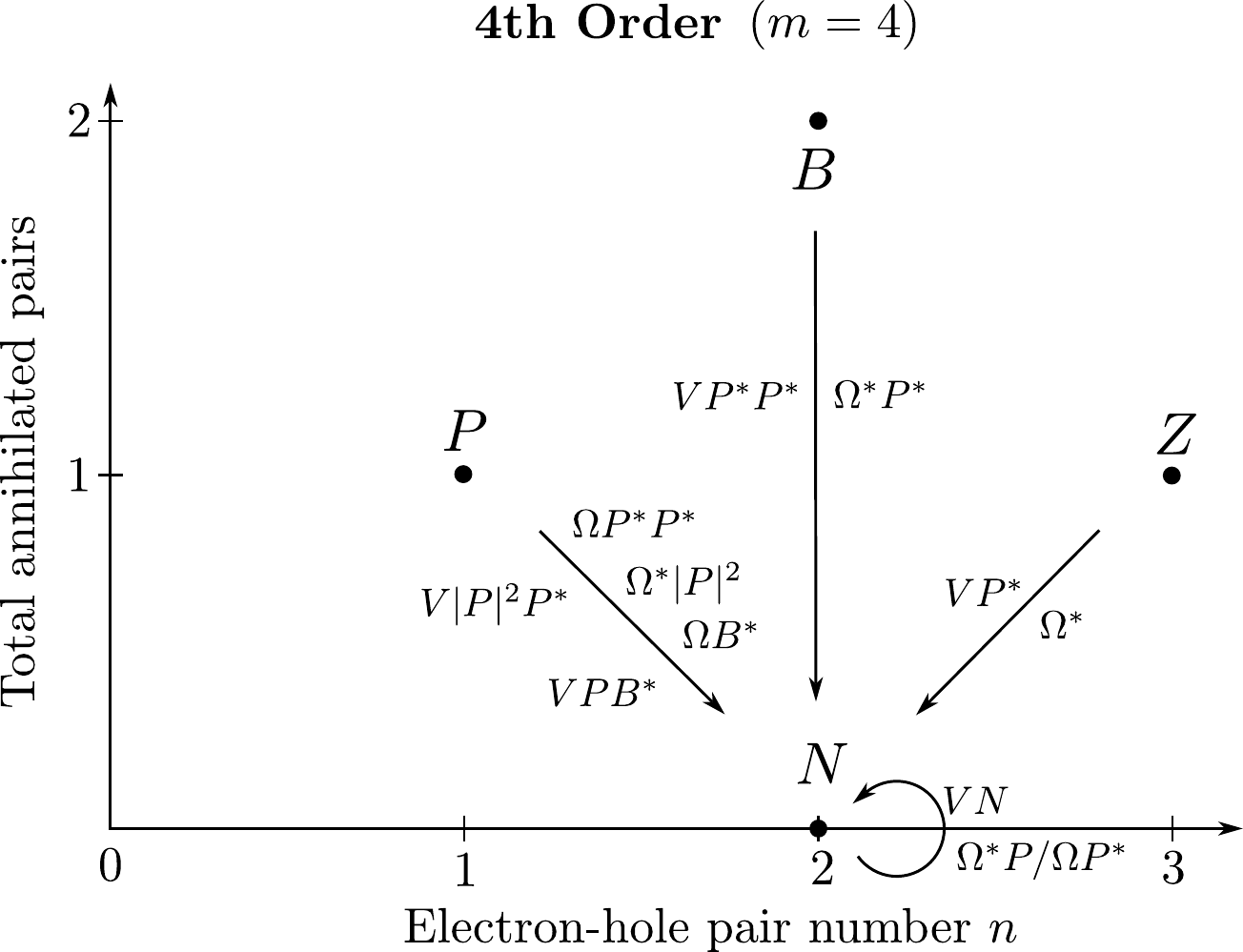}
    \caption{Scheme of the interaction processes via optical ($\Omega\overset{\wedge}{=}\Omega^{cv}$) and Coulomb ($V$) interaction on fourth order DCT for the correlated expectation values considered in our theory. Note, that we exclude correlated eight-operator expectation values and fourth-order optical and Coulomb contributions to the biexcitons $B$.}
    \label{fig:4th_order_scheme}
\end{figure}

To investigate the optical response beyond the limit of third-order nonlinearities, we have to consider contributions up to the fourth order in the optical field $m=4$.

Next to the distinction in fully coherent (COH), correlated (CORR) and exciton-exciton (XX) contributions, which we introduced in third order, we additionally group the equations in optical interaction ($\Omega$) and Coulomb interaction ($V$).

In Fig.~\ref{fig:4th_order_scheme}, we schematically depict all fourth-order interaction processes considered in this manuscript.

The fully coherent contributions to the intraexcitonic coherences/excitonic occupations $\Nex{\mathbf Q}$ via optical interaction read, cf.~process ``$\Omega P^*P^*$'' in Fig.~\ref{fig:4th_order_scheme}:
\begin{multline}
        \mathrm i\hbar\partial_t \Nextwo{\sigma,\lambda,\mathbf Q}{\xi,\xi^{\prime}}\Big|_{\mathbf E^4,\Omega,\text{COH}} =   \sum_{\mu,\nu,\eta}\left(\hbar\Omega^{vc,\xi}D_{3,\sigma,\lambda,\mu,\nu,\eta,\mathbf Q}^{\xi,\xi^{\prime}}\Polstartwo{\mu}{\xi}\right.\\
        \left.+ \hbar\Omega^{vc,\xi^{\prime}}D_{4,\sigma,\lambda,\mu,\nu,\eta,\mathbf Q}^{\xi,\xi^{\prime}}\Polstartwo{\mu}{\xi^{\prime}} \right)\Poltwo{\nu}{\xi^{\prime}}\Poltwo{\eta}{\xi}- \begin{pmatrix}
    \sigma \leftrightarrow \lambda
    \end{pmatrix}^*,
        \label{eq:N_EOM_opt_COH_fourth_order}
\end{multline}
which describes the coherent excitation of intraexcitonic coherences/excitonic occupations $\Nex{\mathbf Q}$ via coherent excitonic transitions $P$ mediated by the Pauli-blocking matrix elements $D_{3/4}$, cf.~Eq.~(S30) and Eq.~(S31). While optical excitation at normal incidence with $\mathbf E_{\mathbf Q} = \mathbf E\delta_{\mathbf Q,\mathbf 0}$ only excites coherent excitonic transitions at $\mathbf Q=\mathbf 0$ (and a certain distribution of relative momenta $\mathbf q\geq\mathbf 0$), cf.\ Eq.~\eqref{eq:P_EOM_first_order}, interactions in fourth order of the optical field in Eq.~\eqref{eq:N_EOM_opt_COH_fourth_order} cause the coherent excitation of excitonic occupations with a certain center-of-mass distribution out of the light cone $\mathbf Q\geq \mathbf 0$. This behavior can be understood via the fermionic substructure, as individual electron and hole collisions via Coulomb interaction redistribute the center-of-mass momenta $\mathbf Q$ and relative momenta within the exciton gas. These individual electron and hole collisions are encoded in the many-body matrix elements $D_{3/4}$.

Moreover, contributions to intraexcitonic coherences/excitonic occupations $\Nex{\mathbf Q}$ via optical interaction beyond the fully coherent limit occur, cf.~process ``$\Omega^* P$/$\Omega P^*$'' in Fig.~\ref{fig:4th_order_scheme}:
\begin{multline}
\mathrm i\hbar\partial_t \Nextwo{\sigma,\lambda,\mathbf Q}{\xi,\xi^{\prime}}\Big|_{\mathbf E^4,\Omega,\text{CORR}} = 
        - \sum_{\mu,\nu}\left(\hbar\Omega^{vc,\xi}D_{1,\mu,\sigma,\nu,\mathbf Q}^{*,\xi,\xi^{\prime}}
        \Poltwo{\mu}{\xi}\right.\\
        \quad\quad\left.+ \hbar\Omega^{vc,\xi^{\prime}}D_{2,\mu,\sigma,\nu,\mathbf Q}^{*,\xi,\xi^{\prime}}
        \Poltwo{\mu}{\xi^{\prime}}\right)\Nextwo{\nu,\lambda,\mathbf Q}{\xi,\xi^{\prime}}\\
         + \sum_{\mathbf Q^{\prime},\mu,\nu,\eta}\hbar\Omega^{vc,\xi}
        D_{5,\sigma,\lambda,\mu,\nu,\eta,\mathbf Q,\mathbf Q^{\prime}}^{\xi}
        \Poltwo{\mu}{\xi}\Nextwo{\nu,\eta,\mathbf Q^{\prime}}{\xi,\xi}\delta_{\xi,\xi^{\prime}}\\
         + \sum_{\mathbf Q^{\prime},\mu,\nu,\eta,\xi^{\prime\prime}}\left(\hbar\Omega^{vc,\xi}
        D_{6,\sigma,\lambda,\mu,\nu,\eta,\mathbf Q,\mathbf Q^{\prime}}^{\xi,\xi^{\prime},\xi^{\prime\prime}}
        \Poltwo{\mu}{\xi}\Nextwo{\nu,\eta,\mathbf Q^{\prime}}{\xi^{\prime\prime},\xi^{\prime}}\right.\\
        \quad\quad\left. + \hbar\Omega^{vc,\xi^{\prime}}
        D_{7,\sigma,\lambda,\mu,\nu,\eta,\mathbf Q,\mathbf Q^{\prime}}^{\xi,\xi^{\prime},\xi^{\prime\prime}}
        \Poltwo{\mu}{\xi^{\prime}}\Nextwo{\nu,\eta,\mathbf Q^{\prime}}{\xi,\xi^{\prime\prime}}\right)
        - \begin{pmatrix}
    \sigma \leftrightarrow \lambda
    \end{pmatrix}^*.
        \label{eq:N_EOM_opt_CORR_fourth_order}
\end{multline}
Here, the first two lines describe an attenuation of the coherent excitation of intraexcitonic coherences/excitonic occupations, as soon as excitonic states at equal center-of-mass momenta are occupied, 
mediated by the Pauli-blocking matrix elements $D_{1/2}$, cf.~Eq.~(S28) and Eq.~(S29). The last three lines describe a build-up of the coherent excitation of intraexcitonic coherences/excitonic occupations at COM momentum $\mathbf Q$ in the presence of intraexcitonic coherences/excitonic occupations at all other COM momenta $\mathbf Q^{\prime}$ via intravalley (third line) or intervalley correlations (fourth and fifth line) via the many-body matrix elements $D_{5/6/7}$, cf.~Eqs.~(S32)--(S34).

Additionally, Coulomb interaction induces a coherent excitation of intraexcitonic coherences, cf.~process ``$V|P|^2P$'' in Fig.~\ref{fig:4th_order_scheme}:
\begin{multline}
    \mathrm i\hbar\partial_t \Nextwo{\sigma,\lambda,\mathbf Q}{\xi,\xi^{\prime}}\Big|_{\mathbf E^4,V,{\text{COH}}} = \sum_{\mu,\nu,\rho,\eta} \Polstartwo{\mu}{\xi}\Poltwo{\nu}{\xi}\Polstartwo{\rho}{\xi^{\prime}}\Poltwo{\eta}{\xi^{\prime}}\\
    \times  \left(U_{5,\sigma,\lambda,\mu,\nu,\rho,\eta,\mathbf Q}^{\xi}\delta_{\xi,\xi^{\prime}} + U_{6,\sigma,\lambda,\mu,\nu,\rho,\eta,\mathbf Q}^{\xi,\xi^{\prime}} \right)
    - \begin{pmatrix}
    \sigma \leftrightarrow \lambda
    \end{pmatrix}^*,
    \label{eq:N_EOM_coul_COH_fourth_order}
\end{multline}
in the presence of excitonic transitions $P$ mediated via the coupling elements $U_{5/6}$, cf.~Eq.~(S42) and Eq.~(S43). If only excitonic transitions $\Pol{\mu}$ of a single excitonic state $\mu=\nu=\rho=\eta$ are present, this contribution vanishes for excitonic occupations with $\sigma=\lambda$:
\begin{align}
    \mathrm i\hbar\partial_t \Nextwo{\sigma,\sigma,\mathbf Q}{\xi,\xi^{\prime}}\Big|_{\mathbf E^4,V,\text{COH}} = 0,
\end{align}
while intraexcitonic coherences with $\sigma\neq\lambda$ can still be induced:
\begin{align}
    \mathrm i\hbar\partial_t \Nextwo{\sigma,\lambda,\mathbf Q}{\xi,\xi^{\prime}}\Big|_{\mathbf E^4,V,\text{COH}} \neq 0.
\end{align}
For excitonic transitions of an arbitrary admixture of states, intraexcitonic coherences with $\sigma\neq\lambda$ and also excitonic occupations with $\mu=\lambda$ are induced.

Beyond a fully coherent limit, Coulomb interaction causes a coupling between intraexcitonic coherences and excitonic occupations, cf.~process ``$VN$'' in Fig.~\ref{fig:4th_order_scheme}:
\begin{multline}
    \mathrm i\hbar\partial_t \Nextwo{\sigma,\lambda,\mathbf Q}{\xi,\xi^{\prime}}\Big|_{\mathbf E^4,V,\text{CORR}} = \\
    \sum_{\mu,\nu,\rho,\eta,\mathbf Q^{\prime}} \left(
    \vphantom{\sum_{\mathbf Q^{\prime\prime},\xi^{\prime\prime},\xi^{\prime\prime\prime}}U_{12,\sigma,\lambda,\mu,\nu,\rho,\eta,\mathbf Q}^{\xi,\xi^{\prime},\xi^{\prime\prime},\xi^{\prime\prime\prime}}\Nextwo{\mu,\nu,\mathbf Q^{\prime}}{\xi^{\prime\prime},\xi^{\prime}}\Nextwo{\rho,\eta,\mathbf Q^{\prime\prime}}{\xi,\xi^{\prime\prime\prime}}}
    U_{7,\sigma,\lambda,\mu,\nu,\rho,\eta,\mathbf Q,\mathbf Q^{\prime}}^{\xi,\xi^{\prime}}\Nextwo{\mu,\nu,\mathbf Q}{\xi,\xi^{\prime}}\overline{\Nextwo{\rho,\eta,\mathbf Q^{\prime}}{\xi,\xi^{\prime}}} \right.\\
     + \sum_{\xi^{\prime\prime}}U_{8,\sigma,\lambda,\mu,\nu,\rho,\eta,\mathbf Q,\mathbf Q^{\prime}}^{\xi,\xi^{\prime},\xi^{\prime\prime}}\Nextwo{\mu,\nu,\mathbf Q}{\xi,\xi^{\prime}}\overline{\Nextwo{\rho,\eta,\mathbf Q^{\prime}}{\xi^{\prime\prime},\xi^{\prime}}}\\
     + \sum_{\xi^{\prime\prime}}U_{9,\sigma,\lambda,\mu,\nu,\rho,\eta,\mathbf Q,\mathbf Q^{\prime}}^{\xi,\xi^{\prime},\xi^{\prime\prime}}\Nextwo{\mu,\nu,\mathbf Q}{\xi,\xi^{\prime}}\overline{\Nextwo{\rho,\eta,\mathbf Q^{\prime}}{\xi,\xi^{\prime\prime}}}\\
     + \sum_{\mathbf Q^{\prime\prime},\xi^{\prime\prime}}U_{10,\sigma,\lambda,\mu,\nu,\rho,\eta,\mathbf Q,\mathbf Q^{\prime},\mathbf Q^{\prime\prime}}^{\xi,\xi^{\prime},\xi^{\prime\prime}}\overline{\Nextwo{\mu,\nu,\mathbf Q^{\prime}}{\xi,\xi^{\prime}}\Nextwo{\rho,\eta,\mathbf Q^{\prime\prime}}{\xi^{\prime\prime},\xi^{\prime}}}\\
     + \sum_{\mathbf Q^{\prime\prime},\xi^{\prime\prime}}U_{11,\sigma,\lambda,\mu,\nu,\rho,\eta,\mathbf Q,\mathbf Q^{\prime},\mathbf Q^{\prime\prime}}^{\xi,\xi^{\prime},\xi^{\prime\prime}}\overline{\Nextwo{\mu,\nu,\mathbf Q^{\prime}}{\xi,\xi^{\prime}}\Nextwo{\rho,\eta,\mathbf Q^{\prime\prime}}{\xi,\xi^{\prime\prime}}}\\
     + \sum_{\mathbf Q^{\prime\prime}}U_{12,\sigma,\lambda,\mu,\nu,\rho,\eta,\mathbf Q,\mathbf Q^{\prime},\mathbf Q^{\prime\prime}}^{\xi,\xi^{\prime}}\overline{\Nextwo{\mu,\nu,\mathbf Q^{\prime}}{\xi,\xi^{\prime}}\Nextwo{\rho,\eta,\mathbf Q^{\prime\prime}}{\xi,\xi^{\prime}}}\\
     \left.+ \sum_{\mathbf Q^{\prime\prime},\xi^{\prime\prime},\xi^{\prime\prime\prime}}U_{13,\sigma,\lambda,\mu,\nu,\rho,\eta,\mathbf Q,\mathbf Q^{\prime},\mathbf Q^{\prime\prime}}^{\xi,\xi^{\prime},\xi^{\prime\prime},\xi^{\prime\prime\prime}}\overline{\Nextwo{\mu,\nu,\mathbf Q^{\prime}}{\xi^{\prime\prime},\xi^{\prime}}\Nextwo{\rho,\eta,\mathbf Q^{\prime\prime}}{\xi,\xi^{\prime\prime\prime}}}\right)\\
     - \begin{pmatrix}
    \sigma \leftrightarrow \lambda
    \end{pmatrix}^*,
     \label{eq:N_EOM_coul_CORR_fourth_order}
\end{multline}
mediated by the exciton-exciton matrix elements $U_{7/8/9/10/11/12/13}$, cf.~Eqs.~(S44)--(S50). 
In Eq.~\eqref{eq:N_EOM_coul_CORR_fourth_order}, we use a short-hand notation as follows:
\begin{align}
    \overline{N \vphantom{N}_{\mu,\nu,\mathbf Q}^{\xi,\xi^{\prime}} } = \Nextwo{\mu,\nu,\mathbf Q}{\xi,\xi^{\prime}} + \Polstartwo{\mu}{\xi}\Poltwo{\nu}{\xi}\delta_{\mathbf Q,\mathbf 0}^{\xi,\xi^{\prime}},
\end{align}
and:
\begin{multline}
    \overline{N \vphantom{N}_{\mu,\nu,\mathbf Q}^{\xi,\xi^{\prime}}  N \vphantom{N}_{\rho,\eta,\mathbf Q^{\prime}}^{\xi^{\prime\prime},\xi^{\prime\prime\prime}}} = \Nextwo{\mu,\nu,\mathbf Q}{\xi,\xi^{\prime}}\Nextwo{\rho,\eta,\mathbf Q^{\prime}}{\xi^{\prime\prime},\xi^{\prime\prime\prime}}\\
    + \Nextwo{\mu,\nu,\mathbf Q}{\xi,\xi^{\prime}} \Polstartwo{\rho}{\xi^{\prime\prime}}\Poltwo{\eta}{\xi^{\prime\prime}}\delta_{\mathbf Q^{\prime},\mathbf 0}^{\xi^{\prime\prime},\xi^{\prime\prime\prime}} + \Nextwo{\rho,\eta,\mathbf Q^{\prime}}{\xi^{\prime\prime},\xi^{\prime\prime\prime}} \Polstartwo{\mu}{\xi}\Poltwo{\nu}{\xi}\delta_{\mathbf Q,\mathbf 0}^{\xi,\xi^{\prime}}.
\end{multline}
If we only consider diagonal excitonic occupations $\sigma=\lambda$, $\mu=\nu$ and $\rho=\eta$ on both sides, Eq.~\eqref{eq:N_EOM_coul_CORR_fourth_order} vanishes:
\begin{align}
    \mathrm i\hbar\partial_t \Nextwo{\sigma,\sigma,\mathbf Q}{\xi,\xi^{\prime}}\Big|_{\mathbf E^4,V,\text{CORR}} = 0.
\end{align}
Therefore, there does not exist a direct coupling between excitonic occupations, but excitonic occupations at quantum number $\mu$ can induce intraexcitonic coherences at quantum number $\mu,\nu$, which in turn couple back to excitonic occupations at quantum number $\nu$ creating potential excitonic oscillations. 
In Refs.~\cite{wang2007excitonic,wang2009excitonic}, similar expressions have been derived in the limit of vanishing center-of-mass momentum, where the interaction of excitonic occupations and intraexcitonic coherences involving higher-lying excitonic states has been shown to cause an oscillatory behavior in the $1s$ excitonic occupation in time.

At last, contributions, which describe the coupling between two-exciton transitions and exciton-two-exciton transitions via optical ($\Omega$) and Coulomb ($V$) interaction, occur on fourth order. The optical interaction process, cf.~process ``$\Omega B^*$'' denoted by an arrow from $P$ to $N$ or ``$\Omega^* P^*$'' denoted by an arrow from $B$ to $N$ (both denote the same process) in Fig.~\ref{fig:4th_order_scheme}, exhibits the following structure:
\begin{align}
    \mathrm i\hbar\partial_t \Nextwo{\sigma,\lambda,\mathbf Q}{\xi,\xi^{\prime}}\Big|_{\mathbf E^4,\Omega,\text{XX}} \sim \hbar\Omega^{vc}\Polstar{}B + \hbar\Omega^{vc}Z - \begin{pmatrix}
        \sigma\leftrightarrow\lambda
    \end{pmatrix}^*.
    \label{eq:N_EOM_opt_XX_fourth_order}
\end{align}
The Coulomb interaction process, cf.~process ``$VP B^*$'' denoted by an arrow from $P$ to $N$ or ``$VP^* P^*$'' denoted by an arrow from $B$ to $N$ (both denote the same process) in Fig.~\ref{fig:4th_order_scheme}, exhibits the following structure:
\begin{align}
    \mathrm i\hbar\partial_t \Nextwo{\sigma,\lambda,\mathbf Q}{\xi,\xi^{\prime}}\Big|_{\mathbf E^4,V,\text{XX}} \sim V\Polstar{}\Polstar{}B + V\Polstar{}Z - \begin{pmatrix}
        \sigma\leftrightarrow\lambda
    \end{pmatrix}^*.
    \label{eq:N_EOM_coul_XX_fourth_order}
\end{align}
Eq.~\eqref{eq:N_EOM_opt_XX_fourth_order} and Eq.~\eqref{eq:N_EOM_coul_XX_fourth_order} exhibit very lengthy expressions, so that we refer to Eq.~(S66) and Eq.~(S67) in the SM for their explicit forms. Again, similar to Eq.~\eqref{eq:N_EOM_opt_COH_fourth_order}, Eq.~\eqref{eq:N_EOM_opt_XX_fourth_order} and Eq.~\eqref{eq:N_EOM_coul_XX_fourth_order} induce incoherent, optically dark excitonic occupations out of the light cone due to optical and Coulomb interaction processes via intermediate two-exciton transitions or exciton-two-exciton transitions. 
Similar contributions to the coherent formation of excitonic occupations outside the light cone at $\mathbf Q>\mathbf 0$ have been found in Ref.~\cite{yang2004ultrafast}, where a fully bosonic excitonic description has been applied. In particular, the fully coherent contributions in Eq.~\eqref{eq:N_EOM_coul_XX_fourth_order} (first terms $\sim \Polstar{}\Polstar{}\Bex{}$) resemble the terms found in Ref.~\cite{yang2004ultrafast}. However, in our numerical simulations, it turns out, that the most important contributions to the coherent excitation of excitonic occupations out of the light cone are the fermionic Pauli-blocking contributions in Eq.~\eqref{eq:N_EOM_opt_COH_fourth_order} and Eq.~\eqref{eq:N_EOM_opt_CORR_fourth_order}. Eq.~\eqref{eq:N_EOM_coul_XX_fourth_order} constitute only minor corrections, which we found to \textit{attenuate} rather than increase the overall coherently excited excitonic occupation out of the light cone. This observation is in line with the argumentation in Ref.~\cite{axt2001evidence}, where similar expressions in a Wannier basis have been derived.

We note, that, in fourth order, additional contributions to the two-exciton transitions $\Bex{}$ with respect to optical interaction and Coulomb interaction appear. However, since we always eliminate the two-exciton and exciton-two-exciton continua within a Markov approximation in our simulations, these fourth-order contributions would yield equations of motion already beyond fourth order. Hence, we disregard them. Thus, a Markovian elimination of the two-exciton continua, cf.~Eq.~\eqref{eq:B_EOM_second_order}, and exciton-two-exciton continua, cf.~Eq.~\eqref{eq:Z_EOM_third_order}, corresponds to a second-order Born approximation on third order DCT, but is already beyond a Born approximation in fourth-order DCT, as the coupling to the excitonic occupations/intraexcitonic coherences $N$ contains mixed light-matter and Coulomb interaction contributions. Hence, a second-order Born approximation is only exact up to third-order DCT, as a fourth-order DCT gives rise to mixed interactions.

Moreover, in a full treatment without a Markovian elimination, the additionally appearing fourth-order terms in the equations of motion of $B$ give rise to optical Pauli-blocking and Coulomb energy renormalizations of the biexcitons and a coupling between biexcitons $B$ and biexcitonic occupations, if correlated eight-operator expectation values are explicitly considered as well. These contributions could be necessary to explain power-dependent biexcitonic photoluminescence signatures \cite{ye2018efficient} opening up the possibility of a theoretical understanding of the biexciton gain \cite{hayamizu2007biexciton}. However, their evaluation is beyond the scope of this manuscript.

We also note, that incoherent exciton scattering processes such as Auger scattering \cite{steinhoff2021microscopic,erkensten2021dark} and density-dependent exciton-phonon scattering \cite{katzer2024fermionic,katzer2023exciton} are also fourth-order processes, which can be straightforwardly included in our theory by adding the respective interaction Hamiltonians in Eq.~\eqref{eq:hamiltonian_total}.

\section{Excitonic Rabi Oscillations}
\label{sec:rabi_oscillations}
We assume an undoped confined semiconductor and 
examine the two-dimensionally confined carrier dynamics via the total electron density $N$:
\begin{align}
    N 
    = \frac{1}{\mathcal A}\sum_{\mathbf k,\xi}\langle \cdagtwo{\mathbf k}{\xi}\cndagtwo{\mathbf k}{\xi}\rangle,% \equiv N_{\text{SBE}},
    \label{eq:N_TOT_SBE}
\end{align}
in units of one over area. 
Within an expansion in electron-hole operator pairs \cite{katsch2018theory,ivanov1993self}, cf.~Eq.~\eqref{eq:unit_operator_method_example}, the total electron density $N$ can also be expressed as:
\begin{align}
    N = \frac{1}{\mathcal A}\sum_{\mu,\mathbf Q,\xi,\xi^{\prime}}\left(|\Poltwo{\mu}{\xi}|^2\delta_{\mathbf Q,\mathbf 0}\delta_{\xi,\xi^{\prime}}+\Nextwo{\mu,\mathbf Q}{\xi,\xi^{\prime}}\right),
    \label{eq:N_TOT_EXC}
\end{align}
where $\Poltwo{\mu}{\xi}$ is the coherent excitonic transition from Eq.~\eqref{eq:P} and $\Nextwo{\mu,\mathbf Q}{\xi,\xi^{\prime}}$ is the incoherent excitonic occupation from Eq.~\eqref{eq:N}. Note, that we truncated the expansion in electron-hole pairs, cf.~Eq.~\eqref{eq:unit_operator_method_example} on a doublet level. A more complete treatment in fourth-order DCT would involve higher-order correlations up to quadruplets.

To study the exciton-density dynamics in Eq.~\eqref{eq:N_TOT_EXC}, consider two cases, a MoSe$_2$ ML (strong Coulomb coupling) and a GaAs QW (less strong Coulomb coupling). Both systems are resonantly excited at $\hbar\omega_{\text{P}} = E_{1s}$ using a circularly polarized 2-ps pulse. We solve the excitonic equations of motion for the excitonic transitions $\Pol{1s}$ and excitonic occupations $\Nex{1s,\mathbf Q}$. The equations of motion for the two-exciton continua $\Bex{+,\zeta,1s}$ and exciton-two-exciton continua $\Rex{+,\zeta,1s,\mathbf Q}$ for $\xi=\xi^{\prime} = K,\uparrow$ (MoSe$_2$ ML) or $\xi=\xi^{\prime} = \text{heavy-hole~exciton}$ (GaAs QW) are eliminated via a Markov approximation, i.e., we assume an instantaneous build-up of exciton-exciton interaction and neglect memory effects \cite{axt2004estimating}, which is reasonable given the comparably long pulse duration. We take only $\mu=1s$ excitonic transitions and occupations into account and neglect any intraexcitonic coherences $\Nex{\mu,\nu\neq\mu,\mathbf Q}$, but include $1s$, $2s$ and $3s$ excitonic states in the calculation of the symmetric two-exciton continua $\Bex{+,\zeta}$ and symmetric exciton-two-exciton continua $\Rex{+,\zeta}$. Due to the circular excitation, antisymmetric ``$-$''-configurations, which host a bound biexciton/exciton-biexciton, are not induced.

For comparison with the excitonic theory, we also solve the well-established semiconductor Bloch equations in Hartree-Fock limit \cite{lindberg1988effective} to obtain the density of uncorrelated electrons in Eq.~\eqref{eq:N_TOT_SBE}:
\begin{align}
\begin{split}
&\mathrm i \hbar \partial_{t} \Pol{\mathbf{k}} =  (E_\mathbf{k} -\hbar\omega_{\text{P}}
- \mathrm i\hbar\gamma_{\text{nrad}}) \Pol{\mathbf{k}} - \sum_{\mathbf q}V_{\mathbf q-\mathbf k}\Pol{\mathbf q}\\
&\,-\hbar\Omega^{cv}\left(1-2n_\mathbf{k}\right) + 2\sum_{\mathbf q}V_{\mathbf q-\mathbf k}\left(n_{\mathbf k}\Pol{\mathbf q} - n_{\mathbf q}\Pol{\mathbf k}\right),\\
&\mathrm i\hbar \partial_{t} n_\mathbf{k} =  \hbar\Omega^{vc}\Pol{\mathbf k} + \sum_{\mathbf q}V_{\mathbf q-\mathbf k}\Polstar{\mathbf q}\Pol{\mathbf k} - \text{c.c.},
\end{split}
\label{eq:SBE_EOM}
\end{align}
where $\Pol{\mathbf k} = \langle\vdag{\mathbf k}\cndag{\mathbf k}\rangle$ is the interband transition, cf.\ also Eq.~\eqref{eq:P}, and $n_{\mathbf k}^{\xi}$ are the electron occupations:
\begin{align}
    n_{\mathbf k}^{\xi} = \langle \cdagtwo{\mathbf k}{\xi}\cndagtwo{\mathbf k}{\xi}\rangle.
    \label{eq:n_el}
\end{align}
All material parameters are listed in Tab.~\ref{tab:parameters}. To ensure comparable conditions, we assume equal nonradiative dephasings $\gamma_{\text{nrad}}$ for the MoSe$_2$ ML and GaAs QW and neglect any contributions due to reradiation \cite{knorr1996theory}. In a MoSe$_2$ ML, a total (half) linewidth of 1\,meV is not far from the radiatively-limited case, which can be achieved by suppressing inhomogeneous broadening via h-BN encapsulation \cite{shree2018observation,cadiz2017excitonic,ajayi2017approaching}, while in a GaAs QW, a total (half) linewidth of 1\,meV is much larger than the radiative linewidth of approximately tens of \textmu eV  \cite{schafer2025distinct} and should be regarded as an effective linewidth for a fair comparison between the different materials. Incorporating inhomogeneous broadening, as typically causing the linewidth in GaAs QWs at temperatures around 4\,K \cite{thranhardt2003interplay}, would require an ensemble average.

\begin{table}[h!]
\caption{Parameters used in the numerical calculations.}
    \label{tab:parameters}
    \centering
    \begin{ruledtabular}
    \begin{tabular}{ll}
    Intensity FWHM & 2000$\,$fs\\
    Nonradiative (half) linewidth $\hbar\gamma_{\text{nrad}}$ & $1\,$meV\\
    \hline
    GaAs QW & \\
    \hline
    Real well width $d$ & 8$\,$nm\\
    Effective well width $d_{\text{eff}}$ & $1.3d$ \cite{liu1994local}\\
    %$1s$ exciton energy $E_{1s}$ & 1.6$\,$eV\\
    %$1s$ excitonic binding energy $E_b$ (calculated) & $-8.8$\,meV\\
    \begin{tabular}{l}
    \hspace{-0.2cm}
    $1s$ excitonic binding energy \\
    \hspace{-0.2cm}
    $E_{1s}-\tilde E_{\text{gap}}$ (calculated) 
    \end{tabular} & $-8.8$\,meV\\
    Static dielectric constant of GaAs & 12.46 (5$\,$K) \cite{moore1996infrared}\\
    Effective electron mass $m_e$ & 0.0665$m_0$ \cite{binder1991theory}\\
    Effective heavy-hole mass $m_{h}$ & 0.1106$m_0$ \cite{bataev2022heavy}\\
    \hline
    h-BN-encapsulated MoSe$_2$ ML & \\
    \hline
    Monolayer width $d$ & 0.6527$\,$nm \cite{kylanpaa2015binding}\\
    \begin{tabular}{l}
    \hspace{-0.2cm}
    $1s$ excitonic binding energy \\
    \hspace{-0.2cm}
    $E_{1s}-\tilde E_{\text{gap}}$ (calculated) 
    \end{tabular} & $-341$\,meV\\
    \begin{tabular}{l}
    \hspace{-0.2cm}
    Biexcitonic binding energy\\ 
    \hspace{-0.2cm}
    $E_{B,-,b}-2E_{1s}$ (calculated)
    \end{tabular}&$-19.3$\,meV\\
    Static dielectric constant of bulk MoSe$_2$ & $\sqrt{8.2\cdot 17.7}$ \cite{laturia2018dielectric}\\
    Effective electron mass $m_e$ & 0.5$m_0$ \cite{kormanyos2015k}\\
    Effective hole mass $m_{h}$ & 0.6$m_0$ \cite{kormanyos2015k}\\
    Plasmon energy of MoSe$_2$ $\hbar\omega_{\text{pl}}$ & 22\,eV \cite{kumar2012tunable}\\
    TF screening parameter of MoSe$_2$ $\alpha_{\text{TF}}$ &
    \begin{tabular}{l}
    \hspace{-0.2cm}
    1.9 (adj.\ to \\
    %\hspace{-0.2cm}
    CMR \cite{andersen2015dielectric})
    \end{tabular}\\
    Static dielectric constant of bulk h-BN & 4.8 \cite{latini2015excitons}\\
    Interlayer gap $h$ & 0.3\,nm \cite{florian2018dielectric,druppel2017diversity}
    \end{tabular}
    \end{ruledtabular}
    \end{table}

\begin{figure}
    \centering
    \begin{tabular}{c@{\hspace{-0.1cm}} c@{\hspace{0.05cm}} c@{\hspace{0.0cm}}}
    \begin{tabular}{c}
      \begin{turn}{90} Density ($10^{10}\,$cm$^{-2}$)\end{turn}
      \end{tabular}
      & 
        \begin{tabular}{c@{\hspace{-0.1cm}}c@{\hspace{0cm}}}
         GaAs QW & MoSe$_2$ ML \\
        \includegraphics[height=0.394\linewidth]{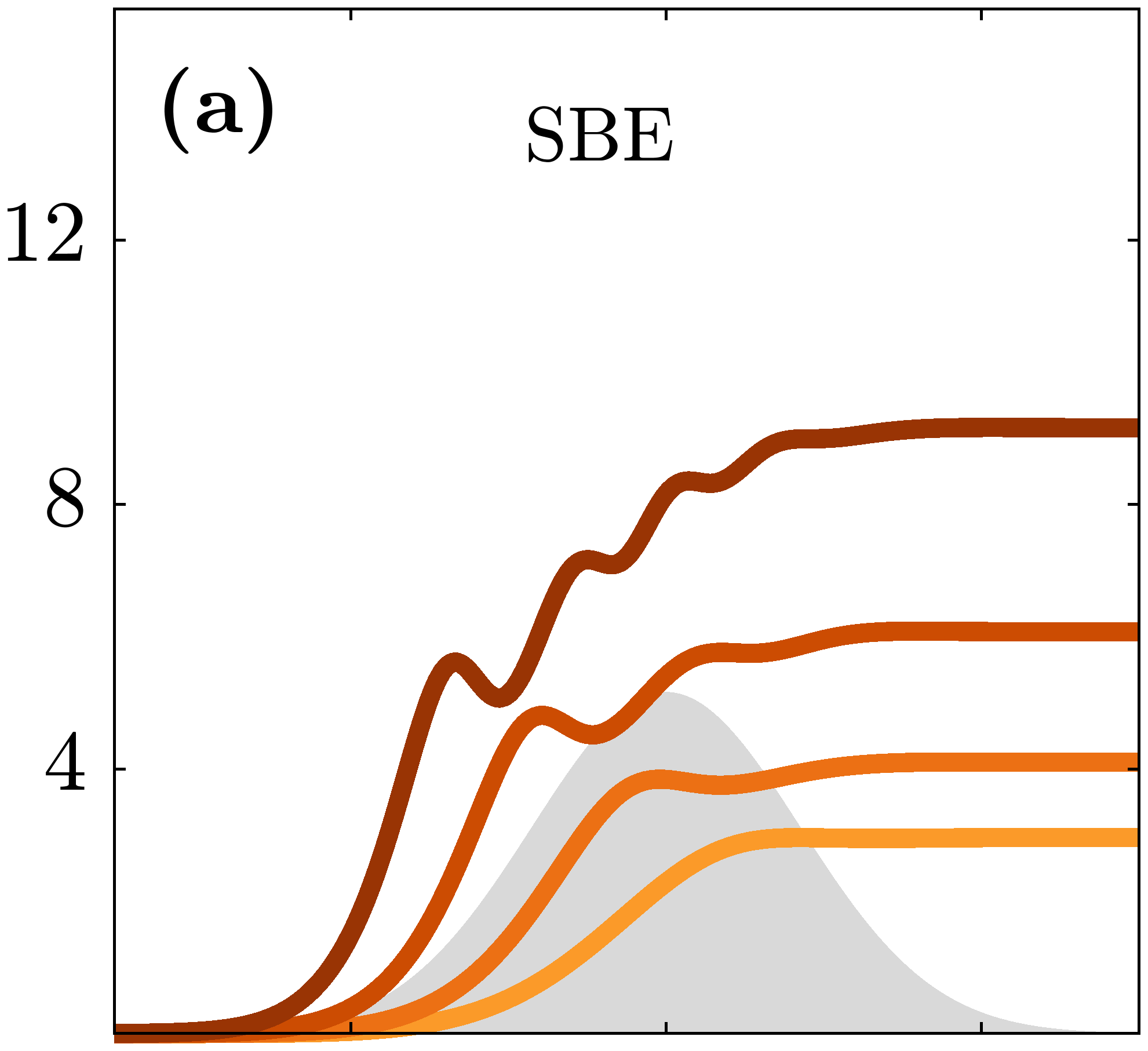}
        &
        \includegraphics[height=0.3945\linewidth]{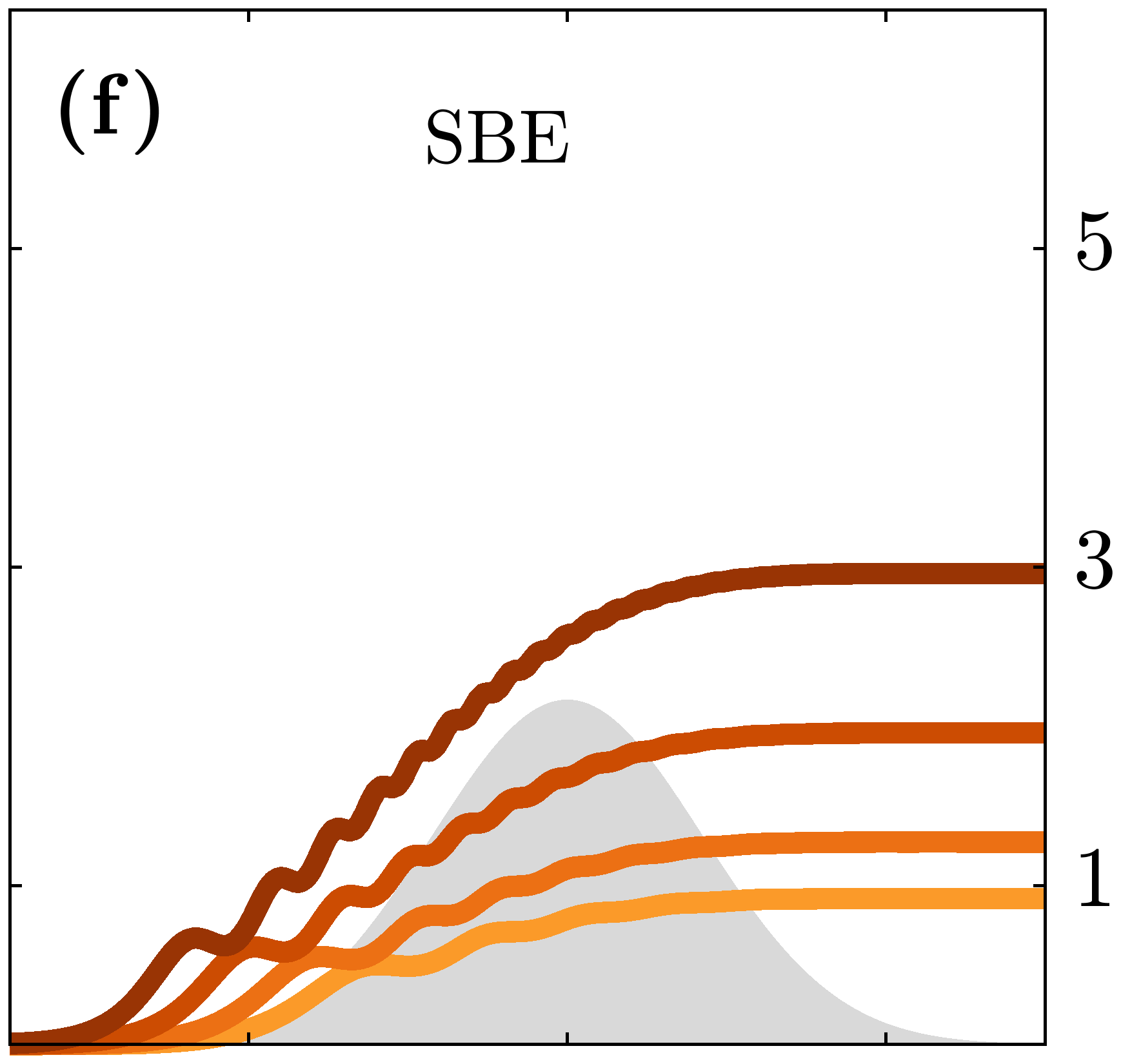}
        \\[-1mm]
        \includegraphics[height=0.39\linewidth]{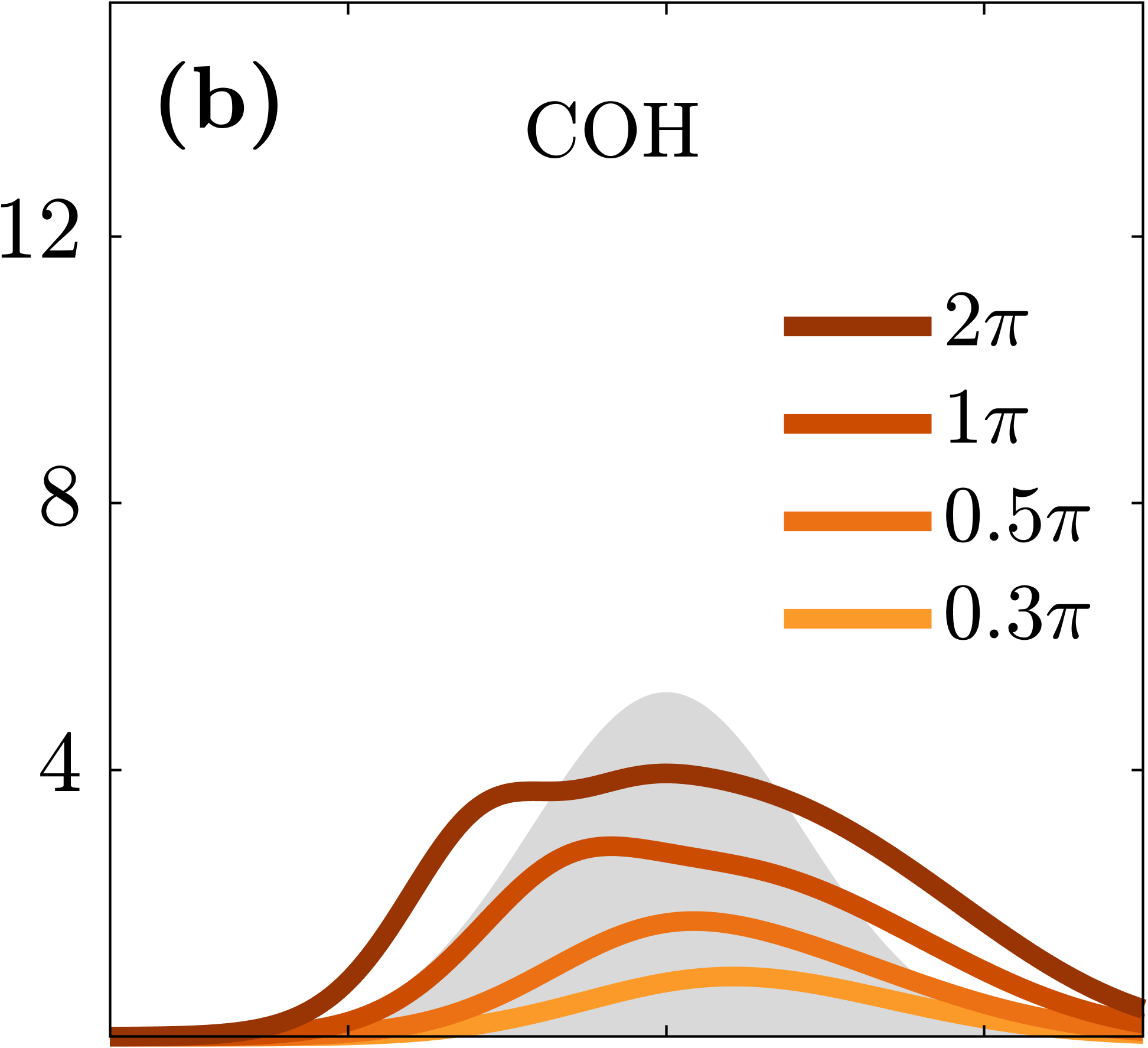}
        &
        \includegraphics[height=0.39\linewidth]{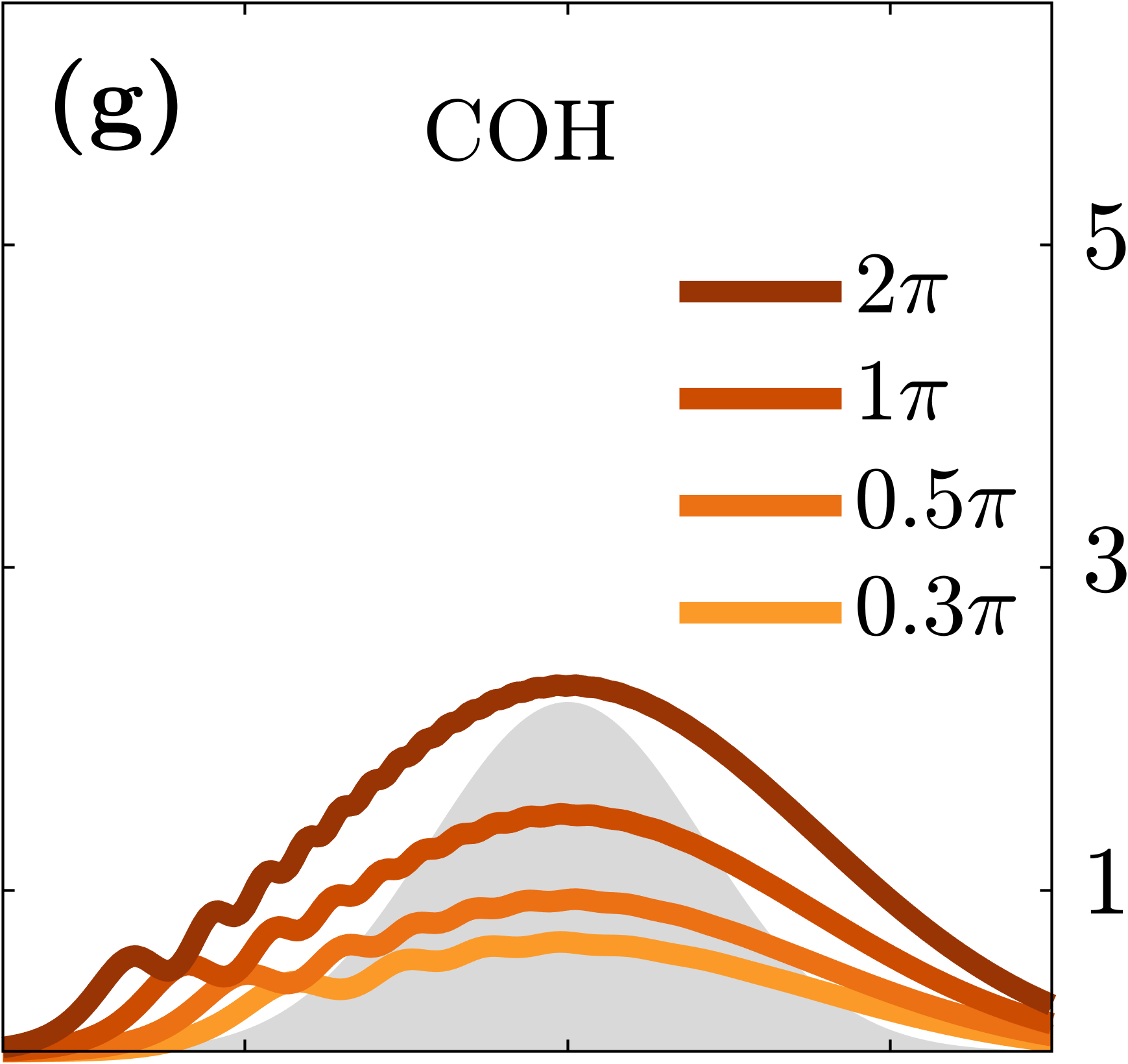}
        \\[-1mm]
        \includegraphics[height=0.39\linewidth]{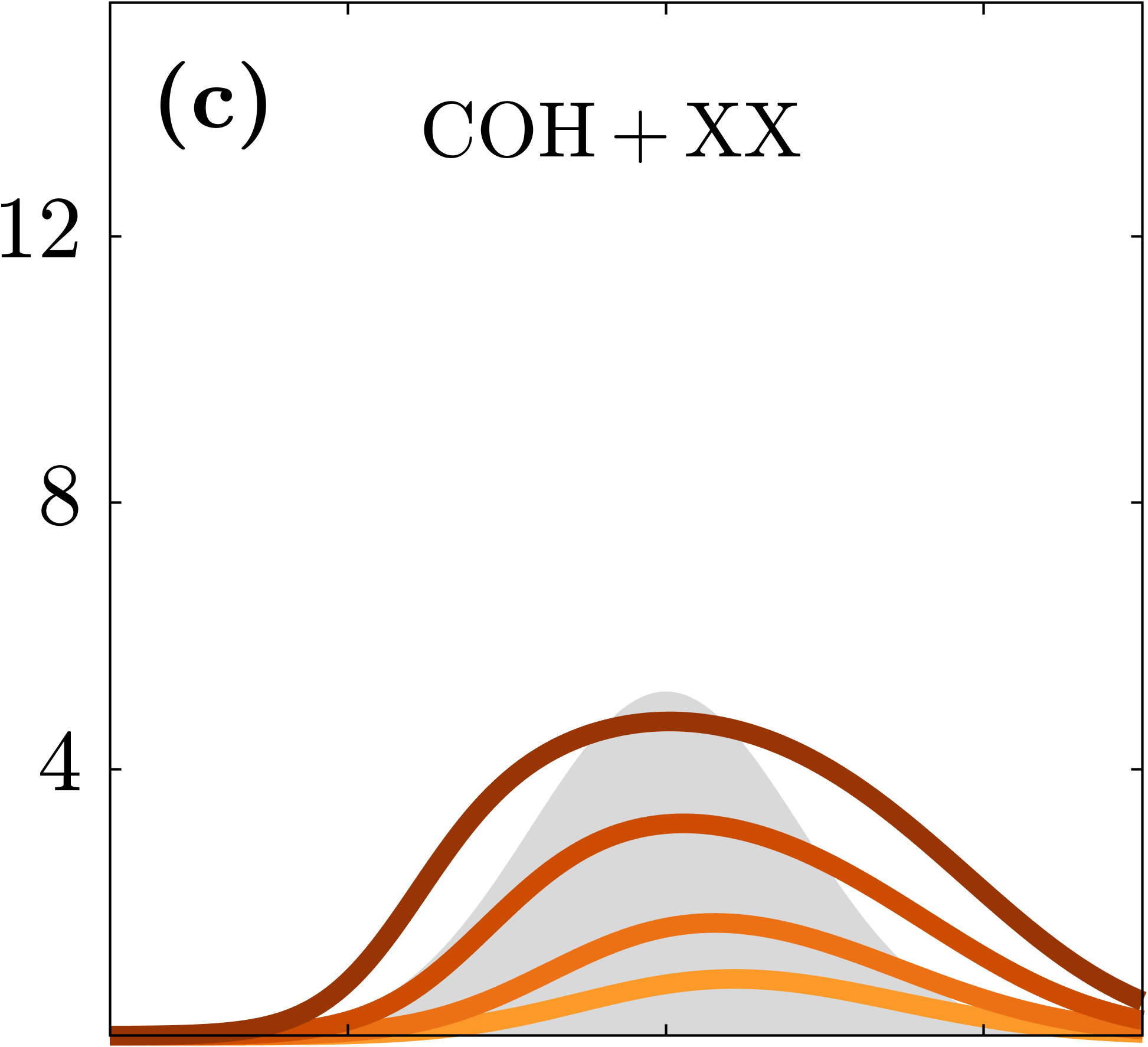}
        &
        \includegraphics[height=0.39\linewidth]{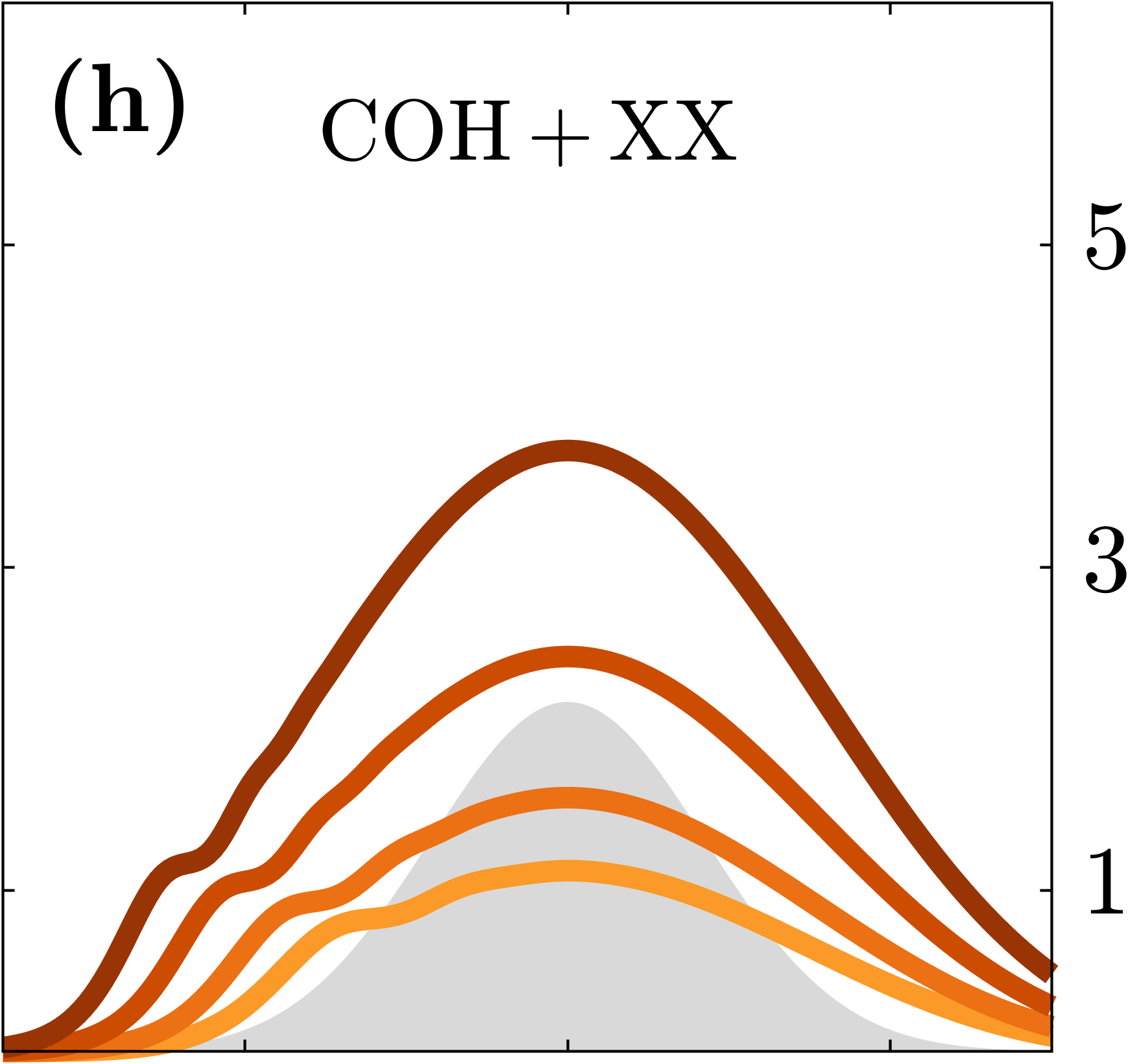}
        \\[-1mm]
        \includegraphics[height=0.39\linewidth]{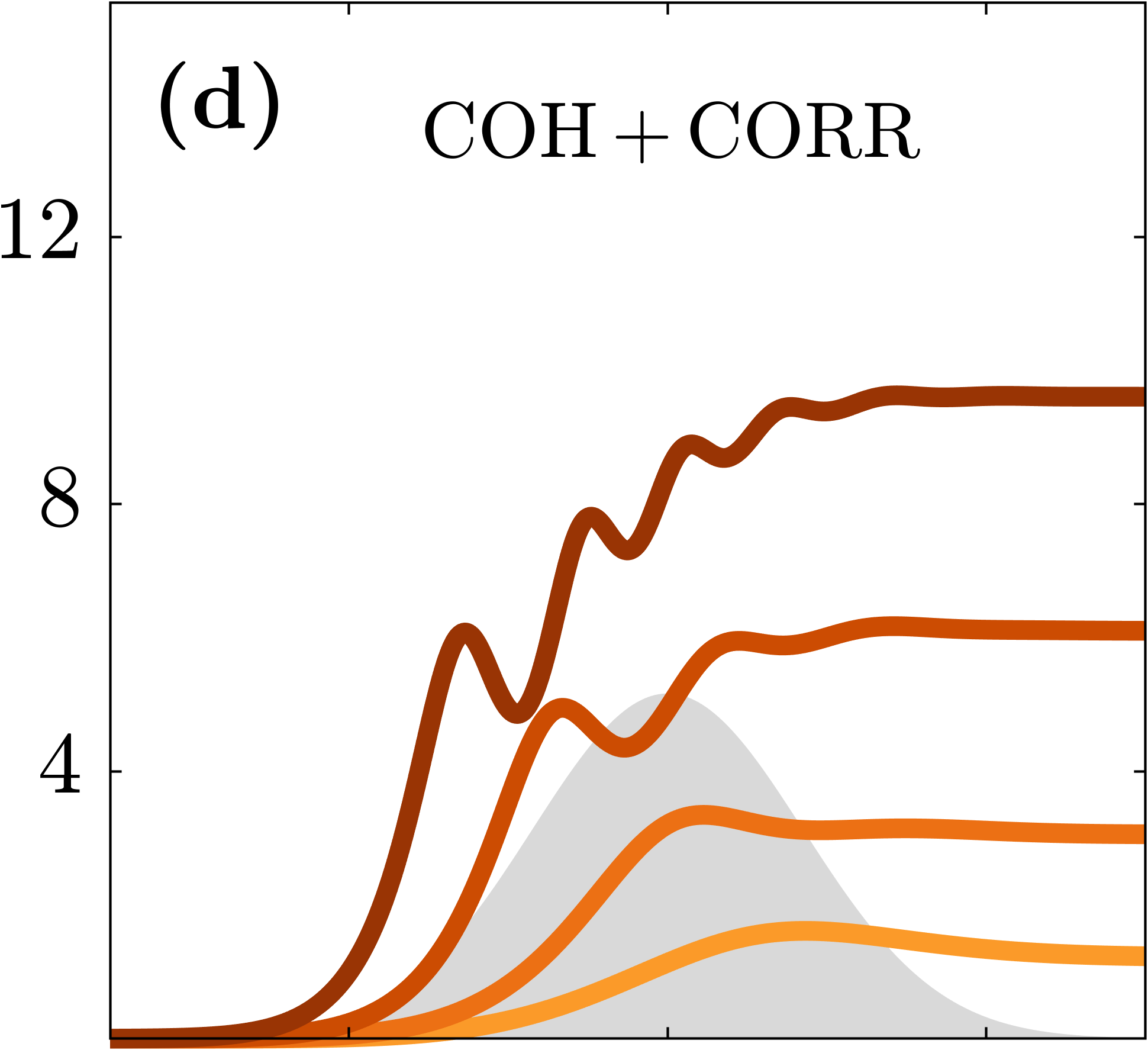}
        &
        \includegraphics[height=0.39\linewidth]{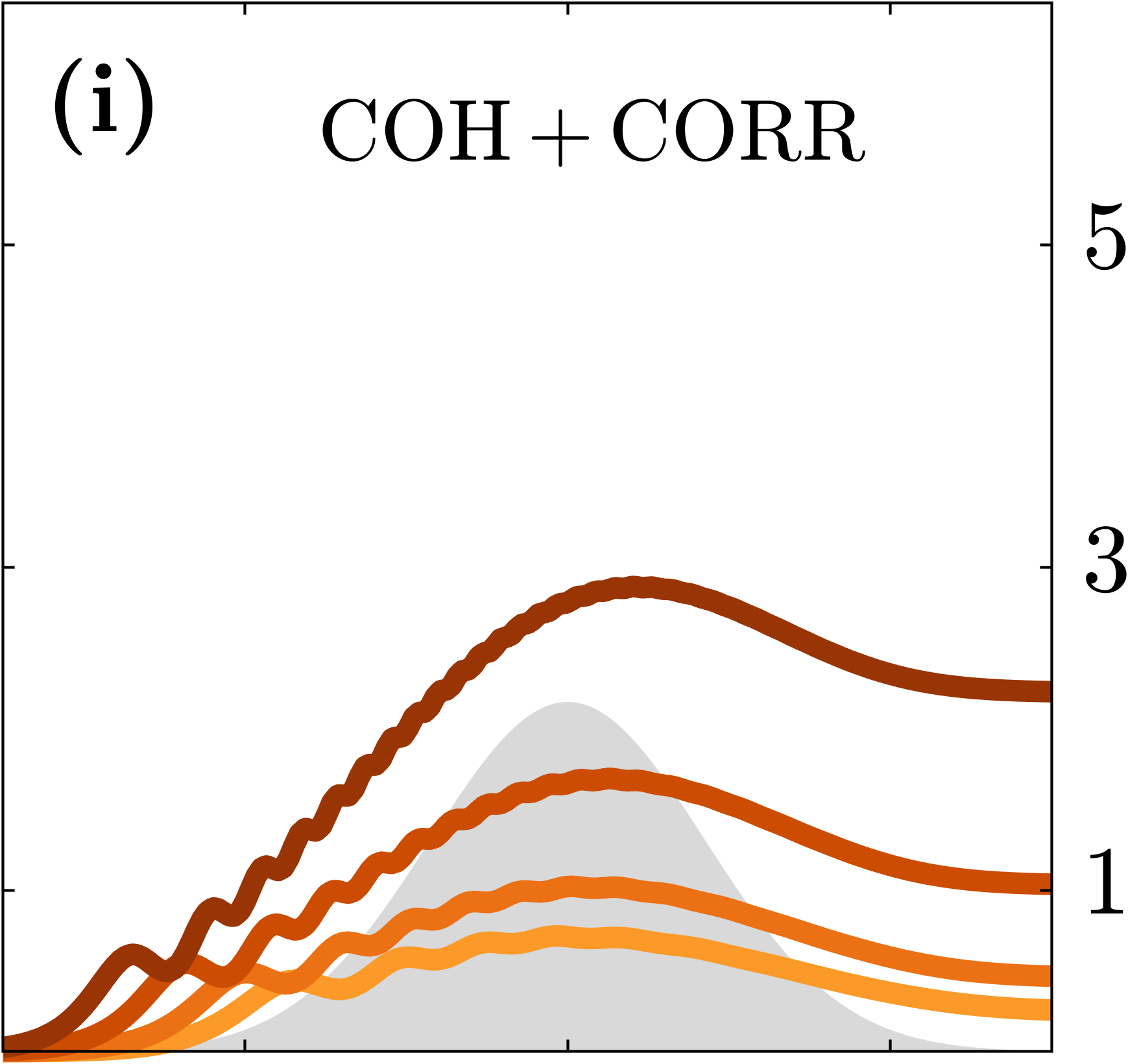}
        \\[-1mm]
        \begin{tabular}{c}
            \includegraphics[height=0.4788\linewidth]{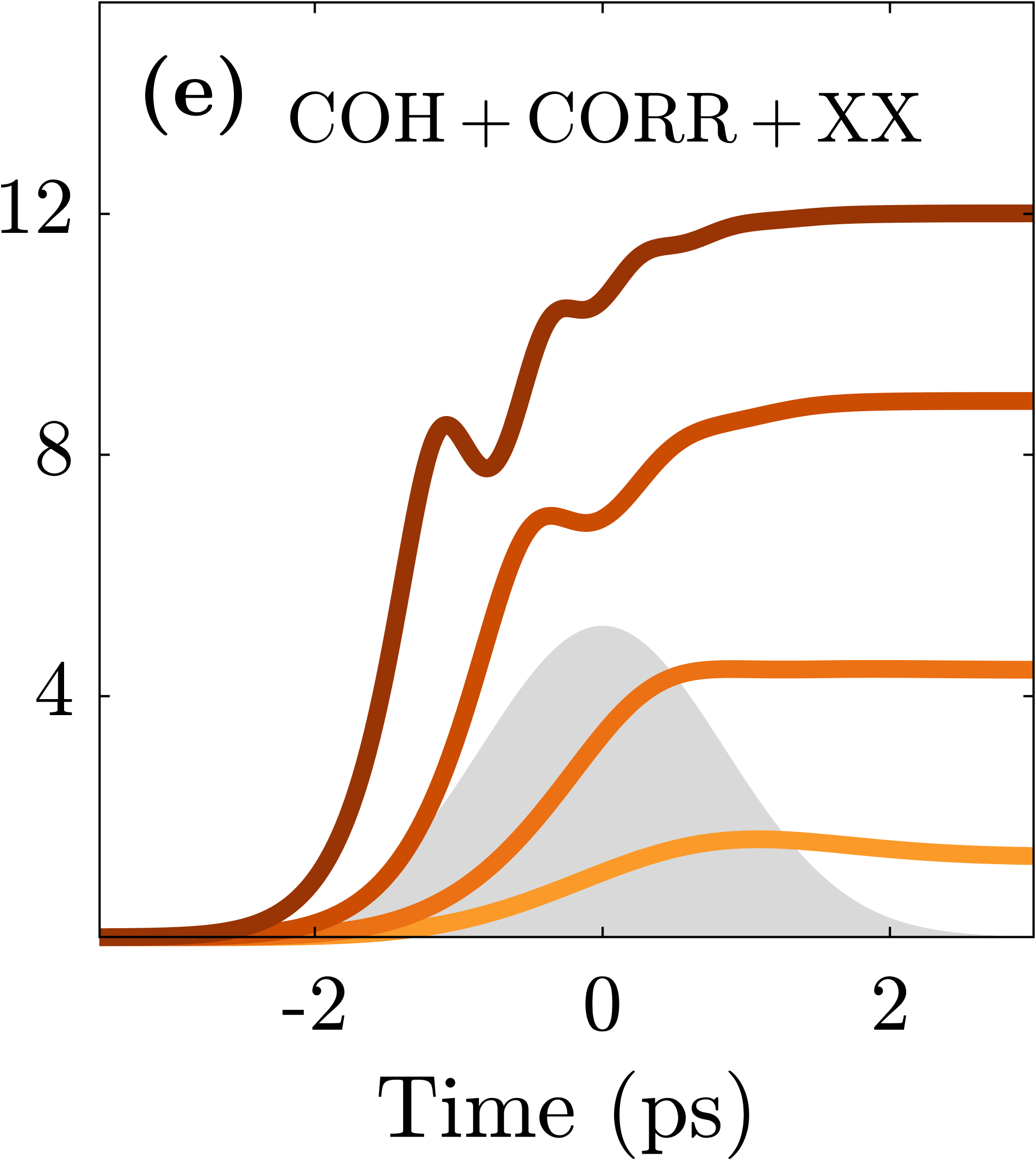}
        \end{tabular}
        &
        \begin{tabular}{c}
            \includegraphics[height=0.482\linewidth]{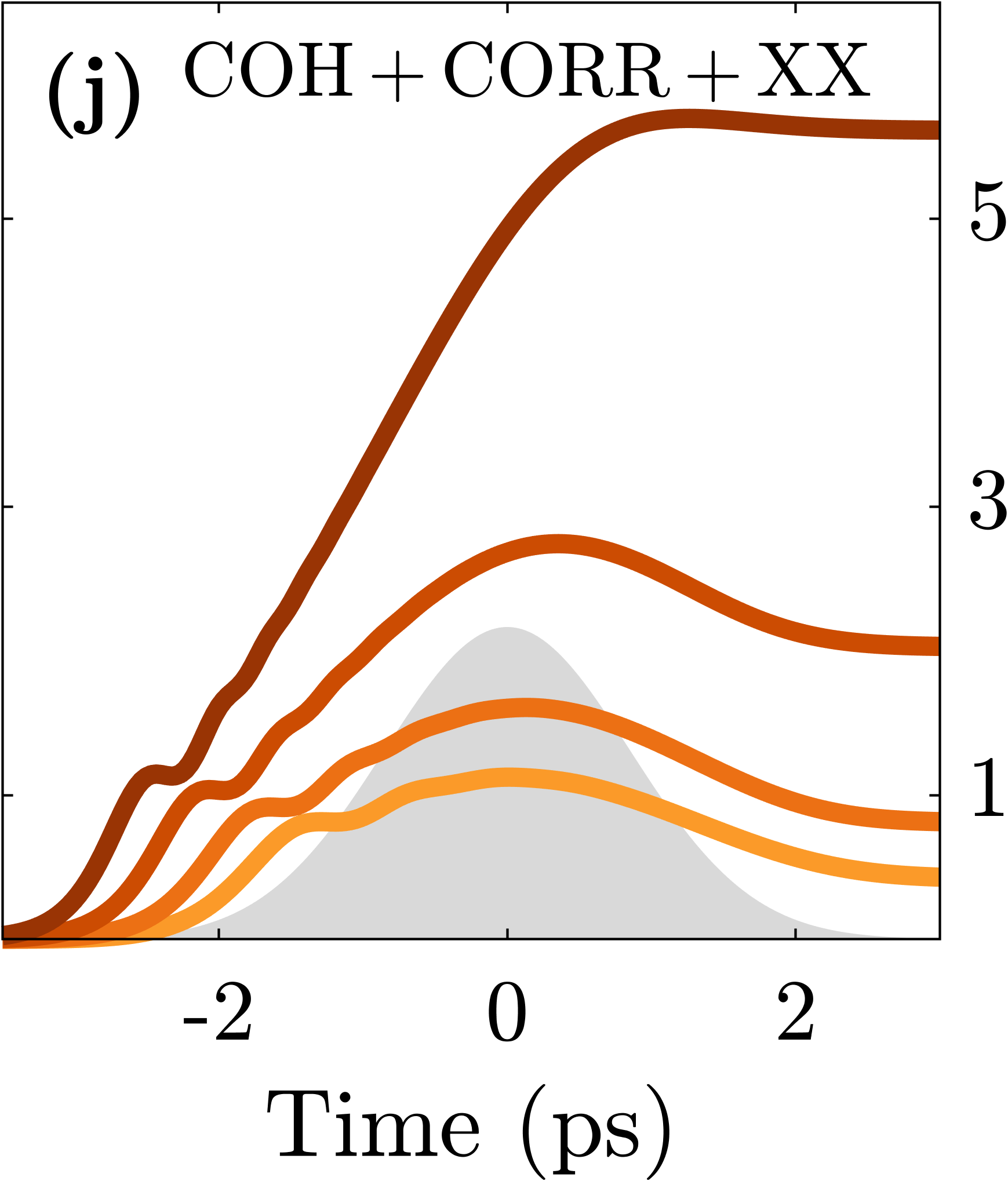}
        \end{tabular}
        \end{tabular}
     &
     \begin{tabular}{c}
     \begin{turn}{-90} Density ($10^{12}\,$cm$^{-2}$)\end{turn}
     \end{tabular}
    \end{tabular}
    \caption{Dynamics of the total electron density $N$ from Eq.~\eqref{eq:N_TOT_SBE} as solution of the semiconductor Bloch equations (SBE) (a,f) compared to the total exciton density $N$ from Eq.~\eqref{eq:N_TOT_EXC} as a solution of the excitonic Bloch equations in the coherent limit (COH) (b,g), 
    with additional exciton-exciton scattering contributions (COH$\,+\,$XX) (c,h) and with additional correlated contributions (COH$\,+\,$CORR) (d,i) 
    and with all contributions (COH$\,+\,$CORR$\,+\,$XX) (e,j) at various pulse areas $\Theta$, cf.~Eq.~\eqref{eq:PulseArea}, for a 8-nm GaAs QW (left column) and for a MoSe$_2$ ML (right column). The grey-shaded area depicts the normalized pulse intensity.}
    \label{fig:dynamics_exc}
\end{figure}

In the following, we dissect the corresponding interaction contributions to the exciton dynamics and compare the emerging dynamics of a MoSe$_2$ ML and a GaAs QW for circularly-polarized excitation (\textit{GaAs QW vs.\ MoSe$_2$ ML}). Then, we identify the role of the coherent and (optically dark) incoherent exciton density (\textit{Coherent vs.\ incoherent exciton density}) and examine the momentum distribution of optically dark excitons (\textit{Momentum distribution of optically dark excitons}). At last, we compare circular and linear excitation in a MoSe$_2$ ML (\textit{Circular vs.\ linear excitation}).

\textit{GaAs QW vs.\ MoSe$_2$ ML:} 
In Fig.~\ref{fig:dynamics_exc}, we depict the dynamics of the electron density, cf.\ Eq.~\eqref{eq:N_TOT_SBE}, by solving the semiconductor Bloch equations (SBEs) in Eq.~\eqref{eq:SBE_EOM} and the dynamics of the total excitonic density, cf.\ Eq.~\eqref{eq:N_TOT_EXC}, as solutions of the full excitonic Bloch equations in Eqs.~\eqref{eq:P_EOM_first_order}--\eqref{eq:N_EOM_coul_XX_fourth_order} within the limits discussed above at various pulse areas, defined in Eq.~\eqref{eq:PulseArea}. The left column depicts simulations for a GaAs QW and the right column depicts the simulations for a h-BN-encapsulated MoSe$_2$ ML.

In Fig.~\ref{fig:dynamics_exc}(a,f), which shows the results from solving the SBEs, Eq.~\eqref{eq:SBE_EOM}, Rabi flops are observed, whose period increases with increasing pulse area. In Fig.~\ref{fig:dynamics_exc}(a) (GaAs QW), up to four Rabi flops can be observed at a pulse area of $2\pi$, while Fig.~\ref{fig:dynamics_exc}(f) (MoSe$_2$ ML) shows a multitude of Rabi flops. As Rabi oscillations are enhanced due to the occurrence of Coulomb-induced internal fields, cf.\ last term in the first line in Eq.~\eqref{eq:SBE_EOM}, and the Coulomb interaction in a MoSe$_2$ ML is much stronger compared to a GaAs QW due to the reduced screening, more Rabi flops occur in a MoSe$_2$ ML. 

In Fig.~\ref{fig:dynamics_exc}(b,g), we depict the coherent (COH) limit of the excitonic Bloch equations, cf.\ Eq.~\eqref{eq:P_EOM_first_order} and Eq.~\eqref{eq:P_EOM_COH_third_order}, which corresponds to the excitonic description in Ref.~\cite{knorr1994asymptotic}, which does not include nonradiative dephasing. Here, the exciton density decays within 3\,ps, as no long-lived incoherent occupations $\Nex{1s,\mathbf Q}$ are induced. In the GaAs QW, Fig.~\ref{fig:dynamics_exc}(b), one Rabi flop occurs at the largest pulse area, as the nonradiative dephasing $\hbar\gamma_{\text{nrad}}$ suppresses them. In a MoSe$_2$ ML, Fig.~\ref{fig:dynamics_exc}(g), Rabi flops occur in a similar number compared to the SBE-case in Fig.~\ref{fig:dynamics_exc}(f), since the stronger Coulomb-induced field renormalization counterbalances the nonradiative dephasing on short timescales.

In Fig.~\ref{fig:dynamics_exc}(c,h), we depict the coherent limit with exciton-exciton interaction corresponding to two-exciton continua (COH$\,+\,$XX), cf.\ Eq.~\eqref{eq:P_EOM_first_order}, Eq.~\eqref{eq:P_EOM_COH_third_order}, Eq.~\eqref{eq:P_EOM_XX_third_order} and Eq.~\eqref{eq:B_EOM_second_order}, which corresponds to the model developed in Refs.~\cite{katsch2019theory,katsch2020exciton,katsch2020optical}. In this limit, the two-exciton-induced dephasing, which corresponds to the coupling of excitonic transitions to two-exciton continua (excitation-induced dephasing) totally suppresses the (coherent) Rabi oscillations in the GaAs QW and greatly suppresses them in a MoSe$_2$ ML, but not completely, as two Rabi flops of small amplitude still remain. The overall induced coherent exciton density is larger compared to Fig.~\ref{fig:dynamics_exc}(b,g), as the exciton-exciton interaction reduces the overall excitation-dependent blue shift due to bandgap reduction and phase-space filling in Eq.~\eqref{eq:P_EOM_COH_third_order} (first line) enabling a more efficient optical excitation of electron-hole pairs. This is especially relevant in a MoSe$_2$ ML, as the enhanced Coulomb interaction causes a stronger exciton-exciton interaction and hence a stronger excitation-induced dephasing and -attenuation of the blue shift.

In Fig.~\ref{fig:dynamics_exc}(d,i), we turn off the exciton-exciton (XX) interaction in Eq.~\eqref{eq:P_EOM_XX_third_order} and Eq.~\eqref{eq:B_EOM_second_order}, but include the correlated contributions in Eq.~\eqref{eq:P_EOM_CORR_third_order}, Eq.~\eqref{eq:N_EOM_opt_COH_fourth_order} and Eq.~\eqref{eq:N_EOM_opt_CORR_fourth_order}, i.e., we allow incoherent, optically dark excitonic occupations $\Nex{1s,\mathbf Q}$ to be induced via Coulomb-enhanced Pauli-blocking. Here, the exciton dynamics for the GaAs QW matches remarkably well the electron dynamics in the SBE in Fig.~\ref{fig:dynamics_exc}(a) with minor deviations at small pulse areas below $1\pi$. In a MoSe$_2$ ML, Fig.~\ref{fig:dynamics_exc}(i), the Rabi flopping dynamics is very similar to the SBE-case in Fig.~\ref{fig:dynamics_exc}(f) until the pulse maximum is reached, from which on deviations occur, as the coherent excitation of incoherent and optically dark excitonic occupations is less effective compared to the GaAs QW.

In Fig.~\ref{fig:dynamics_exc}(e,j), we consider all contributions, i.e., we turn on again Eq.~\eqref{eq:P_EOM_XX_third_order} as well as Eq.~\eqref{eq:B_EOM_second_order} and include Eq.~\eqref{eq:Z_EOM_third_order}, Eq.~\eqref{eq:N_EOM_opt_XX_fourth_order} and Eq.~\eqref{eq:N_EOM_coul_XX_fourth_order}. Here, the Rabi-flopping dynamics for a GaAs QW are slightly suppressed compared to Fig.~\ref{fig:dynamics_exc}(d) due to the excitation-induced dephasing, which occurs via the two-exciton continuum $\Bex{+,\zeta}$, driven nonlinearly via $\Pol{1s}\Pol{1s}$, cf.~Eq.~\eqref{eq:B_EOM_second_order}, as well as via the exciton-two-exciton continuum $\Rex{+,\zeta}$ driven by $\Pol{1s}\Nex{1s,\mathbf Q}$, cf.~Eq.~\eqref{eq:Z_EOM_third_order}. For the MoSe$_2$ ML, almost all Rabi flops are suppressed, while only a few in the very beginning around $-2\,$ps remain. In both cases, the GaAs QW and the MoSe$_2$ ML, the inclusion of exciton-exciton interaction enhances the overall optically injected exciton density with the most pronounced contribution in the MoSe$_2$ ML at the largest pulse areas. This amplification is caused by the red-shift contribution causing a more effective optical excitation of excitonic transitions, which, in turn, as the pulse area increases, leads to an enhanced coherent excitation of excitonic occupations in fourth order via $\hbar\Omega|\Pol{1s}|^2\Pol{1s}$, cf.~Eq.~\eqref{eq:N_EOM_opt_COH_fourth_order}.

\begin{figure}[h!]
    \centering
    \begin{tabular}{c@{\hspace{-0.1cm}} c@{\hspace{0.05cm}} c@{\hspace{0.0cm}}}
    \begin{tabular}{c}
      \begin{turn}{90} Occupation (1)\end{turn}
      \end{tabular}
      &  
        \begin{tabular}{c@{\hspace{-0.1cm}}c@{\hspace{0cm}}}
         Single TLS & SBE without Coulomb \\
        \begin{tabular}{c}
            \includegraphics[height=0.405\linewidth]{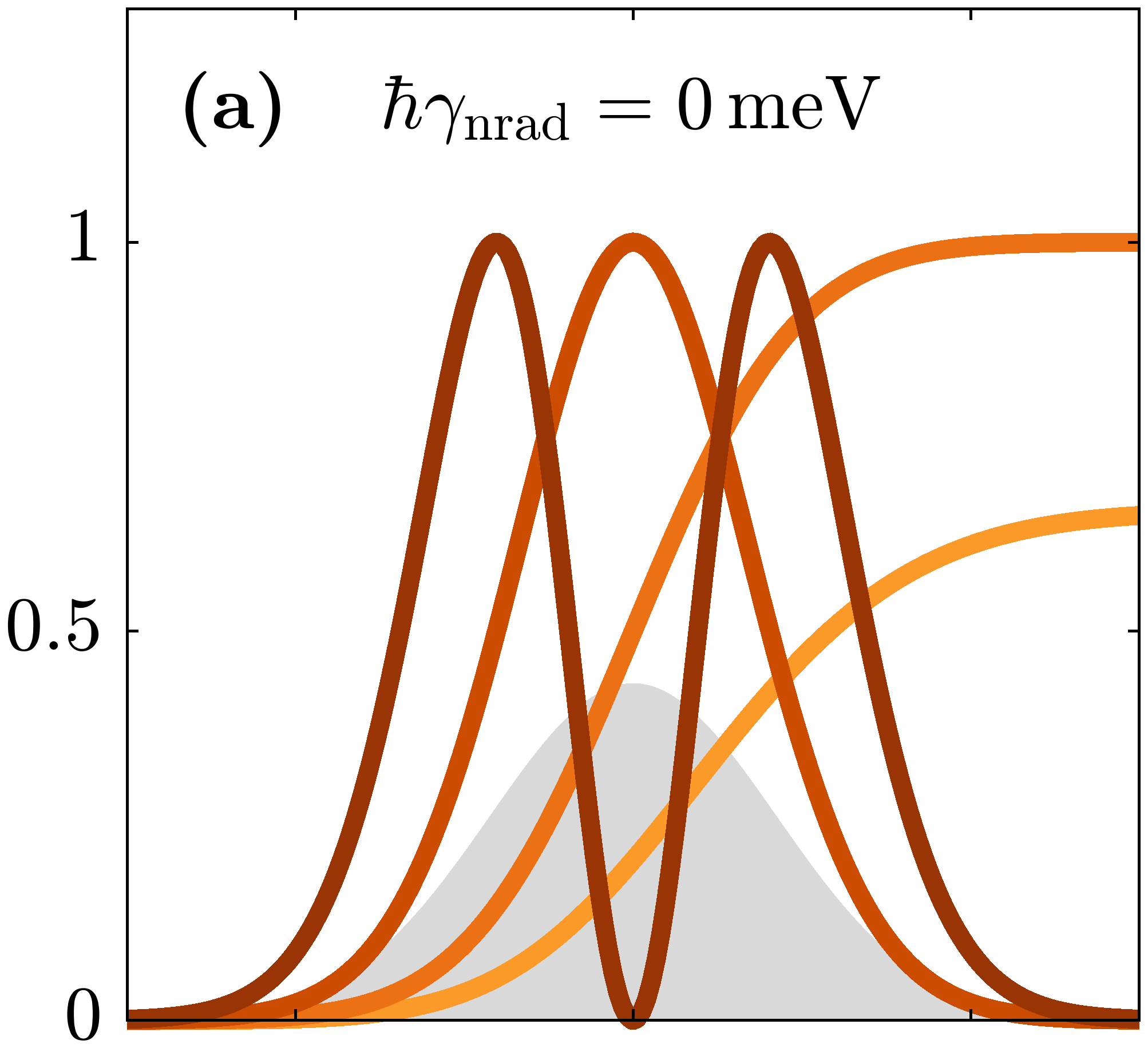}
        \end{tabular}
        &
        \hspace{-0.28cm}
        \begin{tabular}{c}
        \hspace{0.03cm}
        \vspace{0.09cm}
            \includegraphics[height=0.418\linewidth]{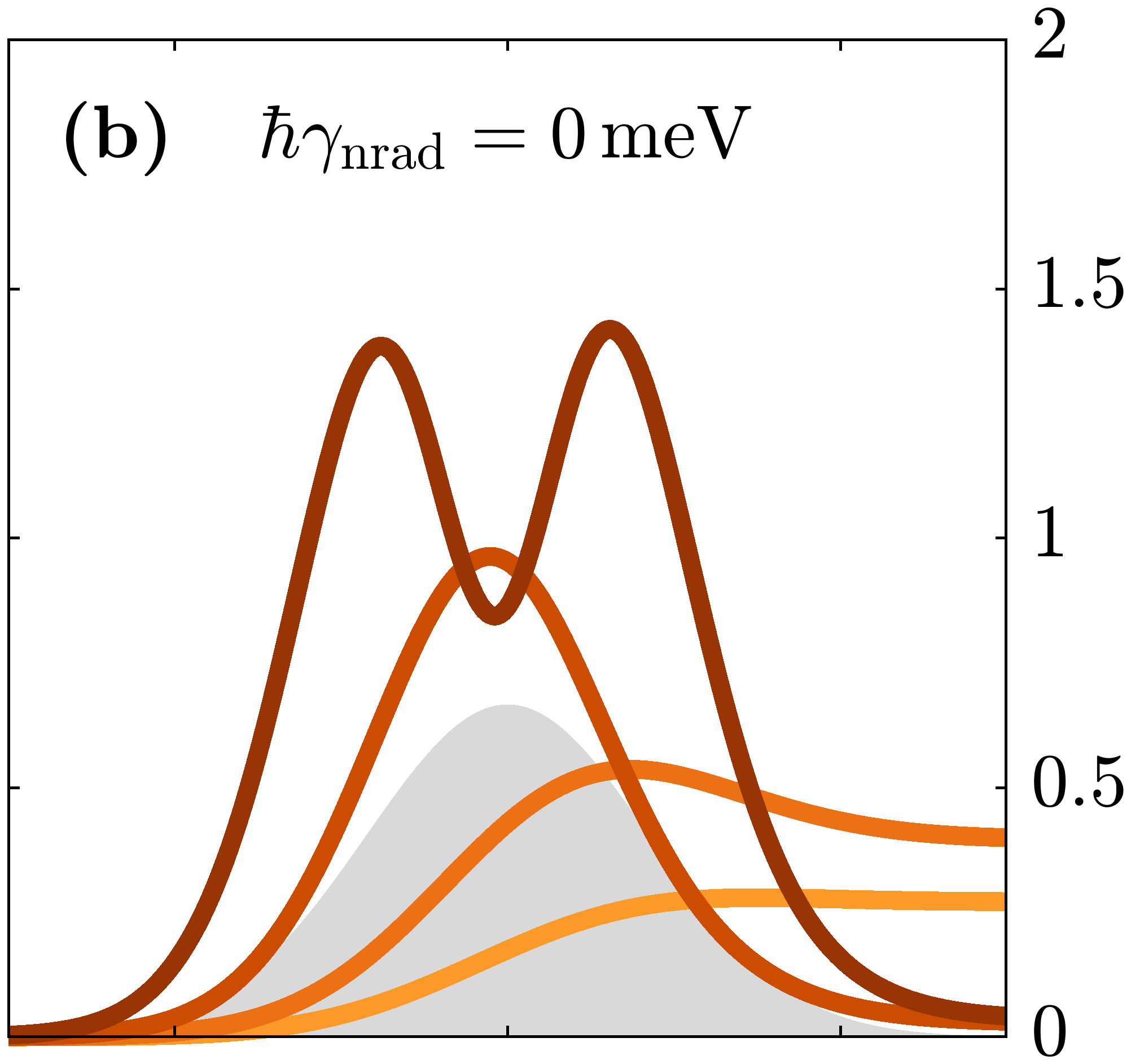}
        \end{tabular}
        \\[-2.3mm]
        \begin{tabular}{c}
            \includegraphics[height=0.479\linewidth]{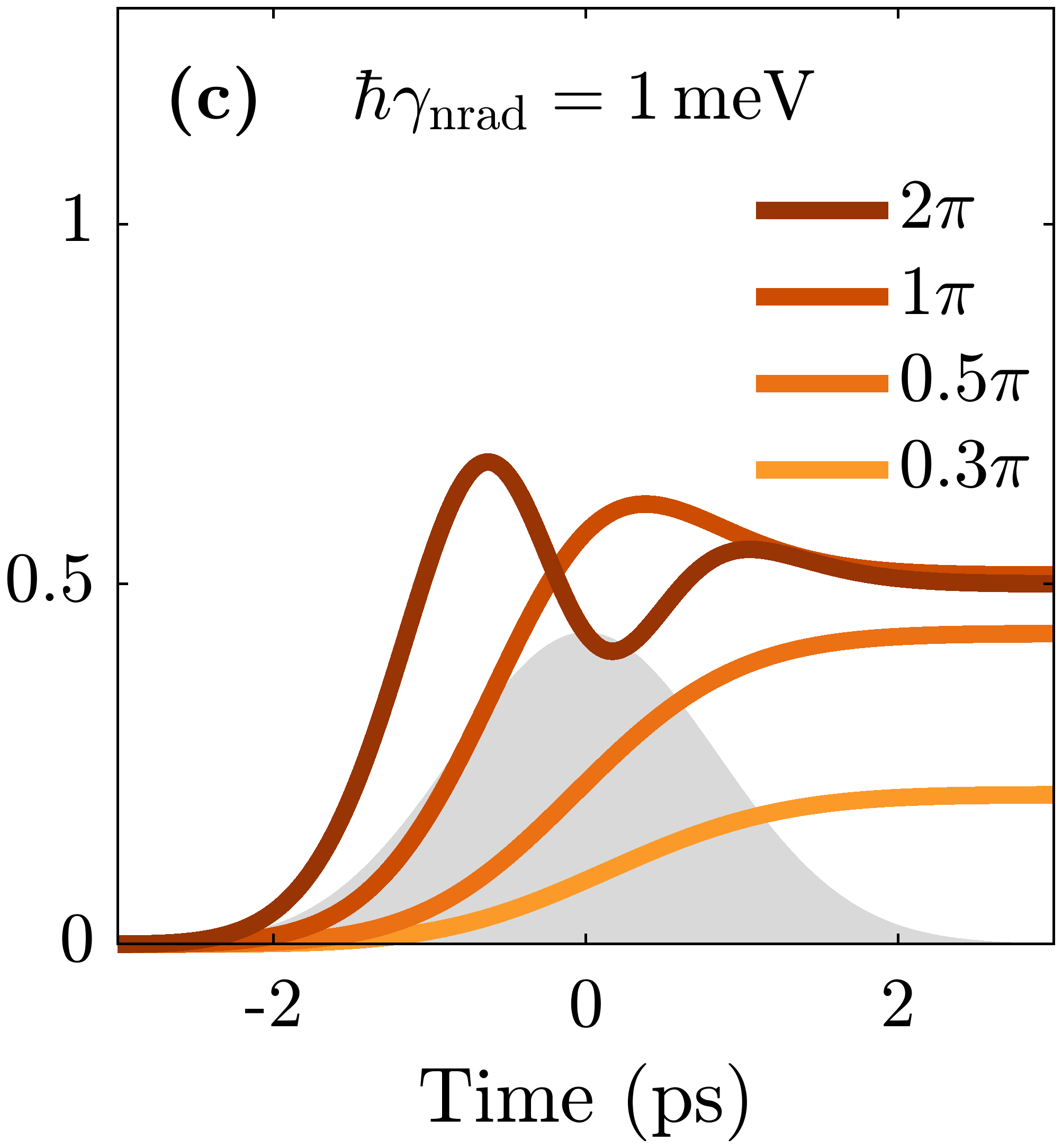}
        \end{tabular}
        &
        \hspace{-0.32cm}
        \begin{tabular}{c}
            \includegraphics[height=0.479\linewidth]{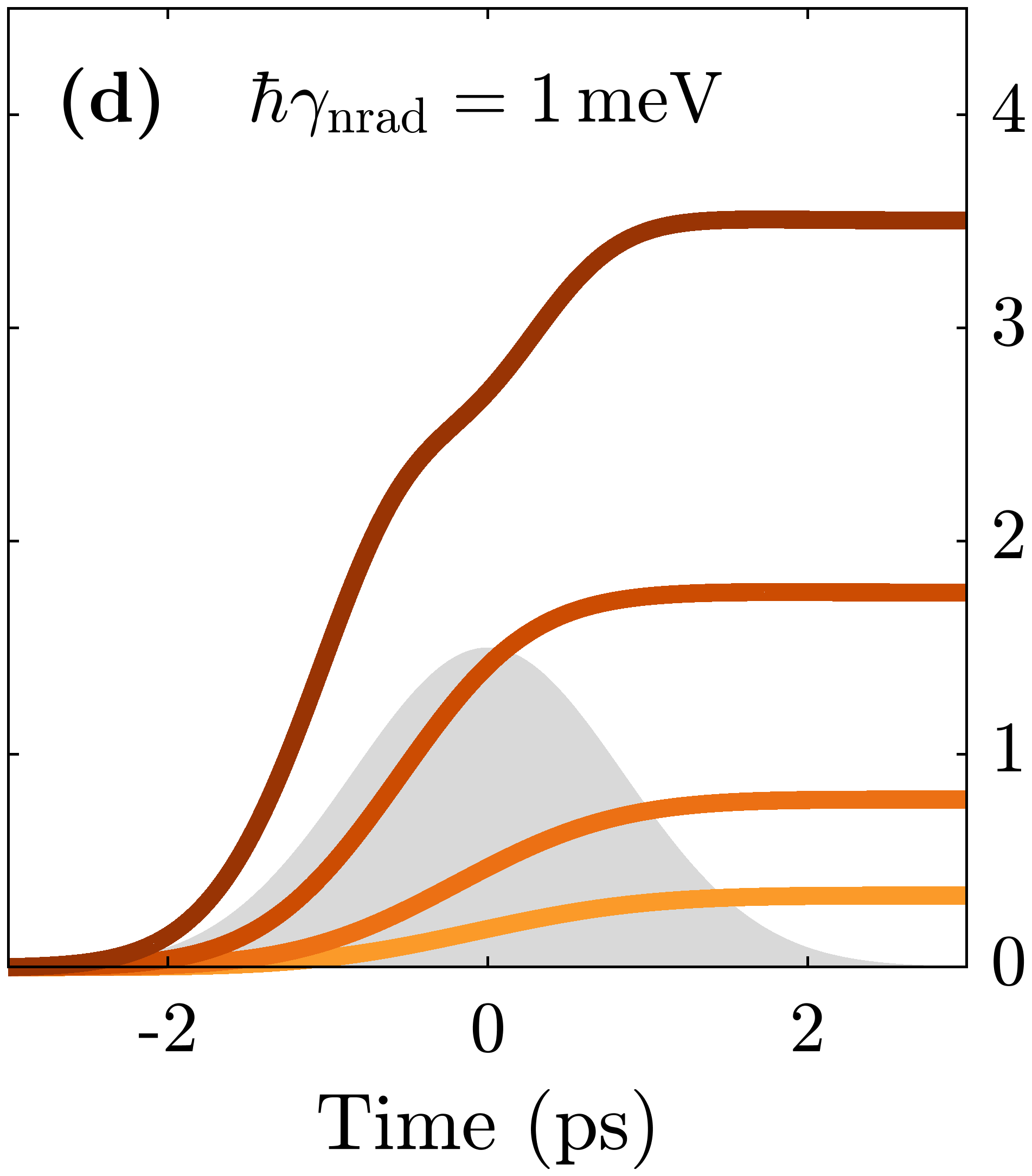}
        \end{tabular}
        \end{tabular}
     &
     \begin{tabular}{c}
     \begin{turn}{-90} Density ($10^{10}\,$cm$^{-2}$)\end{turn}
     \end{tabular}
    \end{tabular}
    \caption{Dynamics of the excited occupation of single two-level system as a solution of the single-state limit of the SBE without Coulomb interaction, the optical Bloch equations \cite{allen2012optical} (a,c) and dynamics of the electron density $N$ from Eq.~\eqref{eq:N_TOT_SBE} as a solution of the SBE without Coulomb interaction (b,d) for a 8-nm GaAs QW for various pulse areas. The grey-shaded area depicts the normalized pulse intensity.}
    \label{fig:dynamics_tls_sbenocoulomb}
\end{figure}

We note that, in general, excitonic Rabi oscillations in semiconductors are -- in contrast to an undamped ideal two-level system, cf.~Fig.~\ref{fig:dynamics_tls_sbenocoulomb}(a) -- never fully modulated, even if we neglect any damping, cf.~Fig.~\ref{fig:dynamics_tls_sbenocoulomb}(b). This behavior already occurs without any Coulomb interaction and is due to the fact, that exciting a semiconductor at the band gap with $\mathbf k=\mathbf 0$ creates a multitude of electron-hole pairs, i.e., an ensemble of two-level systems, which experience a slightly different Rabi energy each with increasing difference at increasing quasi-momentum $\mathbf k$. The Rabi-flopping dynamics of the resulting total electron density are then governed by the average over many two-level systems oscillating with slightly different Rabi frequencies causing a superposition resulting in a partial modulation of the oscillations only. In case of nonzero damping, cf.~Fig.~\ref{fig:dynamics_tls_sbenocoulomb}(c,d), Rabi oscillations in a semiconductor without Coulomb interaction are even more damped. In general, the Coulomb interaction not only retrieves the Rabi flops but also enhances their number by a factor of approximately two \cite{schulzgen1999direct,binder1999many}, cf.~Fig.~\ref{fig:dynamics_exc}, due to excitonic effects on Hartree-Fock level. However, a full modulation of Rabi-oscillations due to two-level-like Pauli-blocking -- even for strict resonant excitation of the exciton -- cannot occur, since exciton-exciton interactions obscure the Pauli-blocking property.

All in all, our excitonic theory reproduces the key physics encoded in the semiconductor Bloch equations, which have been proven to reliably model Rabi flops in ultrafast experiments in the moderate Coulomb-interaction regime (GaAs QW) \cite{schulzgen1999direct,binder1999many,cundiff1994rabi}, while showing major deviations from the SBEs due to exciton-exciton interaction in the strong Coulomb-interaction regime (MoSe$_2$ ML). Here, in circular excitation, Rabi oscillations in the incoherent exciton density are effectively suppressed, while a few Rabi flops in the coherent exciton density can be eventually observed, if the linewidth does not exceed $2\hbar\gamma_{\text{nrad}} = 2\,$meV.\\

\begin{figure}[h!]
    \centering
    \begin{tabular}{c}
        \begin{tabular}{c}
            \includegraphics[width=0.95\linewidth]{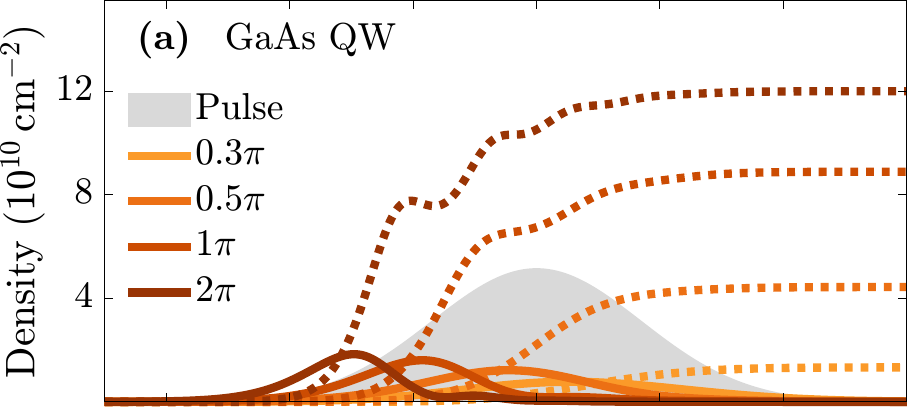}
        \end{tabular}\\[-0.5mm]
        \begin{tabular}{c@{\hspace{-0.13cm}}}
        \includegraphics[width=0.936\linewidth]{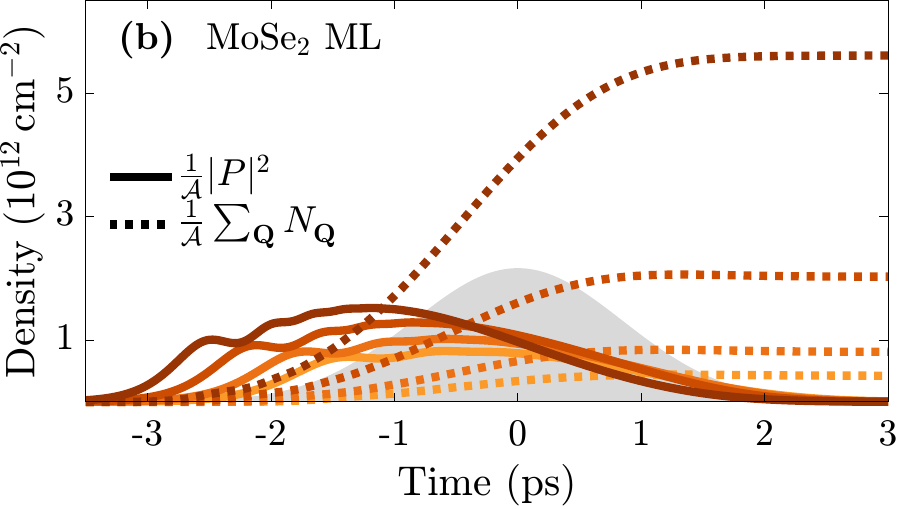}
        \end{tabular}
    \end{tabular}
    \caption{Dynamics of the coherent exciton density $\frac{1}{\mathcal A}|P|^2$ (solid lines) and the incoherent exciton density $\frac{1}{\mathcal A}\sum_{\mathbf Q}\Nex{\mathbf Q}$ (dotted lines) for the full simulations of a GaAs QW and a h-BN-encapsulated MoSe$_2$ ML from Fig.~\ref{fig:dynamics_exc}(e,j).}
    \label{fig:dynamics_exc_P2_N}
\end{figure}

\textit{Coherent vs.\ incoherent exciton density:} 
To examine the different dynamical behavior of the GaAs QW and the MoSe$_2$ ML in more detail, we depict the individual coherent ($|\Pol{1s}|^2$, solid lines) and incoherent ($\Nex{1s,\mathbf Q}$, dotted lines) contributions in Fig.~\ref{fig:dynamics_exc_P2_N}. It becomes clear, that the Rabi flopping in a GaAs QW occurs in the incoherent part of the total excitonic density $\frac{1}{\mathcal A}\sum_{\mathbf Q}\Nex{1s,\mathbf Q}$, while in a MoSe$_2$ ML, it occurs in the coherent part of the total exciton density $\frac{1}{\mathcal A}|\Pol{1s}|^2$. This is a consequence of the weaker Coulomb interaction in the GaAs QW, %compared to the MoSe$_2$ ML, 
which does not significantly suppress fermionic Pauli-blocking effects in Eq.~\eqref{eq:P_EOM_CORR_third_order} (last line) leading to an efficient formation of incoherent, optically dark excitonic occupations $\Nex{1s,\mathbf Q}$ in fourth order, cf.~Eq.~\eqref{eq:N_EOM_opt_COH_fourth_order} and Eq.~\eqref{eq:N_EOM_opt_CORR_fourth_order}. In a MoSe$_2$ ML, due to the stronger Coulomb interaction, fermionic Pauli-blocking effects are reduced, while, at the same time, the Coulomb-induced field renormalization, i.e., the excitonic Rabi energy, is enhanced. This leads to a regime, where Rabi oscillations preferably occur in the coherent exciton density and don't translate to the incoherent, optically dark exciton density.
\\

\begin{figure}[h!]
    \centering
    \subfigure[]{\includegraphics[width=0.9\linewidth]{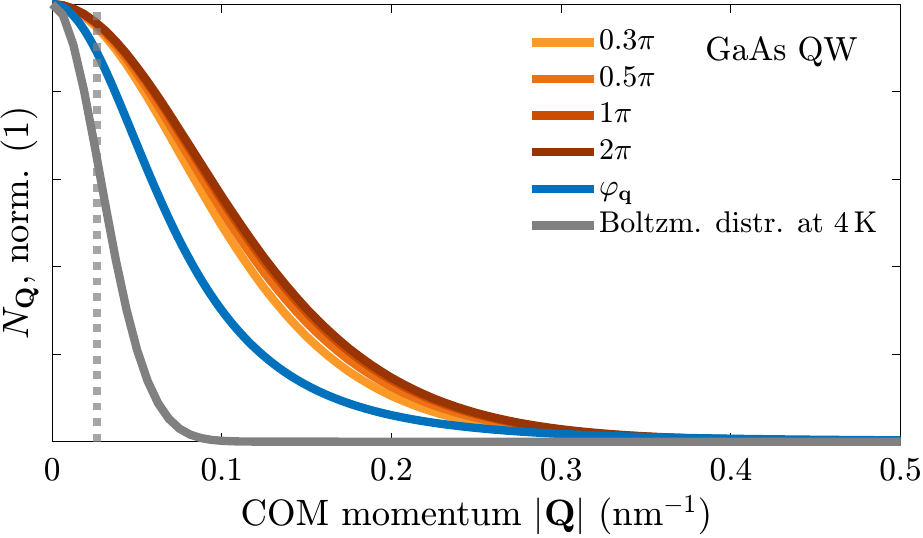}}
    \subfigure[]{\includegraphics[width=0.9\linewidth]{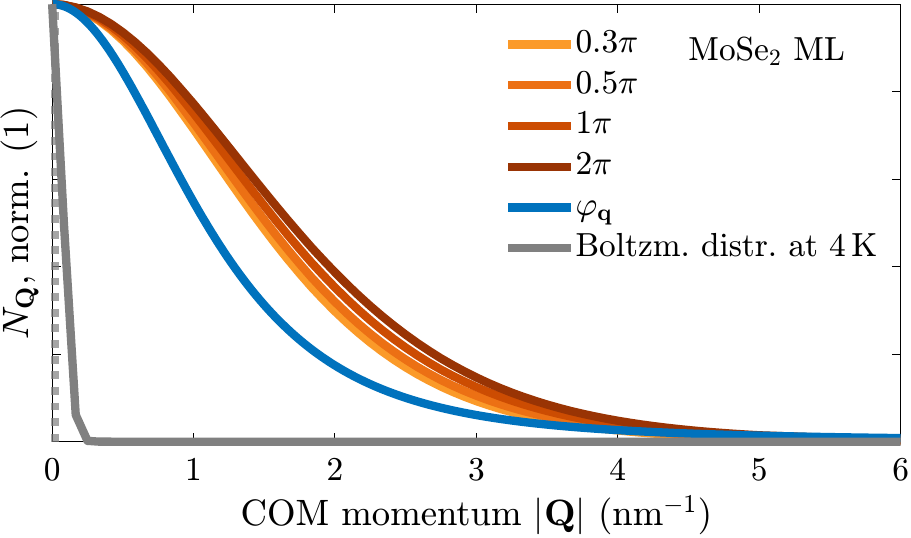}}
    \caption{Momentum distribution of the incoherent excitonic occupation $\Nex{1s,\mathbf Q}$ of a GaAs QW (a) and a h-BN-encapsulated MoSe$_2$ ML (b) after optical excitation at various pulse areas. Vertical dashed lines denote the light cone.}
    \label{fig:excitonic_occupations_q_distr}
\end{figure}

\textit{Momentum distribution of optically dark excitons:} 
In Fig.~\ref{fig:excitonic_occupations_q_distr}, we depict the momentum distribution of the coherently excited excitonic occupations $\Nex{1s,\mathbf Q}$ for a GaAs QW and a MoSe$_2$ ML. Similar to the fully bosonic theory in Ref.~\cite{yang2004ultrafast}, our theory predicts a coherent excitation of excitonic occupations at center-of-mass momenta well outside the light cone (grey dotted vertical line). At increasing pulse areas, the distribution becomes slightly broader. While the distribution in a GaAs QW is well localized in momentum space with an energy width of approximately 45\,K, % and resembles a "cold" Boltzmann distribution of a temperature of 45\,K,
the distribution in a MoSe$_2$ ML is broad with an energy width of approximately 1500\,K (note the different COM-momentum scale). %and resembles a "hot" Boltzmann distribution of 1500\,K.
This difference is a direct consequence of the different Coulomb interaction, cf.\ also the corresponding excitonic wave function $\ExWF{\mathbf q}$ (blue solid line): A stronger Coulomb interaction enhances the individual electron and hole collisions due to the fermionic substructure in the exciton gas, leading to broader distributed coherently excited excitonic occupations.
\\

\begin{figure}[h!]
    \centering
    \begin{tabular}{cc}
    \begin{tabular}{c@{\hspace{-0.1cm}}}
     \begin{turn}{90} Density ($10^{12}\,$cm$^{-2}$)\end{turn}
     \end{tabular}
     &
     \begin{tabular}{c@{\hspace{0.0cm}}}
        \begin{tabular}{c}
            \includegraphics[width=0.86\linewidth]{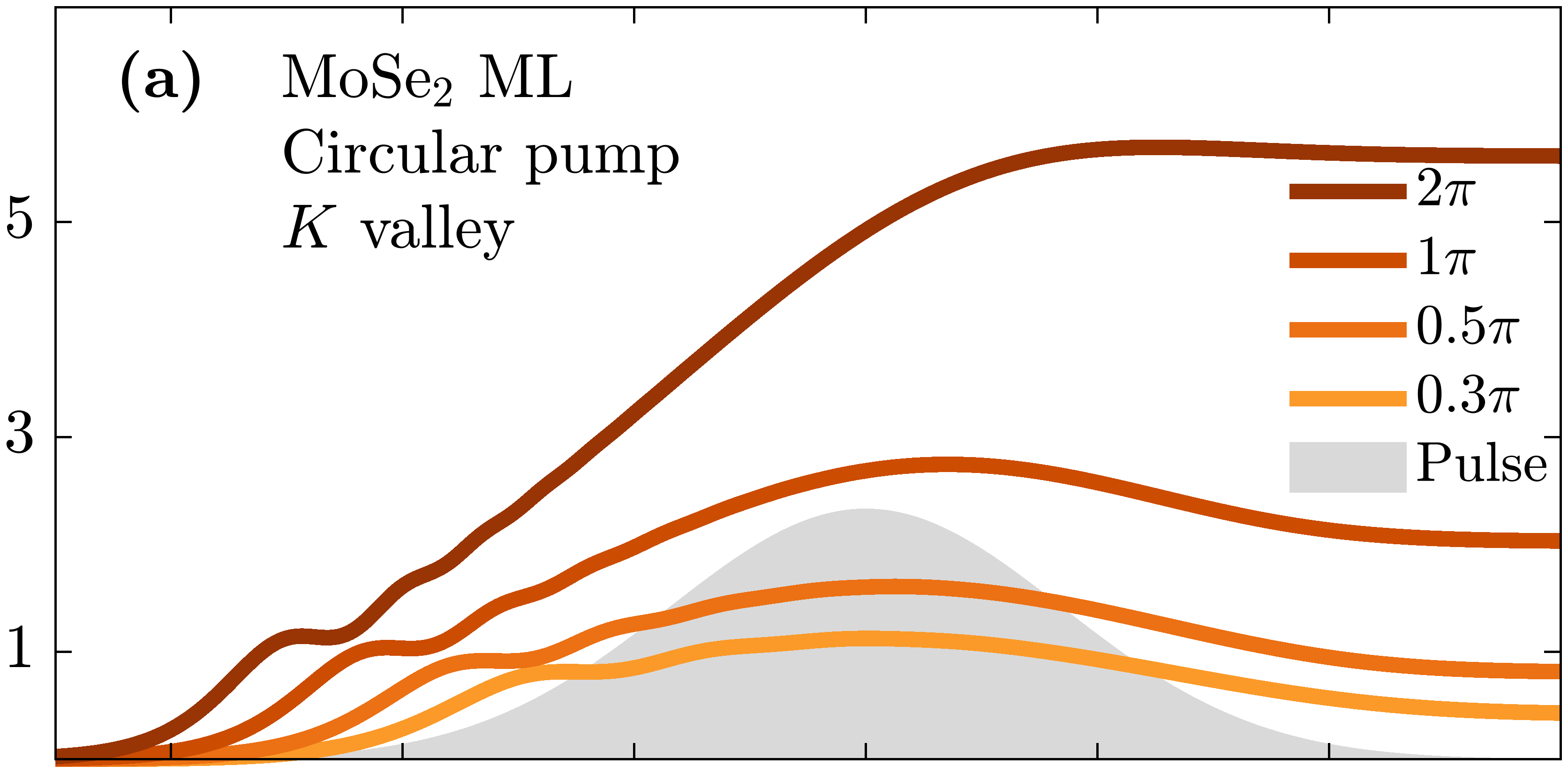}
        \end{tabular}\\[-1.5mm]
        \begin{tabular}{c@{\hspace{0.02cm}}}
        \includegraphics[width=0.8665\linewidth]{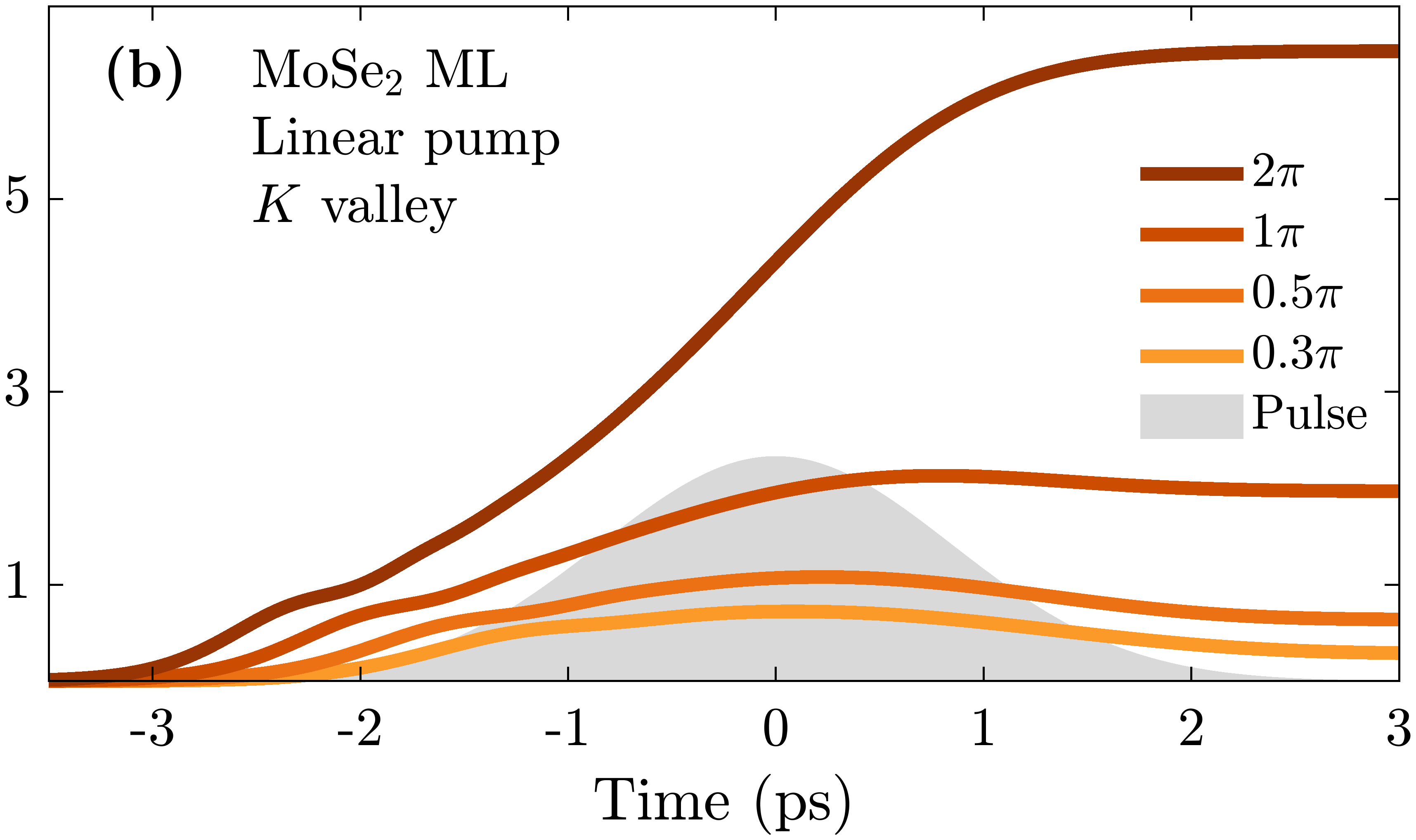}
        \end{tabular}
    \end{tabular}
    \end{tabular}
    \caption{Dynamics of the total exciton density $N = \frac{1}{\mathcal A}|\Poltwo{1s}{K,\uparrow}|^2+\frac{1}{\mathcal A}\sum_{\mathbf Q}\Nextwo{1s,\mathbf Q}{K,\uparrow,K,\uparrow}$ at the $K$ valley for circular excitation (a) and linear excitation (b) in a h-BN-encapsulated MoSe$_2$ ML.}
    \label{fig:dynamics_exc_circ_lin}
\end{figure}

\textit{Circular vs.\ linear excitation:} 
In Fig.~\ref{fig:dynamics_exc_circ_lin}, we depict the dynamics of the total exciton density at the $K$ valley from Eq.~\eqref{eq:N_TOT_EXC}, i.e., $\xi=\xi^{\prime} = K,\uparrow$, for circular (a), cf.\ also Fig.~\ref{fig:dynamics_exc}(j), and linear (b) excitation. For the linear-excitation case, we include coherent transitions at the $K$ and $K^{\prime}$ valleys $\Poltwo{1s}{K}$, $\Poltwo{1s}{K^{\prime}}$ as well as intravalley incoherent occupations $\Nextwo{1s,\mathbf Q}{K,K}$, $\Nextwo{1s,\mathbf Q}{K^{\prime},K^{\prime}}$ and neglect any intervalley occupations $\Nextwo{1s,\mathbf Q}{K,K^{\prime}}$/$\Nextwo{1s,\mathbf Q}{K^{\prime},K}$. While in circular excitation, some Rabi flops of the coherent exciton density remain, cf.\ Fig.~\ref{fig:dynamics_exc_P2_N} and the discussion above, in linear excitation, even the Rabi flops in the coherent exciton density are almost totally suppressed. The reason is the larger two-exciton/exciton-two-exciton phase space: In circular excitation, only symmetric ($+$) intravalley two-exciton continua $\Bextwo{+,\zeta}{K,K,K,K}$ and exciton-two-exciton continua $\Rextwo{+,\zeta,\mathbf Q}{K,K,K,K,K,K}$ are induced. In linear excitation, the $K$ valley and the $K^{\prime}$ valley are equally excited, so that additional symmetric ($+$) and antisymmetric ($-$) intervalley two-exciton continua $\Bextwo{\pm,\zeta}{K,K^{\prime},K^{\prime},K}$/$\Bextwo{\pm,\zeta}{K^{\prime},K,K,K^{\prime}}$ and exciton-two-exciton continua $\Rextwo{\pm,\zeta,\mathbf Q}{K^{\prime},K^{\prime},K,K,K^{\prime},K^{\prime}}$/$\Rextwo{\pm,\zeta,\mathbf Q}{K,K,K^{\prime},K^{\prime},K,K}$ are induced. These yield a dynamic coupling between the excitonic transitions at the $K$ valley $\Poltwo{1s}{K}$ and at the $K^{\prime}$ valley $\Poltwo{1s}{K^{\prime}}$, hence further increasing the excitation-induced dephasing. 
Moreover, the formation of a bound ($b$) biexciton $\Bextwo{-,\zeta=b}{K,K^{\prime},K^{\prime},K}$/$\Bextwo{-,\zeta=b}{K^{\prime},K,K,K^{\prime}}$ and bound exciton-biexciton $\Rextwo{-,\zeta=b,\mathbf Q}{K^{\prime},K^{\prime},K,K,K^{\prime},K^{\prime}}$/$\Rextwo{-,\zeta=b,\mathbf Q}{K,K,K^{\prime},K^{\prime},K,K}$ becomes possible, further enhancing the excitation-induced dephasing. Hence, due to the overall larger number of Coulomb-interaction-induced exciton-exciton correlations and the resulting dynamic coupling of the excitonic transitions $P$ between the $K$ and $K^{\prime}$ valley, the excitation-induced dephasing is larger in linear excitation compared to circular excitation causing a suppression of the Rabi oscillations in the coherent exciton density. 
This theoretical finding is consistent with recent experiments in a MoSe$_2$ ML, where a clear Rabi splitting due to an occupation grating, which is insensible to coherent excitation-induced dephasing, has been measured, but no Rabi oscillations could be observed \cite{schafer2025distinct}.
\\

%\FloatBarrier

\section{Conclusion}
In this work, we developed and discussed a theory for the ultrafast dynamics of Wannier excitons valid from the low-density regime up to an elevated-density regime below the Mott transition, which is unleashed on two distinct materials as proponents of different Coulomb-to-light-matter-interaction regimes: A GaAs QW with moderate Coulomb interaction and a MoSe$_2$ ML with strong Coulomb interaction compared to light-matter interaction.

It has been found, that the excitonic approach reproduces the key physics described by the semiconductor Bloch equations well, which have been proven to reliably model Rabi flops in ultrafast experiments in the moderate Coulomb-interaction regime (GaAs QW) \cite{schulzgen1999direct,binder1999many,cundiff1994rabi}, while showing major deviations from the SBEs in the strong Coulomb-interaction regime (MoSe$_2$ ML) due to exciton-exciton interaction induced by Coulomb-correlated doublets and triplets, in line with recent experiments \cite{schafer2025distinct}. Here, in circular excitation, Rabi oscillations are considerably less strong and vanish almost completely in linear excitation. Our findings strongly indicate, that an excitonic approach has to be chosen to simulate the non-linear response of confined semiconductors, as long as the excitation conditions are below the Mott transition and correlated electron-hole pairs dominate. On the other hand, if the excitation conditions are above the Mott transition \cite{erben2022optical,wang2019optical,lohof2018prospects,steinhoff2017exciton,article:EXP_low_density_limit_FoglerNovoselov2014}, where uncorrelated electrons and holes dominate, the expansion in correlated electron-hole pairs breaks down and the semiconductor Bloch equations need to be applied.

%\acknowledgments

\begin{acknowledgments}
%\section*{Acknowledgments}
Funded by the Deutsche Forschungsgemeinschaft (DFG, German Research Foundation) -- Project No.\ 420760124 (H.M., A.K.); 556436549 (A.K.). H.M.\ acknowledges funding by project 21209528 (``proof of trust'').

H.M.\ and A.K.\ thank Felix Schäfer, Markus Stein and Sangam Chatterjee (Justus-Liebig-Universität Gießen) for fruitful discussions and their measurements, which strongly motivated the theory developed here.

\end{acknowledgments}

\section*{Data Availability Statement}
The data supporting the findings of this work are available from the authors within reasonable request.

\FloatBarrier
%\clearpage
%\onecolumngrid
%\changepage{0cm}{0cm}{0cm}{}{}{}{}{}{}
%\setlength{\textwidth}{15.5cm}
%\setlength{\textwidth}{17.5cm}
%\setlength{\linewidth}{17.5cm}
%\clearpage
%\newgeometry{textwidth=18cm, left = 1.5cm} 
\bibliography{bibliography_full}
\end{document}